\documentclass[11pt,a4paper]{article}
\pdfoutput=1
\usepackage{jheppub}
\usepackage{rotating}
\usepackage{array}
\usepackage{amsmath,amssymb,amstext}
\usepackage[normalem]{ulem}
\usepackage{slashed}
\usepackage{booktabs}
\usepackage[pdftex,table,dvipsnames]{xcolor}
\usepackage{units}
\usepackage{multirow}
\usepackage{url}
\usepackage{color}
\usepackage{xspace}
\usepackage{comment}
\usepackage{enumitem}
\usepackage{multirow}
\preprint{TTK-21-12}

\title{Autoencoders for unsupervised anomaly detection in high energy physics}

\author{Thorben Finke,}
\author{Michael Kr\"amer,}
\author{Alessandro Morandini,}
\author{Alexander M\"uck and}
\author{Ivan Oleksiyuk}

\affiliation{Institute for Theoretical Particle Physics and Cosmology (TTK),\\ RWTH Aachen University, D-52056 Aachen, Germany}

\emailAdd{finke@physik.rwth-aachen.de}
\emailAdd{mkraemer@physik.rwth-aachen.de}
\emailAdd{morandini@physik.rwth-achen.de}
\emailAdd{mueck@physik.rwth-aachen.de}
\emailAdd{ivan.oleksiyuk@rwth-aachen.de}

\abstract{Autoencoders are widely used in machine learning applications, in particular for anomaly detection. Hence, they have been introduced in high energy physics as a promising tool for model-independent new physics searches. We scrutinize the usage of autoencoders for unsupervised anomaly detection based on reconstruction loss to show their capabilities, but also their limitations. As a particle physics benchmark scenario, we study the tagging of top jet images in a background of QCD jet images. Although we reproduce the positive results from the literature, we show that the standard autoencoder setup cannot be considered as a model-independent anomaly tagger by inverting the task: due to the sparsity and the specific structure of the jet images, the autoencoder fails to tag QCD jets if it is trained on top jets even in a semi-supervised setup. Since the same autoencoder architecture can be a good tagger for a specific example of an anomaly and a bad tagger for a different example, we suggest improved performance measures for the task of model-independent anomaly detection. We also improve the capability of the autoencoder to learn non-trivial features of the jet images, such that it is able to achieve both top jet tagging and the inverse task of QCD jet tagging with the same setup. However, we want to stress that a truly model-independent and powerful autoencoder-based unsupervised jet tagger still needs to be developed.}

\begin{document}
\maketitle
\section{Introduction}
Having discovered the Higgs boson in 2012, the experimental foundation of the Standard Model of particle physics has been completed at the Large Hadron Collider (LHC)~\cite{Aad:2012tfa,Chatrchyan:2012ufa}. However, working at the energy frontier, new particles and interactions, not covered by the Standard Model, could be measured by accumulating more and more statistics, or may already hide in the huge amount of data taken by the LHC experiments. From a machine learning perspective, the search for new physics is a quest for finding anomalous data, usually called signal in the physics context, in the vast background of collider events described by Standard Model interactions. New physics would contribute the out-of-distribution data with respect to the data expected in the Standard Model. Hence, it is not surprising that the recent advances in machine learning have also had a huge impact
in the LHC context~\cite{Feickert:2021ajf,Schwartz:2021ftp,Bourilkov:2019yoi,Guest:2018yhq,Albertsson:2018maf,Larkoski_2020}. In addition to improving classical search strategies, machine learning may even open up completely new ways to look for anomalous data, i.e.\ new physics, in a model-independent way. 
\par
Single collider events, i.e.\ data instances, usually cannot simply be labeled as signal and background due to the intrinsically probabilistic nature of quantum mechanics. However, identifying so-called reducible backgrounds, which share some but not all features of the signal events, amounts to a straightforward classification task, where one can try to either employ supervised or unsupervised machine learning techniques to separate the two classes. 
\par
Supervised classifiers are extremely powerful tools but have a limited applicability since labeled data has to be available. In the collider-physics context, they cannot be directly applied to experimental data since collider data is not labeled. However, labeled data is available from the sophisticated simulation of collider events. If the transition from simulation to measured data is understood well enough, supervised machine learning can be successfully applied in particle physics. Still, traditional supervised algorithms are always model-dependent and can only efficiently detect the signals simulated for training. Interesting ideas to overcome those shortcomings have been recently discussed in the literature \cite{Dery:2017fap, Cohen:2017exh, Metodiev:2017vrx, Komiske:2018oaa, Borisyak:2019vbz, Amram:2020ykb, Lee:2019ssx}.
\par  
In unsupervised machine learning, an algorithm does not learn from labeled examples but should understand the structure of the data in some other way in order to sort it into a fixed or variable number of classes. Using unsupervised methods, which can be directly applied to LHC data, is a much more difficult but also an even more rewarding task. Designing an unsupervised tagger which is able to detect anomalies independently of the new physics model is certainly the ultimate vision.
\par  
For anomaly or outlier detection~(see e.g.\ \cite{Ruff_2021, chalapathy2019deep} for reviews), it is usually assumed that only a few anomalous data instances are to be found in the data distribution. In semi-supervised machine learning, which is slightly less ambitious, a data sample consisting only of the background class is provided during training. The task is to tag signal data when testing on a mixed sample. This might be also an option in particle physics applications since signal free data from control regions might be available. Many different unsupervised and semi-supervised algorithms have been applied in the physics context~(see e.g. \cite{Nachman:2020ccu, Kasieczka:2021xcg} for reviews).
\par
One particularly promising unsupervised method for anomaly detection is based on the autoencoder (AE) architecture~\cite{BALDI198953}. An AE consists of two parts: an encoder and a decoder. The encoder is a neural network that compresses the input data into a few latent space variables which are also often called the bottleneck. The decoder is a neural network which reconstructs the initial data from the latent space variables. By choosing a suitable loss function both parts are trained together as one neural network to reconstruct the input data as well as possible.
\par
The main idea is to train an AE on a dataset consisting purely (for semi-supervised) or mostly (for unsupervised anomaly detection) of the background class. By having much fewer variables in the latent space than in the input and output layers, the AE is not able to learn the identity transformation. Instead, to minimize the loss, it is forced to extract in its bottleneck the variables that correspond to the most prominent features of the background class. Ideally it extracts correlations in the input data which allow for an efficient data compression. The AE learns a representation that uses the structure of the training data and is therefore specific for this set. Hence, if an AE encounters data that has features somewhat different from the background class, it should not be able to effectively encode and decode these features. The loss of the reconstruction is expected to be larger. Therefore, a trained AE can be used as an anomaly tagger with its loss function as the anomaly score. This simple idea and its variants have been successfully explored in different machine learning applications (see e.g. ~\cite{bengio2013representation, pang2020deep,Ruff_2021} for reviews). Also in particle physics autoencoders have been successfully used for anomaly detection \cite{Hajer_2020, Crispim_Rom_o_2021, alexander2020decoding, Blance_2019, Cerri_2019,cheng2021variational, bortolato2021bump}, in particular for top tagging~\cite{Heimel_2019,Farina_2020, Roy:2019jae} which will be the benchmark application in this work.
\par
An unsupervised or semi-supervised machine learning algorithm for anomaly detection has an advantage compared to supervised methods if it is as model-independent as possible. In particular it should not be tailored for a specific kind of anomaly, i.e.\ a specific new physics model, but it should work for any, or at least a wide variety of possible signals. Ideally, it should also be able to detect an unexpected new physics signature. 
\par
Specifically, the setup should also be a working tagger if the background data and the anomalous signal interchange their role. This offers a simple test for model-independence. If such an inverse tagger performs significantly worse (or better) than the original tagger, there is a bias in the setup favoring a particular kind of anomaly. This bias will limit the model-independence of the tagging capabilities and, hence, the usefulness of a given unsupervised machine learning method. In machine learning applications outside of particle physics this kind of bias has recently been investigated mostly in the context of anomaly tagging using deep generative models~\cite{nalisnick2019deep, schirrmeister2020understanding, kirichenko2020normalizing, ren2019likelihood, serra2019input, tong2020fixing}. Once an algorithm is known to work as an anomaly tagger in specific examples, it is a natural next step to study and understand potential biases in order to evaluate the performance and to improve the method.
\par
In this work, we explicitly perform this investigation for an autoencoder which is used as a semi-supervised binary classifier to find an unknown signal within a Standard Model background. As a well-known benchmark example~\cite{Almeida:2015jua, Kasieczka:2017nvn, Pearkes:2017hku,Macaluso:2018tck,  Kasieczka_2019, Araz:2021wqm}, we distinguish boosted top jets from QCD jets employing a simple convolutional autoencoder working on jet images. We confirm findings from the literature~\cite{Heimel_2019,Farina_2020} that an autoencoder can indeed find top jets as anomalies without having seen them during training. However, the ability to detect this particular anomaly does not imply that the autoencoder works in general. Indeed, in Section~\ref{sec:problem} we will show that the very same autoencoder fails on the inverse task of finding QCD jets as anomalies when being exposed only to top jets during training; it performs worse than picking anomalies randomly. Our anomaly tagging setup employing the autoencoder is strongly biased to label top jets as anomalous no matter what the AE has seen during training. To understand this behavior we investigate what the autoencoder actually learns and why it fails under certain circumstances. It will become clear why a functional autoencoder can be a bad anomaly tagger and vice versa, in particular in the particle physics application at hand. Using those insights, we propose several improvements for our particular autoencoder architecture, such that it works as an anomaly tagger in both ways to distinguish QCD and top jets.
\par
This work is structured as follows: After defining our setup in Section~\ref{sec:framework}, we scrutinize the autoencoder performance in Sections~\ref{sec:ignored} and \ref{sec:simplicity_bias}, and discuss its tagging performance in Section~\ref{sec:tagging_performance}. In Sections~\ref{subs:preproc} and \ref{subs:FIL} we introduce improved data preprocessings and loss functions. We discuss the impact of these improvements on the autoencoder performance in general and on the tagging performance in particular in Sections~\ref{sec:AE_performance} and \ref{sec:Tag_performance}, respectively. We conclude in Section~\ref{sec:conclusion} and present additional material in the appendices.

%\section{Jet data and autoencoder architecture}
%\input{framework.tex}
\section{Autoencoder limitations}
\label{sec:problem}

In this section, we introduce our AE architecture and investigate its training and performance. We either train the AE on a pure sample of QCD jets and call it a direct tagger, or we train the AE on a pure sample of top jets and call it an inverse tagger. While the former setup is designed to perform the well-known task of tagging top jets as anomalies, the latter setup is designed to perform the inverse task, i.e.\ tagging QCD jets as anomalies in a background sample of top jets. Hence, we always use a semi-supervised setup, i.e.\ we assume that a dataset consisting only of background data is available.

Before we discuss the tagging performance in Section~\ref{sec:tagging_performance}, we scrutinize how and to what extent the AE actually learns to reconstruct jet images during training and how these AE capabilities affect its performance as a model-independent tagger. In particular, we are able to explain the success of the direct tagger and the failure of the inverse tagger by the interplay between an insufficient AE performance and the different complexity in the images of the two jet classes.

\subsection{Jet data and autoencoder architecture}
\label{sec:framework}

Throughout this work, we use top tagging as an example, which is a well established benchmark in collider physics: Simulated QCD jets (initiated by light-flavor quarks or gluons) have to be distinguished from hadronically decaying boosted top-quarks which form single jets. We use the benchmark dataset from Ref.~\cite{Butter_2018}, which is publicly available at~\cite{kasieczka_gregor_2019_2603256}. A subset of 100k and 40k jets of one of the two classes is used for training and validation, respectively, and two 40k subsets of each type for testing.

In the following, we describe our default setup which will be modified and improved in Section~\ref{sec:Methods}. We employ the preprocessing introduced in Ref.~\cite{Macaluso:2018tck}. The four momenta of the jet constituents are converted into two-dimensional jet images with $40\times40$ pixels. The details of the jet generation and the preprocessing are discussed in Appendix~\ref{sec:appendix}. 
\begin{figure}[t]
    \centering
    \includegraphics[width=\linewidth]{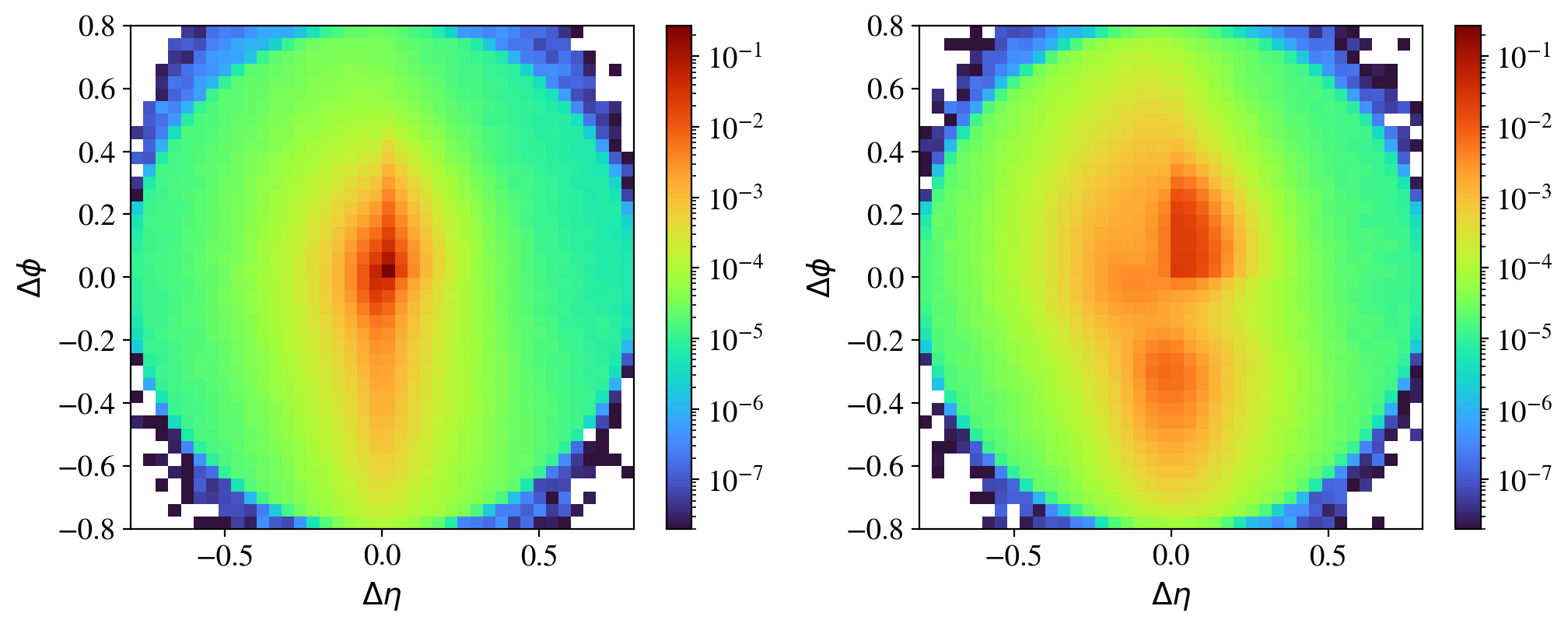}
    \caption[Caption for LOF]{Average of 40k QCD (left) and top jet images (right) in the test dataset, according to the standard preprocessing as described in Appendix~\ref{sec:appendix}.}
    \label{fig:average}
\end{figure}
Fig.~\ref{fig:average} shows the average images of both classes. For QCD jets, most of the intensity is concentrated in the central pixels. For top jets, there is a clearly visible three-prong structure (as expected for top-quark decays after preprocessing). Of course, individual jets are harder to distinguish than their average images may indicate.

As our anomaly detection algorithm, we use a convolutional autoencoder with an architecture similar to the one in Ref.~\cite{Heimel_2019}. We implement our AE with \textsc{Tensorflow 2.4.1}~\cite{tensorflow2015-whitepaper}, relying on the built in version of \textsc{Keras}~\cite{chollet2015keras}. Several convolution layers with $4\times 4$ kernel and average pooling layers with $2\times 2$ kernel are applied before the image is flattened and a fully connected network reduces the input further into the bottleneck latent space with 32 nodes. The Parametric ReLU activation function is used in all layers. The described encoder structure is inverted to form the corresponding decoder which is used to reconstruct the original image from its latent space description. Our architecture is defined in Fig.~\ref{fig:AE3_arch}; the hyperparameter settings are described in more detail in Appendix~\ref{sec:appendix}.

\begin{figure}[t]
    \centering
    \includegraphics[width=\linewidth]{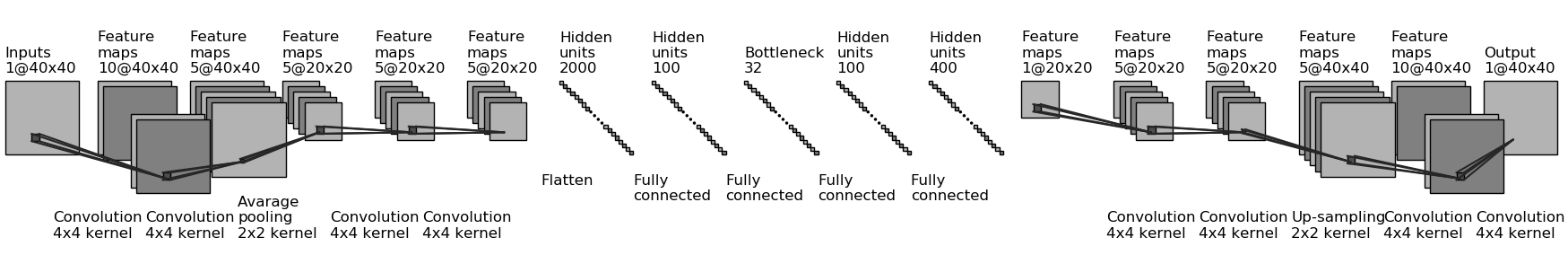}
    \caption[Caption for LOF]{Architecture of our autoencoder, see also Ref.~\cite{Heimel_2019}.}
    \label{fig:AE3_arch}
\end{figure}

Following Ref.~\cite{Heimel_2019}, to evaluate the reconstruction of the input picture we use the mean squared error (MSE), i.e.\ the average of the squared error of each reconstructed pixel with respect to its input value, as a loss function. During testing the value of the loss function is also used as the discriminator between signal and background. An event is tagged as signal/anomaly if the value of the loss function is larger than a given threshold. Changing the threshold value, one obtains the usual receiver operating characteristic (ROC) curve.

\subsection{Limited reconstruction}\label{sec:ignored}

\begin{figure}[t]
    \centering
    \includegraphics[width=\linewidth]{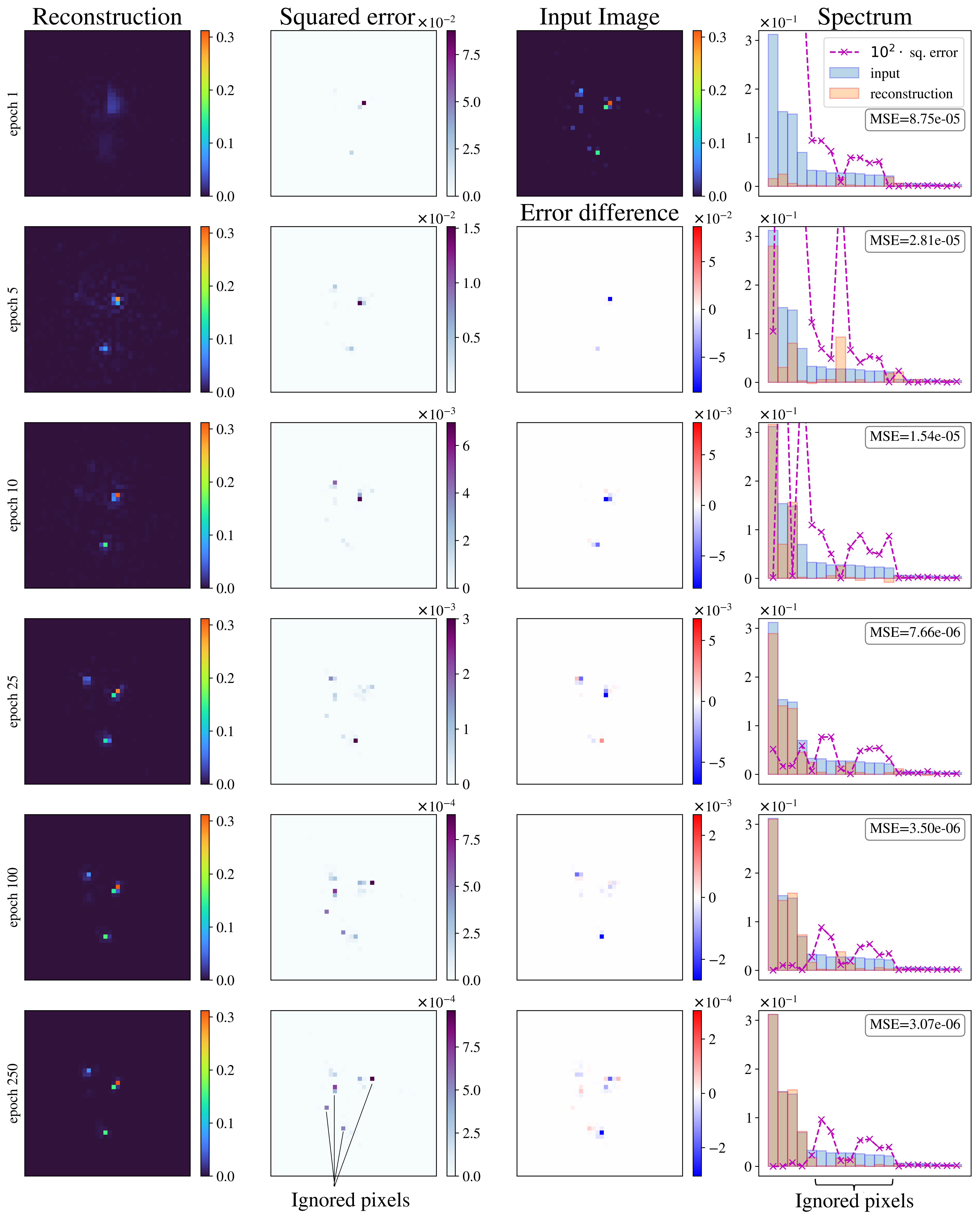}
    \caption{Reconstruction of an exemplary image (1st column) after 1, 5, 10, 25, 100, 250 (top to bottom) epochs of training. We also show the squared error per pixel between input and reconstructed image (2nd column) and its difference w.r.t.\ the previous row (3rd column). The 4th column shows the intensity of the 20 brightest input pixels (blue) together with the reconstructed intensity (orange) and the corresponding squared error (purple crosses).}
    \label{fig:traning1}
\end{figure}

We first investigate what the AE is actually learning, as it is trained for the reconstruction of the input and not as an anomaly tagger. Fig.~\ref{fig:traning1} shows the learning history of an exemplary top jet in terms of its reconstruction after training on top jets for a given number of epochs. To guide the eye, we also show the evolution of the squared error per pixel and highlight the reconstruction of the intensity of the brightest pixels in detail. Moreover, the MSE of the reconstruction, i.e.\ the loss function of the AE, is given as a measure for the overall reconstruction improvement.

During the first epoch, the AE learns to reconstruct the average top jet image, as can be seen by comparing the reconstructed image to the average image in Fig.~\ref{fig:average}. (Note that the images in Fig.~\ref{fig:average} are shown on a logarithmic scale.) In the following epochs, the squared error is dominated by the brightest pixels. The AE improves the loss by improving their reconstruction, so its trainable weights are updated accordingly. After 10 epochs, the AE recovers a smeared reconstruction of the brightest pixels. After roughly 25 epochs, the weights are learned to reconstruct the brightest pixels so well that the corresponding error becomes small compared to the error of the remaining pixels. However, the AE further improves the reconstruction of the brightest pixels (and the surrounding zero-intensity pixels) instead of changing its focus to the dimmer pixels. The latter would probably harm the previously learned reconstruction of the brightest pixels and therefore increase the loss. After around 100 epochs the four  brightest pixels are very well reconstructed, while most of the dimmer pixels are still ignored. The dimmer pixels dominate the total error, as can be seen in the right column of Fig.~\ref{fig:traning1}. The AE is apparently trapped in a local minimum of the loss function, and training longer does not change the picture significantly.

\begin{figure}[t]
\includegraphics[width=0.5\linewidth]{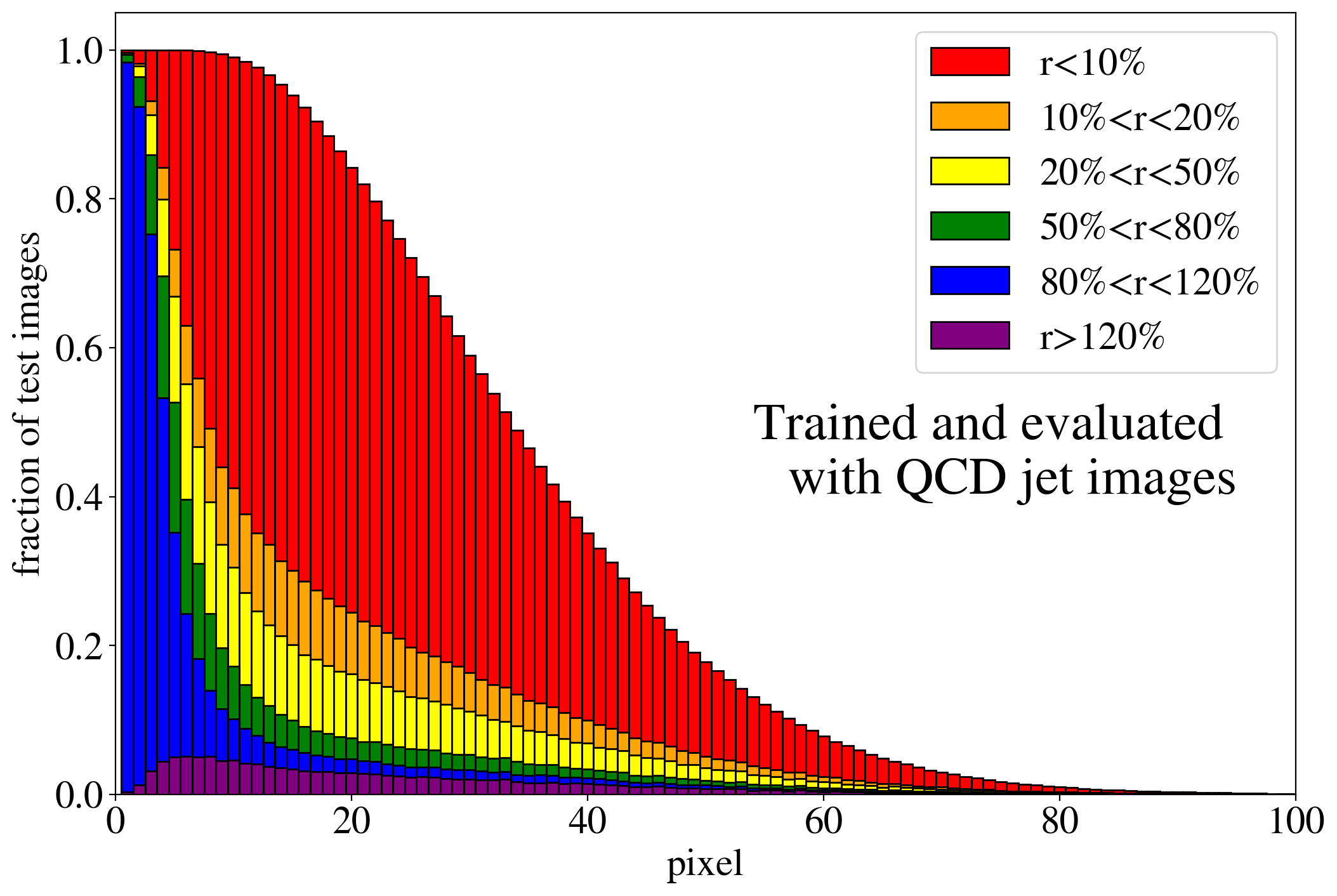}
\includegraphics[width=0.5\linewidth]{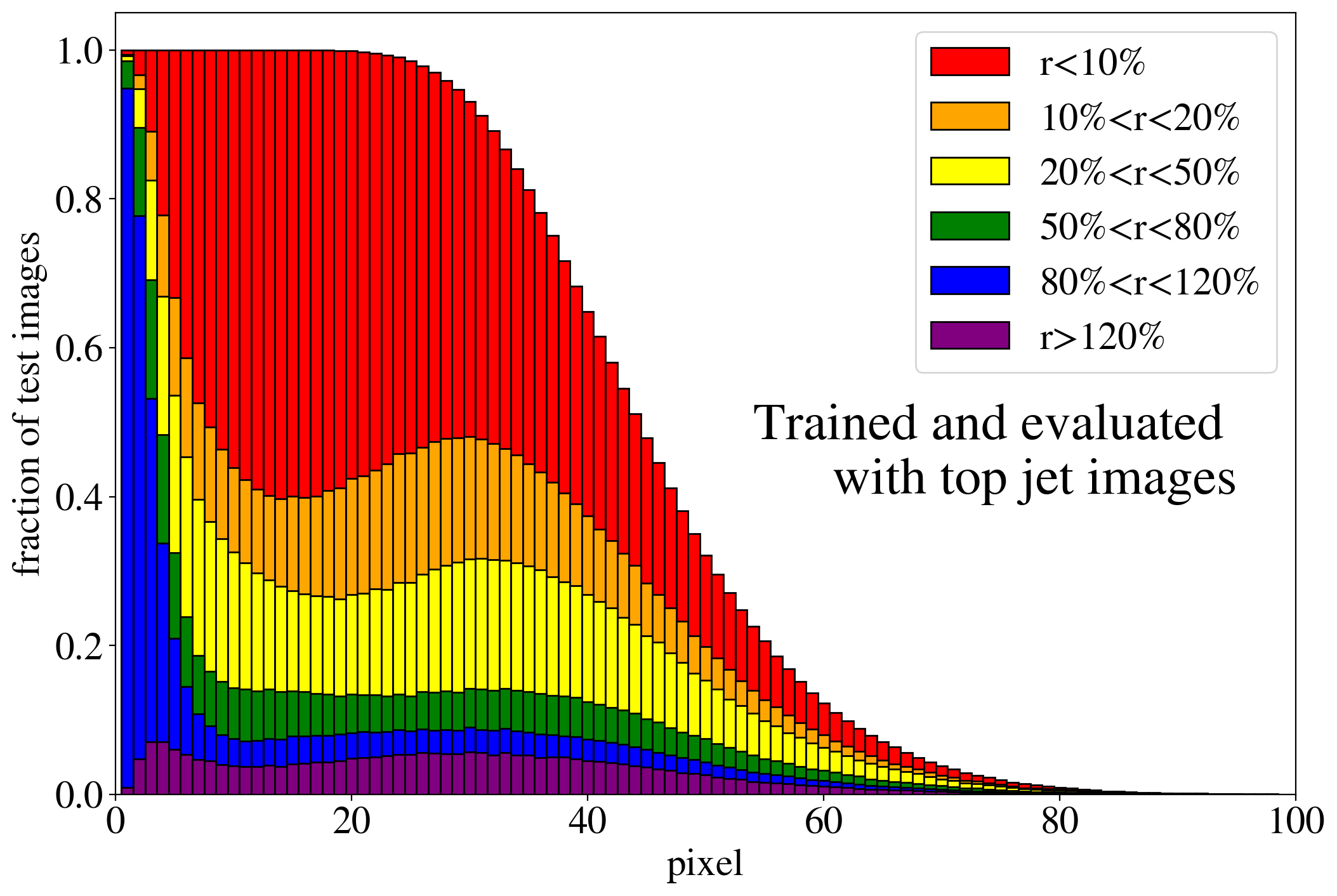}
\caption{Stacked histogram for different categories of the ratio $r$ of the reconstructed and the input intensity of the non-zero pixels. Pixels are ordered by intensity from left to right for each of the 40k test jets. We show results for both QCD jets (left) and top jets (right), where the AE has been trained on the corresponding training set.} \label{fig:igno_pix}
\end{figure}

The previous discussion applies to both training on QCD and top images. To show that the example presented in Fig.~\ref{fig:traning1} is rather generic, the reconstruction capabilities of the AE for the whole test dataset are summarized in Fig.~\ref{fig:igno_pix}, where we quantify the quality of the reconstruction for individual pixels of a jet image. Specifically, for each of the 40k jets of the test data sample, we determine the ratio of the reconstructed and input intensities, $r$, for each pixel, and show the fraction of test jet images where the brightest, next-to-brightest etc.\ pixel (from left to right on the horizontal axis) is reconstructed with a certain quality $r$. In Fig.~\ref{fig:igno_pix} (left) this fraction is shown for the AE trained and tested on QCD jets. For the majority of the jet images, the brightest pixels are reconstructed well (blue histogram, corresponding to a ratio $80\% \le r \le 120\%$). On the other hand, the dimmer a pixel the more likely it will be reconstructed insufficiently, e.g.\ with an intensity of less than 10\% of the input intensity (red histogram).  Note that the overall number of jets in the training data, and thus the fraction of jet images, decreases as one requires an increasing number of non-zero pixels. A qualitatively similar picture emerges for the AE trained and tested on top jets (see Fig.~\ref{fig:igno_pix}, right). In the next section, we will explore why such a limited AE can nevertheless tag top jets as anomalies in a QCD background. 

\subsection{Complexity bias}\label{sec:simplicity_bias}

\begin{figure}[t]
\centering
\includegraphics[width = 0.6\textwidth]{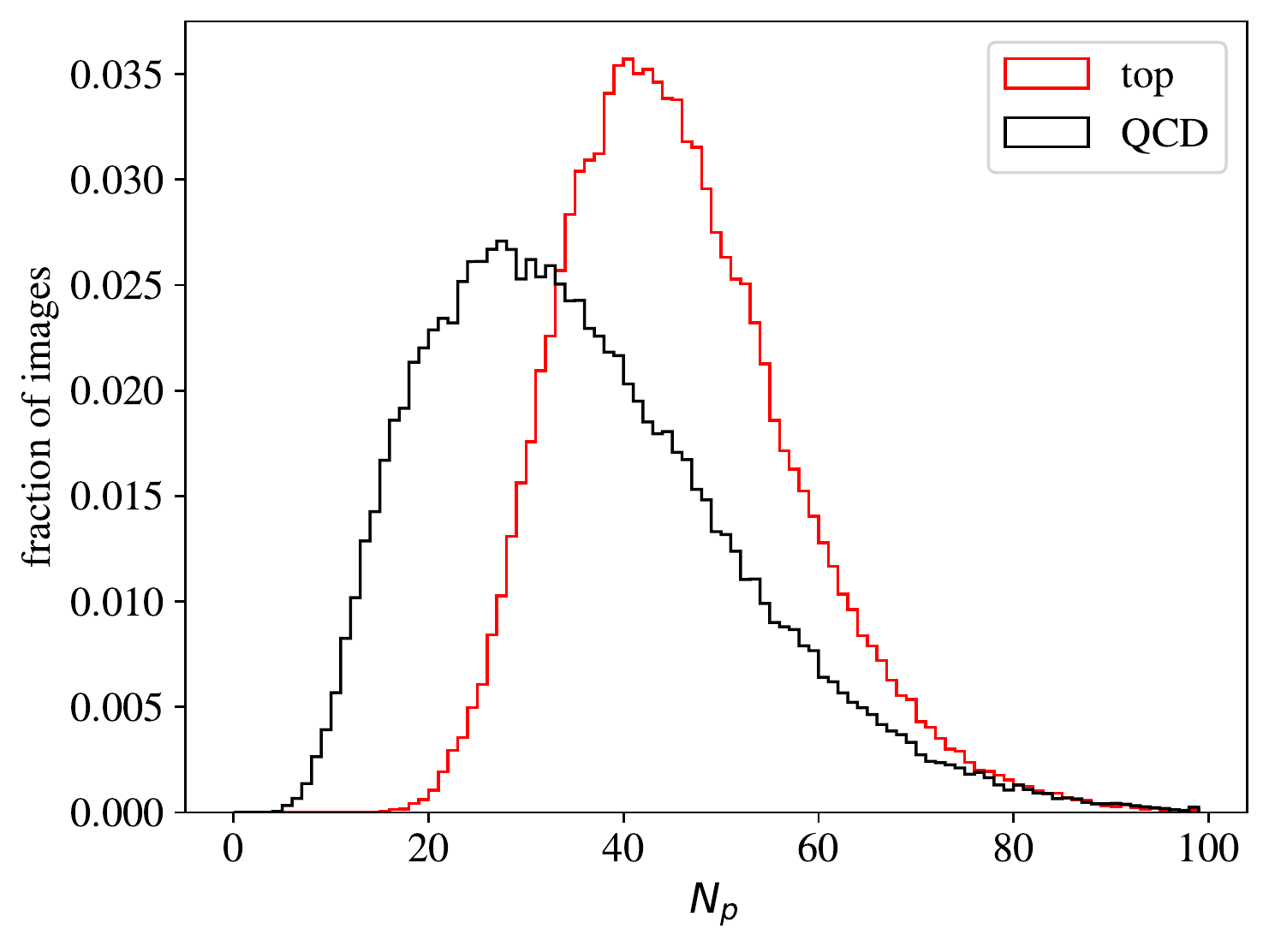}
\caption{Distribution of the QCD and top jet images in the number of non-zero pixels $N_p$.}\label{fig:pixnum_tagger}
\end{figure}

\begin{figure}[t]
\includegraphics[width = 0.5\textwidth]{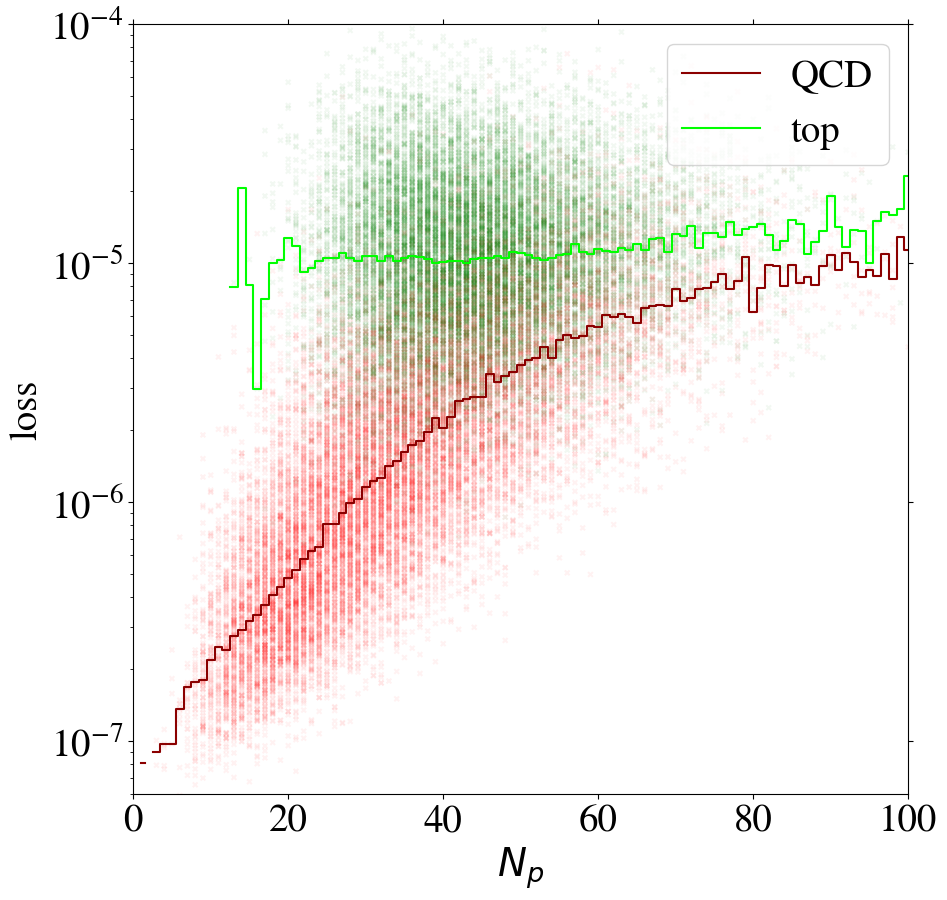}
\includegraphics[width = 0.5\textwidth]{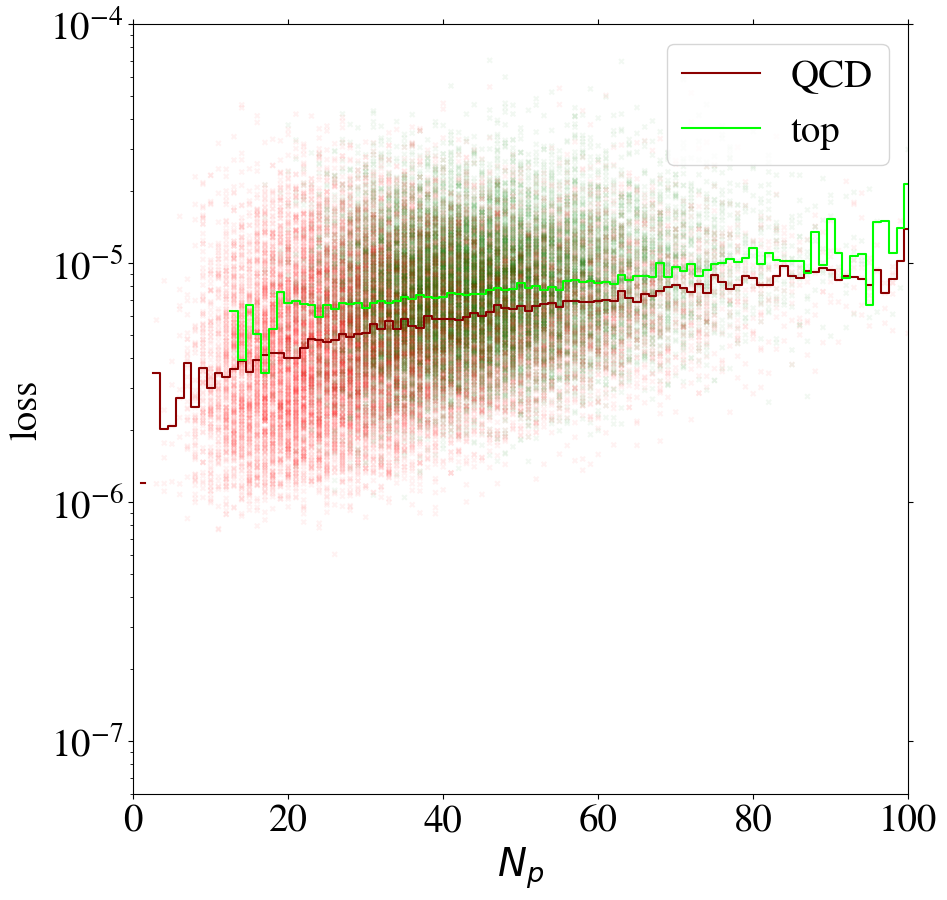}
\caption{Reconstruction loss of individual test jets (only half of the test set is shown as points in the scatter plot) as a function of the number of non-zero pixels $N_p$, for an AE trained on QCD jets (left) and top jets (right). QCD (top) jets are shown in red (green). The solid lines show the median loss for a given number of non-zero pixels.}
\label{fig:pixnum_MSE}
\end{figure}

We can gain additional insight into the AE performance by investigating the reconstruction loss as a function of the number of non-zero pixels, $N_p$, of the jet images. We show the $N_p$ distribution for QCD and top images in Fig.~\ref{fig:pixnum_tagger}. When training on QCD images, the reconstruction loss of QCD images on average increases strongly with $N_p$, as shown in Fig.~\ref{fig:pixnum_MSE}, left. QCD jets with small $N_p$ are simply easier to reconstruct. When training on top images, top jets are also harder to reconstruct for larger $N_p$, see Fig.~\ref{fig:pixnum_MSE}, right. However, in this case the correlation is less pronounced. Since, on average, top images have more non-zero pixels than QCD images, see Fig.~\ref{fig:pixnum_tagger}, the top-trained AE sees less training data with small $N_p$. Given this correlation between the reconstruction loss and the number of non-zero pixels $N_p$, we conclude that $N_p$ at least partially describes the complexity of the images.

However, complexity is not only related to $N_p$. For fixed $N_p$, QCD jets are on average also easier to reconstruct than top jets. This can be understood from the underlying physics. A top jet is naturally composed of three sub-jets initiated by the top-quark decay products. Hence, this intrinsic three-prong structure leads to more complex structures in the jet images.

These results show that the AE has a strong complexity bias: images which would intuitively be labeled as simpler, are reconstructed better. It has been discussed in the literature, especially for natural images, that simpler images may be harder to identify as anomalous \cite{nalisnick2019deep,zong2018deep, gong2019memorizing, ren2019likelihood,serra2019input,schirrmeister2020understanding, kirichenko2020normalizing, tong2020fixing}.  In the context of natural images, it was noted that the algorithm tends to learn features that are not representative of the specific training set, like local pixel correlations. 

The strong complexity bias in our application might be mainly due to the limited reconstruction performance of the AE discussed in Section~\ref{sec:ignored}. It is not surprising that the AE is not able to reconstruct the complex structure of top images when ignoring all but a few pixels. Limitations to the reconstruction of structures in high energy physics have also been noted recently in \cite{batson2021topological, Collins:2021nxn}.

The correlation between the complexity of an image and the reconstruction loss is dangerous if the AE is to be used as a model-independent anomaly tagger. One should not assume that a new physics signal looks more complex than the SM background. An anomaly tagger should only reconstruct well those images it has been trained on. On the other hand, our AE has the tendency to better reconstruct simple images. For a functional model-independent anomaly tagger the first effect should be much more pronounced than the second. For the case at hand, these two effects go in the same direction for the direct tagger, since it is trained on simple QCD images, but go in opposite directions for the inverse tagger, which is trained on more complex top images. This can be seen already in Fig.~\ref{fig:pixnum_MSE},  where we also show the loss of top jets in the test set for the AE trained on QCD jets and vice versa. Indeed, when training on top jets, even QCD images with the same number of non-zero pixels are better reconstructed than top jets, due to the complexity issue. Although the inverse tagger has improved the top reconstruction through its training on top jet images, it cannot overcome the complexity bias. Given the limited reconstruction capabilities, after learning to reproduce top images, the inverse tagger has also learned to interpolate QCD images. On the other hand, the direct tagger is never exposed to the complex features characterizing top images during training and thus fails to extrapolate to the corresponding top jet structures.

\subsection{Tagging performance}
\label{sec:tagging_performance}

\begin{figure}[t]
    \begin{minipage}{0.5\textwidth}
        \includegraphics[width=\linewidth]{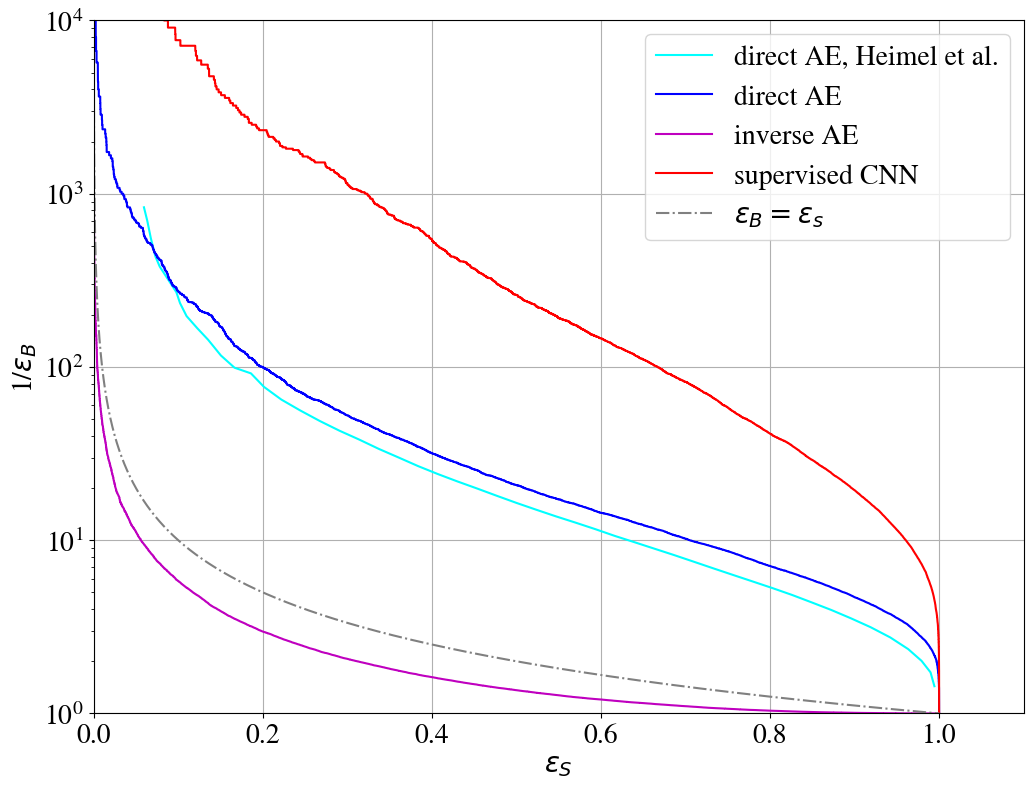}
    \end{minipage}
    \begin{minipage}{0.5\textwidth}
        \includegraphics[width=\linewidth]{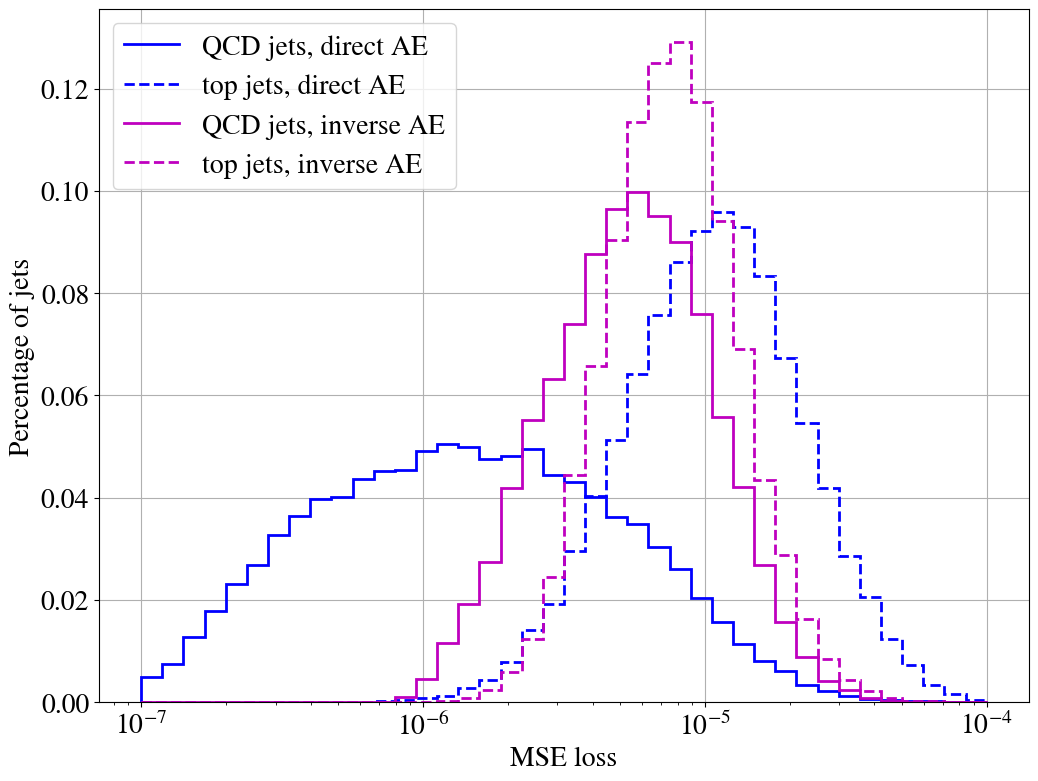}
    \end{minipage}
    \caption{Left: ROC curve of our direct (dark blue) and inverse (purple) tagger. For comparison we also show the performance of a supervised CNN tagger (red), the results of Ref.~\cite{Heimel_2019} for the direct tagger, and a random tagger (grey dashed). Right: Loss distribution for QCD (solid) and top (dashed) images for the direct (blue) and inverse (purple) taggers.}
    \label{fig:stdAE}
\end{figure}

Having investigated the limitations in the AE reconstruction power and the complexity bias, we show the performance of the direct and inverse taggers in Fig.~\ref{fig:stdAE}. The ROC curves are constructed by varying the loss threshold for tagging a jet as anomalous. We also show the ROC curve for a supervised CNN tagger inspired by Ref.~\cite{Kasieczka_2019} (see Appendix~\ref{sec:appendix}). Our direct AE tagger reproduces the results for the direct unsupervised AE tagger in Ref.~\cite{Heimel_2019}, which has used a similar setup. As expected, it performs worse than a supervised tagger, but much better than random guessing. A similar performance for the unsupervised AE tagger has also been obtained in Ref.~\cite{Farina_2020} in a slightly different setup. Hence, our direct tagger works as expected.

As can be seen from Fig.~\ref{fig:stdAE}, the inverse tagger performs worse than randomly tagging jets as anomalous. Our inverse autoencoder fails to tag an anomaly (QCD jets in this case) that is simpler than the background. Through its training on top jets, the inverse tagger learns to reconstruct top jets better than the direct tagger (see right plot in Fig.~\ref{fig:stdAE}), and its performance for the reconstruction of QCD jets is diminished. However, this is not enough to overcome the complexity bias.

To summarize, even with a limited reconstruction capability, an autoencoder can be a good anomaly tagger if there is a bias to reconstruct the background better. However, if the bias works against anomaly detection, the learning capabilities of the AE may not be sufficient to overcome the bias. Only a powerful AE with a background specific data compression in the latent space could potentially be able to overcome such a bias. Improvements of our setup to partly achieve these goals are discussed in Section~\ref{sec:Methods}.

It should be noted that a perfect AE, which is always able to perfectly reconstruct the input via the identity mapping regardless of the input data, would be useless as a tagger as it would always interpolate perfectly from the learned data to the anomalies.

\section{Improving the autoencoder performance}
\label{sec:Methods}

Given the limited performance of the AE setup described in Section~\ref{sec:framework}, we investigate possible improvements. One approach would be to change the AE architecture. There are unlimited possibilities which are worth investigating. However, here we want to point out some generic improvements concerning the complexity bias and the limited learning capabilities which might be helpful for any AE architecture. These improvements are introduced in Sections~\ref{subs:preproc} and \ref{subs:FIL}, and their impact on the AE performance and tagging capabilities is quantified in Sections~\ref{sec:AE_performance} and \ref{sec:Tag_performance}, respectively.

\subsection{Intensity remapping}
\label{subs:preproc}

As shown in Section~\ref{sec:ignored}, the AE focuses on reconstructing only the brightest pixels of an image. The contrast between bright and dim pixels is apparently so large that the dim pixels are not recognized during the early learning epochs. To tackle this problem, we propose several remappings of the pixel intensities, such that the intensities of the dim pixels are enhanced compared to those of the bright pixels. Such a remapping should encourage the AE to ignore less pixels and learn more of the intrinsic structure and features of the given training images. Hence, in particular the complexity of the simplest pictures (having only one or two very bright pixels) is increased, and we expect to also reduce the complexity bias. The intensity remapping can be considered as part of an alternative preprocessing. 

Due to the normalization of the input features in the standard preprocessing described in Appendix~\ref{sec:appendix}, all pixel intensities lie in the interval between 0 and 1. To highlight dim pixels by remapping this interval onto itself, we use the functions
\begin{equation}
\text{R}_1(x)=\sqrt{x} \, , \quad \text{R}_2(x)=\sqrt[4]{x}\, , \quad  \text{R}_3(x)=\frac{\log(\alpha x+1)}{\log(\alpha+1)} \, \quad \text{and} \quad \text{R}_4(x)=\Theta(x)\,,
\label{eq:preprocessings}
\end{equation}
which are displayed in Fig.~\ref{fig:rep_fct} in Appendix~\ref{app:remapping}.
For the mapping $\text{R}_3$ we use $\alpha=1000$; the $\Theta$-function maps each non-zero pixel to one. The original images correspond to the identity remapping $\text{R}_0(x)=x$ and are labeled accordingly in the following. After the remapping the image is normalized again to have a pixel sum equal to one. While remappings $\text{R}_1$ to $\text{R}_3$ are invertible, information is lost through $\text{R}_4$, which should be thought of as a limiting case. The effects of the remappings are exemplified and quantified  further in Appendix~\ref{app:remapping}.

\subsection{Kernel MSE}
\label{subs:FIL}

As discussed in Section~\ref{sec:problem}, it is difficult for the AE to learn the structure of an image. Instead, it only learns the exact position of the few brightest pixels. This is actually not surprising, since the AE is not rewarded by the loss function to reconstruct the image structure. It rather has to reproduce the exact pixel position, which seems to be an unnecessarily tough task. For example, shifting the input image by one pixel would be considered by a human as a valid reconstruction, but is strongly penalized for  the standard MSE loss.

In fact, the similarity of two images is only poorly represented by the MSE loss. To compare two distributions, i.e.\ the intensity distribution in our case, the Wasserstein distance, also called energy or earth mover's distance \cite{10.1023/A:1026543900054}, would be a much better choice. However, its full calculation includes an optimization problem that would need to be solved for each image during training. Approximate estimates of this distance, e.g.\ the sliced Wasserstein distance \cite{bonneel2014sliced}, are also computationally expensive. We thus propose a simple loss function which nevertheless provides a better notion of similarity between two pictures than the MSE loss.

To introduce a notion of neighborhood for each pixel in the loss function we convolve the whole image with a smearing kernel. Hence, neighboring pixels have a partly overlapping smeared distribution which can be recognized by the loss function. The optimal reconstruction, i.e. vanishing loss, is still achieved by reconstructing the original (unsmeared) input image.

On the discrete set of pixels in each image the convolution is defined as
\[
S_{ij} = \sum_{k,l} K_{ij,kl} \, I_{kl} \, ,
\]
where $I_{kl}$ denotes the intensity of the pixel with coordinates $(k,l)$ of the original image, and $S_{ij}$ refers to the smeared pixels. The kernel is defined as 
\[
K_{ij,kl}= \frac{L+1}{L} \left( \frac{1}{1+\sqrt{(i-k)^2+(j-l)^2}} - \frac{1}{1+L} \right)
\]
with the additional constraint that it is set to zero for negative values. Hence, it is only non-zero for an Euclidean distance to the center pixel which is smaller than $L$ pixels and has an approximately circular shape. We choose $L=8$. The normalization is chosen such that the intensity of the central pixel is unchanged. To avoid boundary effects, the original picture is padded with zeros on each side and becomes $54 \times 54$ pixels. There is an infinite amount of possible choices for this kernel. However, we do not expect the following results to depend on its details. The reconstruction loss is defined as the MSE of the smeared input and the reconstructed image. We refer to it as kernel MSE or KMSE in the following. Since the smearing is a standard matrix convolution its computational cost is negligible for training and testing.

\subsection{Autoencoder performance}
\label{sec:AE_performance}

\begin{figure}[t]
    \includegraphics[width=0.5\linewidth]{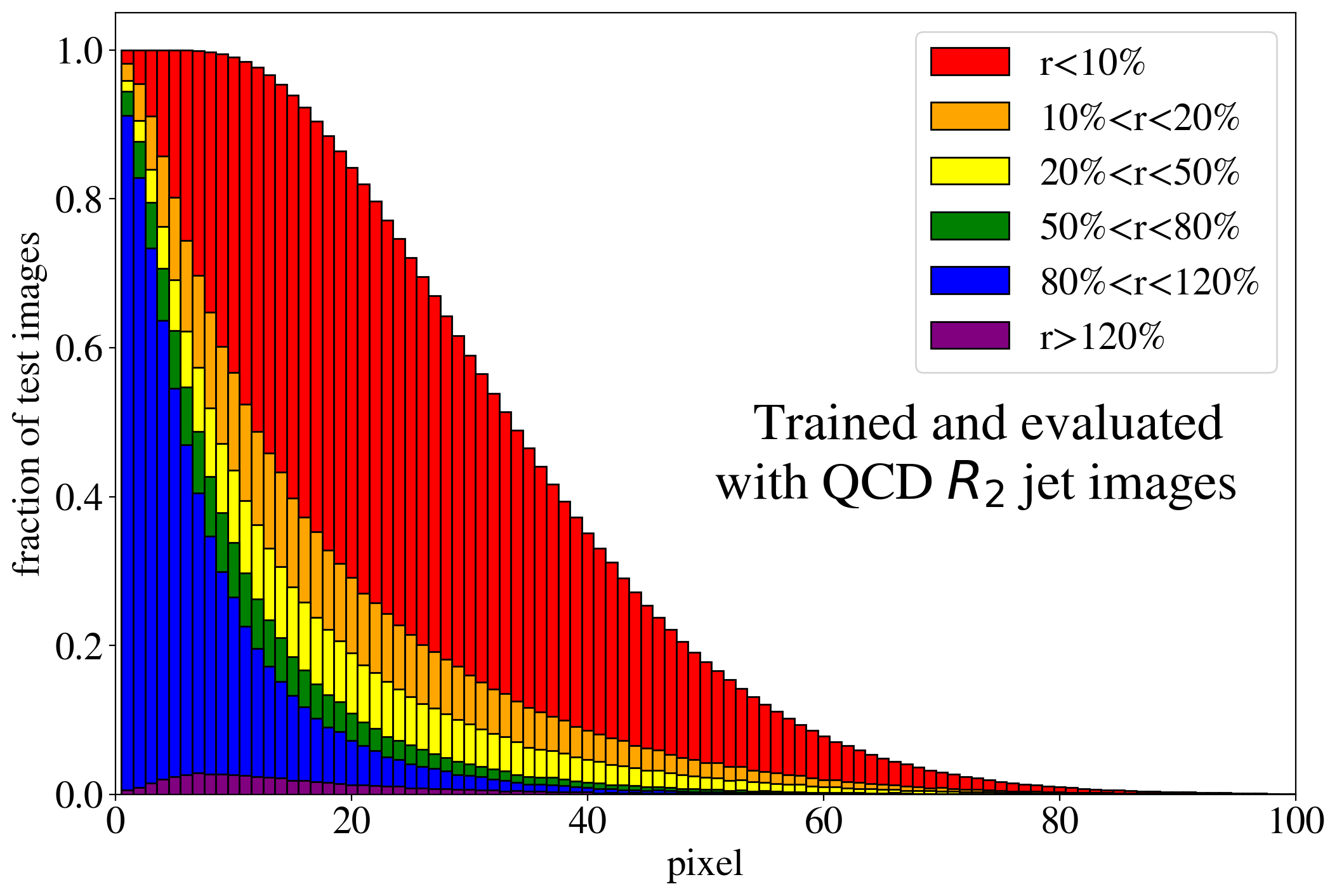}
    \includegraphics[width=0.5\linewidth]{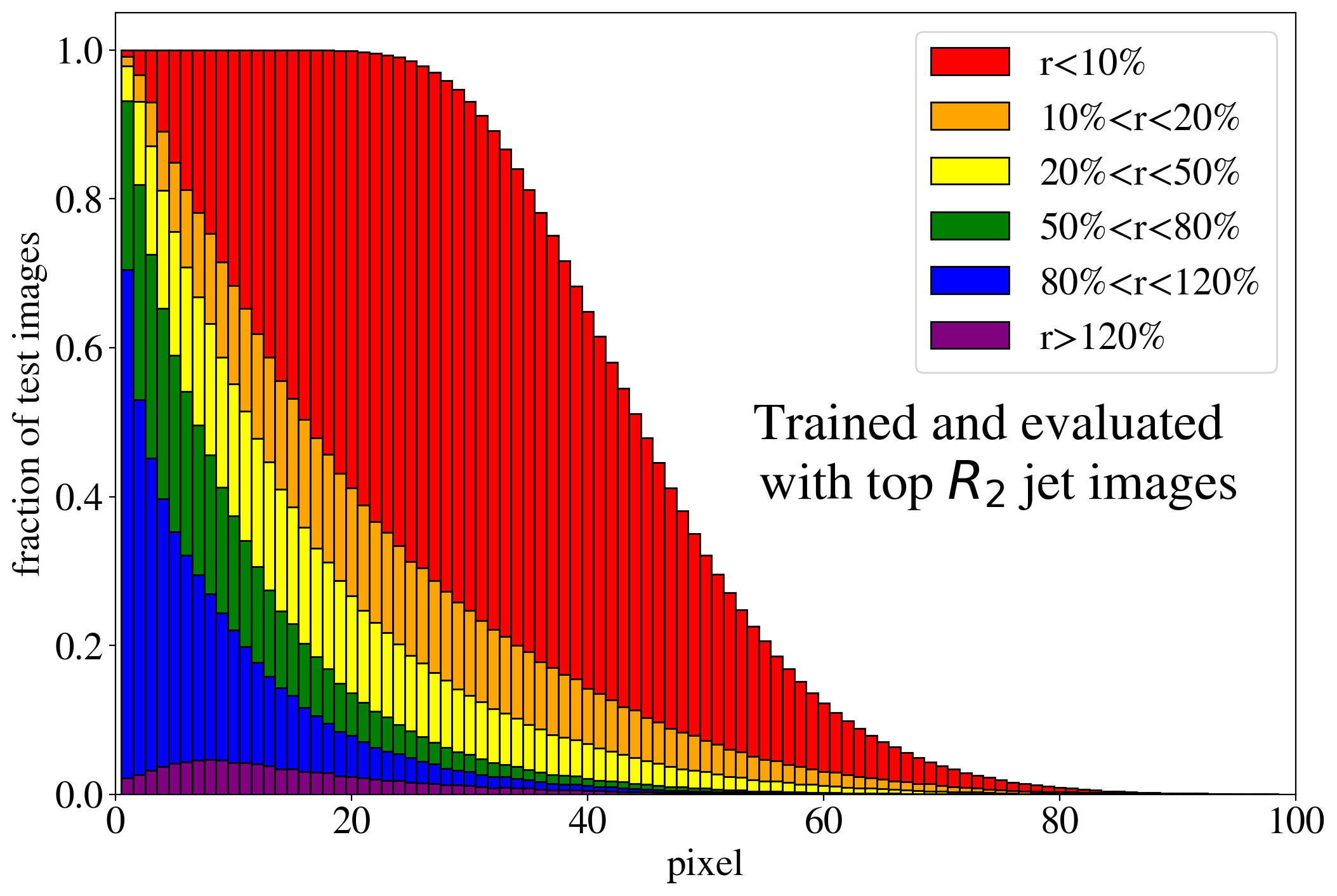}
    \caption{\label{fig:order_sp_all} Same as Fig.~\ref{fig:igno_pix} but using the $\text{R}_2$ intensity remapping.}
\end{figure}

To evaluate the effect of the intensity remapping on the autoencoder performance, we again investigate the distribution of the ratio $r$ of the reconstructed intensity and the input intensity for the leading pixels.
The corresponding histogram, i.e.\ the equivalent of Fig.~\ref{fig:igno_pix}, is shown for the $\text{R}_2$-remapping in Fig.~\ref{fig:order_sp_all}. When training on QCD jets (left figure), the fraction of well-reconstructed pixels is increased by the remapping. This suggests that the AE indeed learns more of the jet features. When training on top jets (right figure), the fraction of well-reconstructed jets is not increased by much, but it extends to more of the brighter pixels, also signaling that more features are potentially accessible. The decrease of performance beyond the 20th pixel (and in particular the disappearance of the peak) compared to the performance without remapping, Fig.~\ref{fig:igno_pix} (right), is not discouraging, since these pixels correspond to very soft particles.\footnote{They are only accidentally well reconstructed for $\text{R}_0$ by a tiny non-vanishing intensity which is typically found in the reconstructed image in the vicinity of bright pixels. Since the $\text{R}_2$-remapping highlights their intensity, this mechanism does not work any more.}

\begin{figure}[t]
	\includegraphics[width=0.5\linewidth]{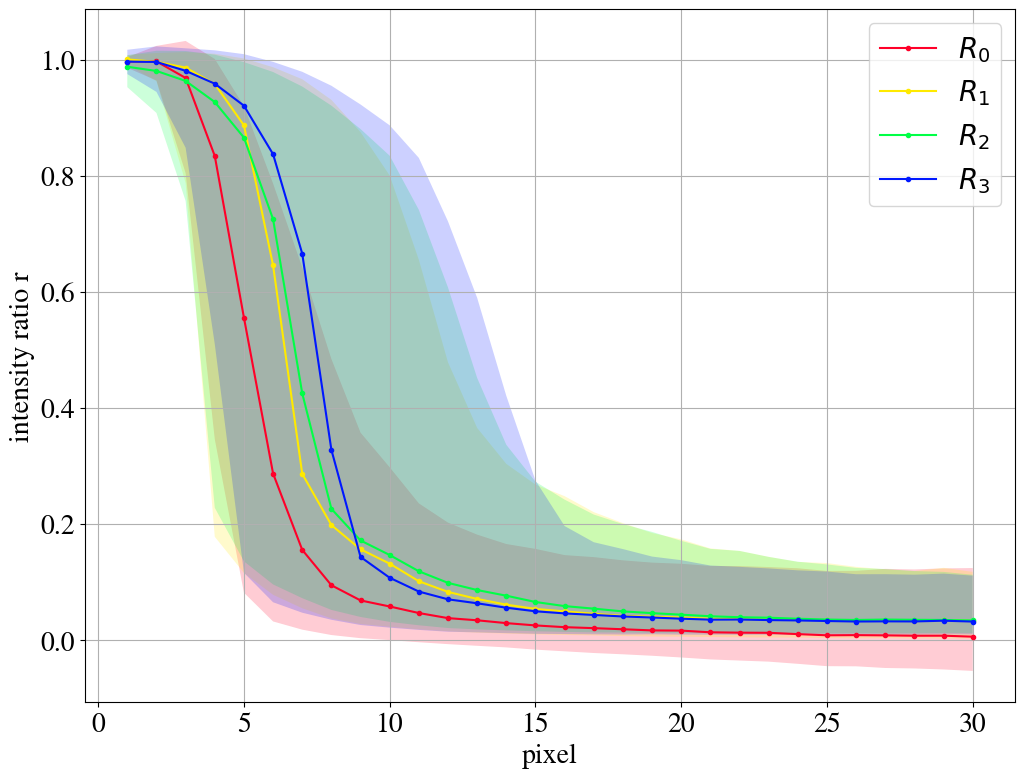}
	\includegraphics[width=0.5\linewidth]{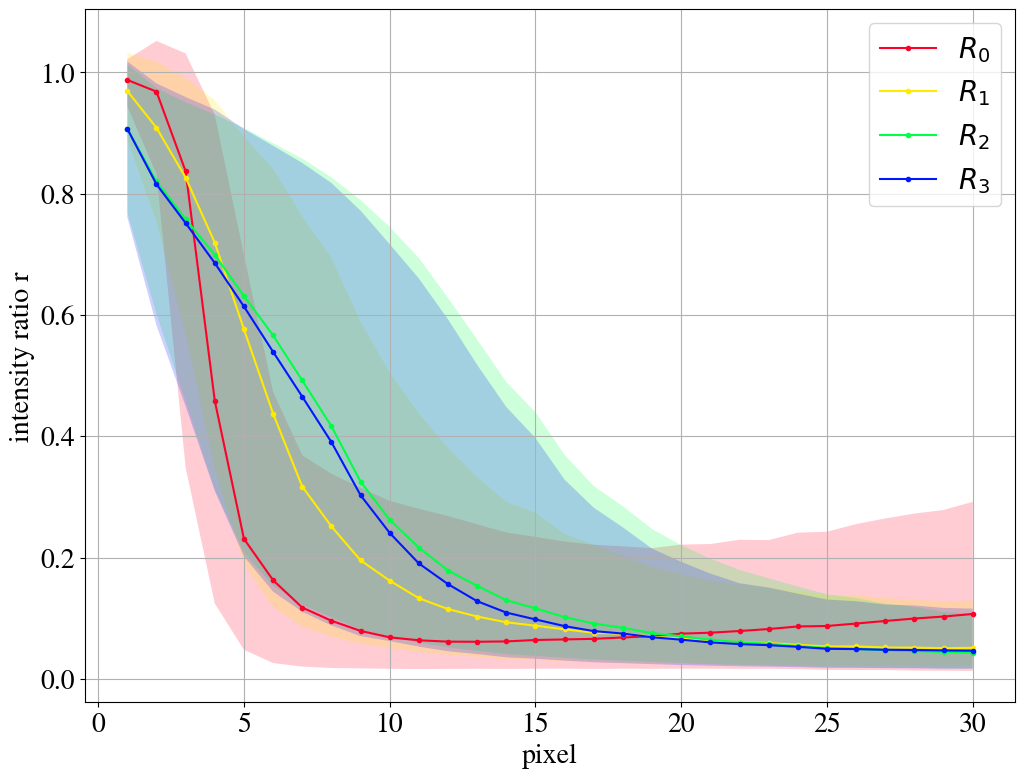}
	\caption{ \label{fig:r_median_MSE} 
		Median and bands for the 25\%/75\% quantiles for the distribution of $r$ for the leading pixels, where $r$ is the ratio of the reconstructed and the input intensity of a given pixel in a given jet. Results are shown for QCD jets (left) and top jets (right). The AE is trained on the respective training set with a given remapping.}
\end{figure}

In Fig.~\ref{fig:r_median_MSE}, we compare the reconstruction quality for the different remappings. The distribution of the intensity ratio $r$ (displayed in Figs.~\ref{fig:igno_pix} and \ref{fig:order_sp_all} by stacked histograms) is represented by its median and the 25\%/75\% quantiles. As long as the distribution is dominated by values of $r$ close to one, the leading pixels are reconstructed well. For both the QCD and the top case, the median curves shift to the right for the remappings, i.e.\ more of the leading pixels are on average taken into account by the AE. Hence, the AE ignores less pixels and thus learns more features of the jet images. For the AE trained and tested on top jets, right figure, this is only achieved by reducing the reconstruction precision of the brightest pixels. 

\begin{figure}[t]
    \includegraphics[width=0.5\linewidth]{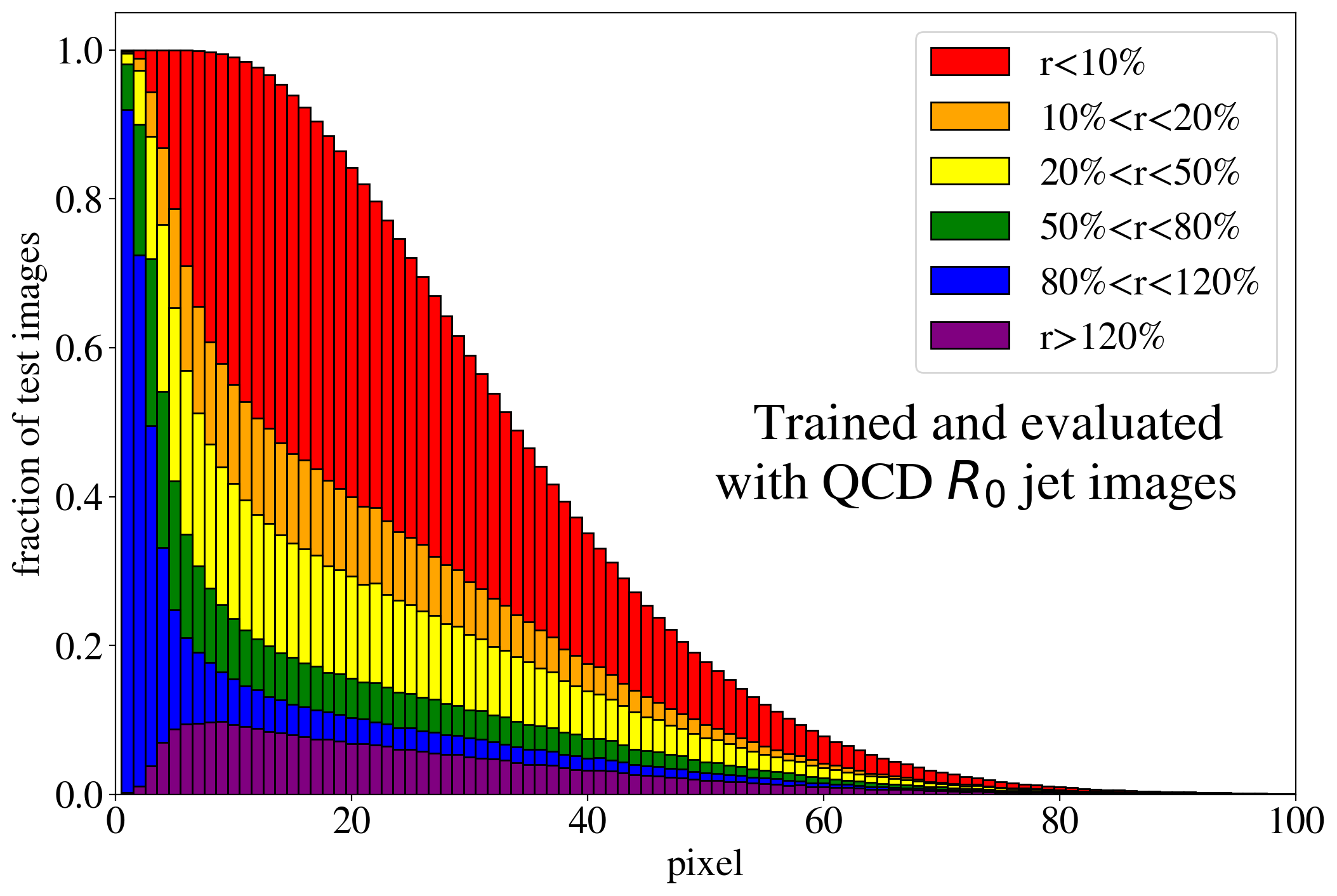}
    \includegraphics[width=0.5\linewidth]{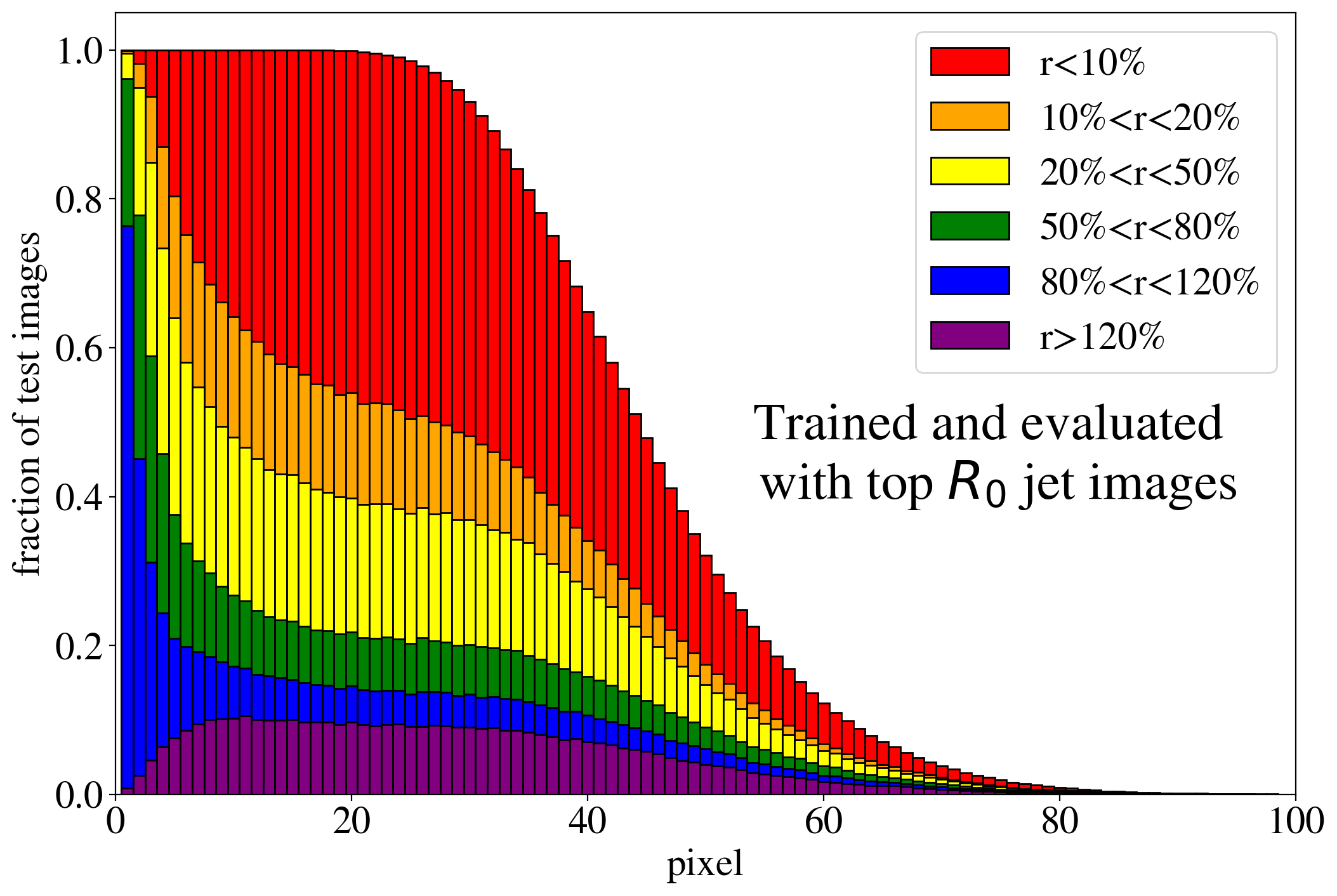}
    \caption{\label{fig:order_sp_all_KMSE} Same as Fig.~\ref{fig:igno_pix} but using KMSE as the loss function during training. No smearing kernel is applied for the data shown during testing.}
\end{figure}

\begin{figure}[t]
	\includegraphics[width=0.5\linewidth]{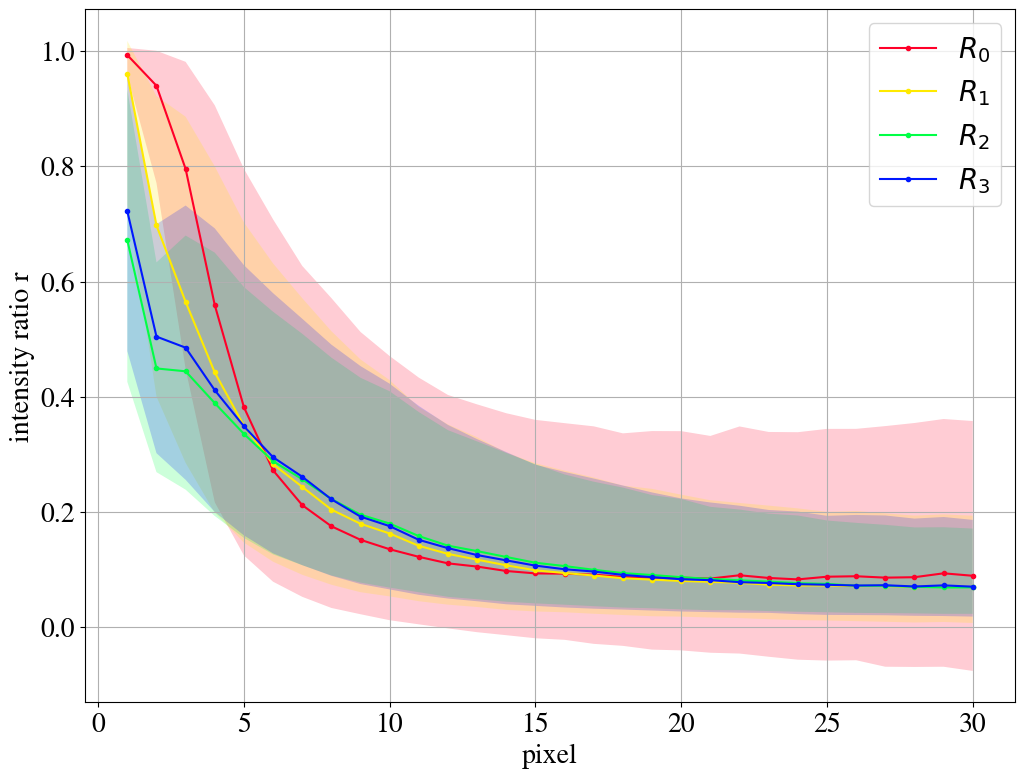}
	\includegraphics[width=0.5\linewidth]{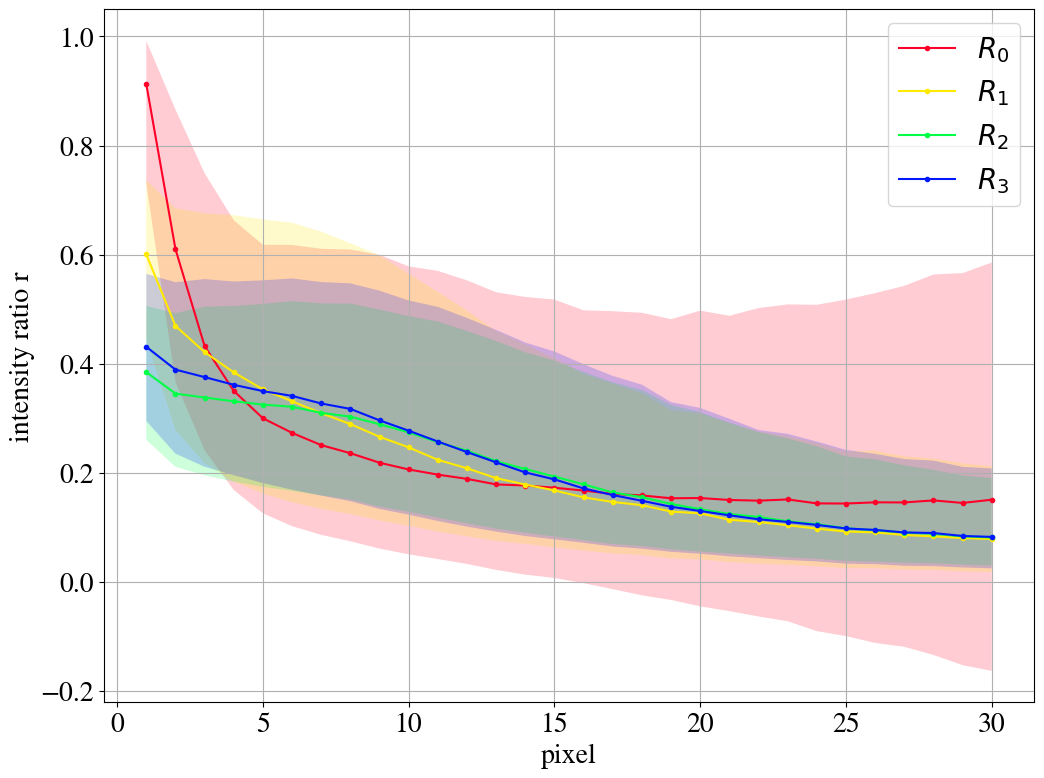}	
	\caption{\label{fig:r_median_FIL}Same as Fig.~\ref{fig:r_median_MSE} but for training with KMSE loss. Results are shown for QCD jets (left) and top jets (right).}
\end{figure}

In Fig.~\ref{fig:order_sp_all_KMSE}, we show the reconstruction performance for training on $\text{R}_0$ images using the KMSE loss introduced in Section~\ref{subs:FIL}. Note that the KMSE loss function is only used during training. To evaluate the pixel reconstruction in the plots no smearing kernel is applied. As one expects, the exact intensity reconstruction of the leading pixels is traded for paying more attention to dimmer pixels. The strong focus to reduce the squared error of the brightest pixels as much as possible is removed.
The same can be seen in Fig.~\ref{fig:r_median_FIL} where we again show the distributions of the intensity ratio $r$ in terms of the median and quantiles for the different remappings. Although the brightest pixels are not very precisely reconstructed, dim pixels often have a significant part of their intensity reconstructed. Also when training on top jets using the KMSE loss functions, the remappings lead to more pixels being taken into account.

The stacked histograms for all remappings with and without KMSE loss function, which contain additional information on the $r$-distribution, are displayed in Appendix~\ref{app:more_results}.

\begin{figure}[t]
    \centering
    \includegraphics[width=1\linewidth]{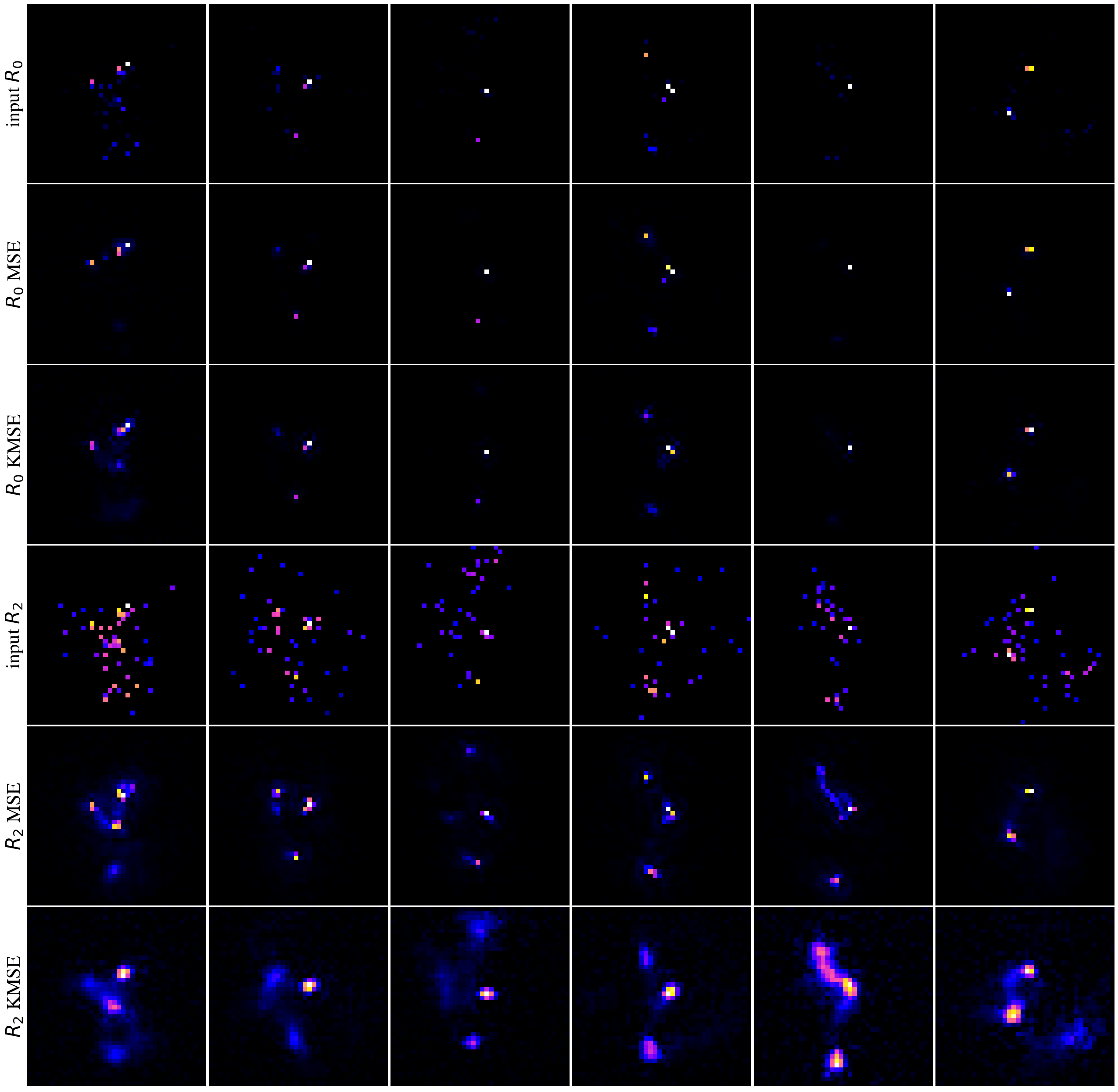}
    \caption{Six top jet input $\text{R}_0$ images (first row) and the corresponding $\text{R}_2$ images (fourth row) together with the AE reconstructions using the MSE (second and fifth row, respectively) or KMSE loss (third and sixth row, respectively). Each image is individually normalized to have the same maximum pixel intensity. Small negative pixel intensities in the reconstructions are set to zero.} 
    \label{fig:examples}
\end{figure}

The reconstruction of top jets (when training on top jets) with the $\text{R}_0$ and $\text{R}_2$ remappings using the standard MSE and the KMSE loss functions is illustrated in Fig.~\ref{fig:examples}. Using the KMSE loss, the AE focuses more on the reconstruction of a continuous distribution of the intensity than on the reconstruction of individual pixels. Even relatively dim regions far from the center receive some attention. Thus, as also shown in Fig.~\ref{fig:r_median_FIL}, the dim pixels are partially reconstructed as part of this continuous distribution. Therefore, we expect the KMSE autoencoder to extract more useful features of the overall jet structure compared to the standard AE that focuses almost exclusively on the brightest pixels.  

\begin{figure}[t]
\includegraphics[width = 0.5\textwidth]{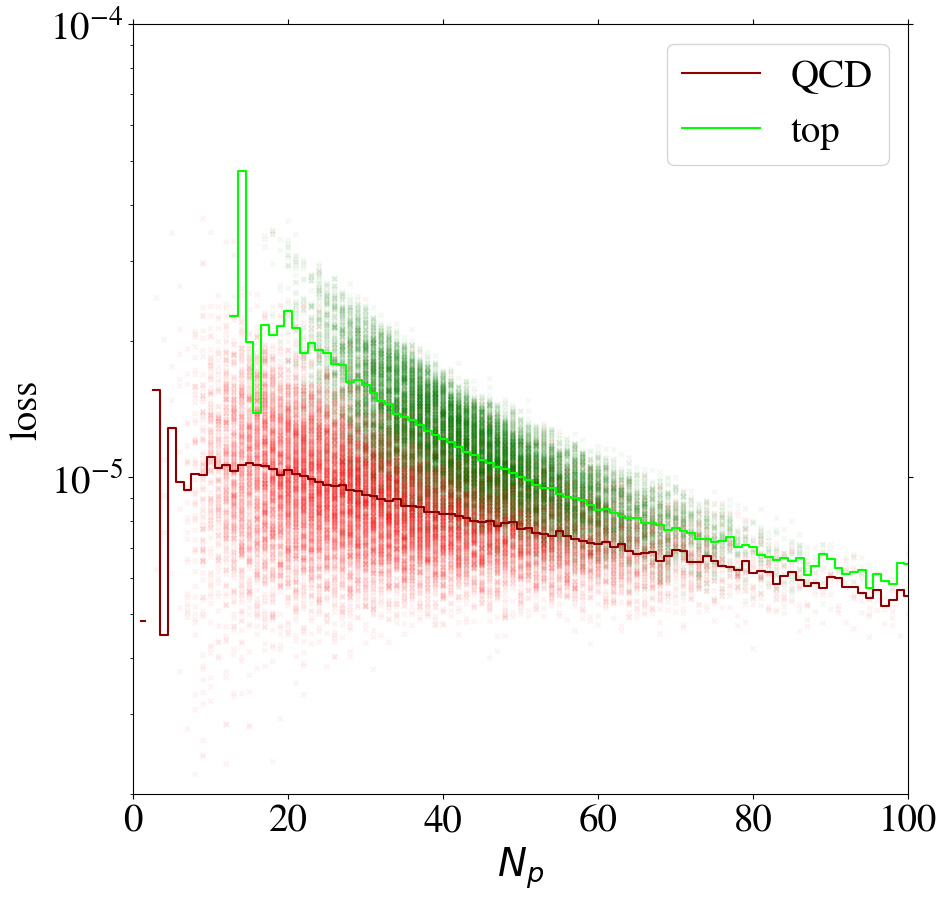}
\includegraphics[width = 0.5\textwidth]{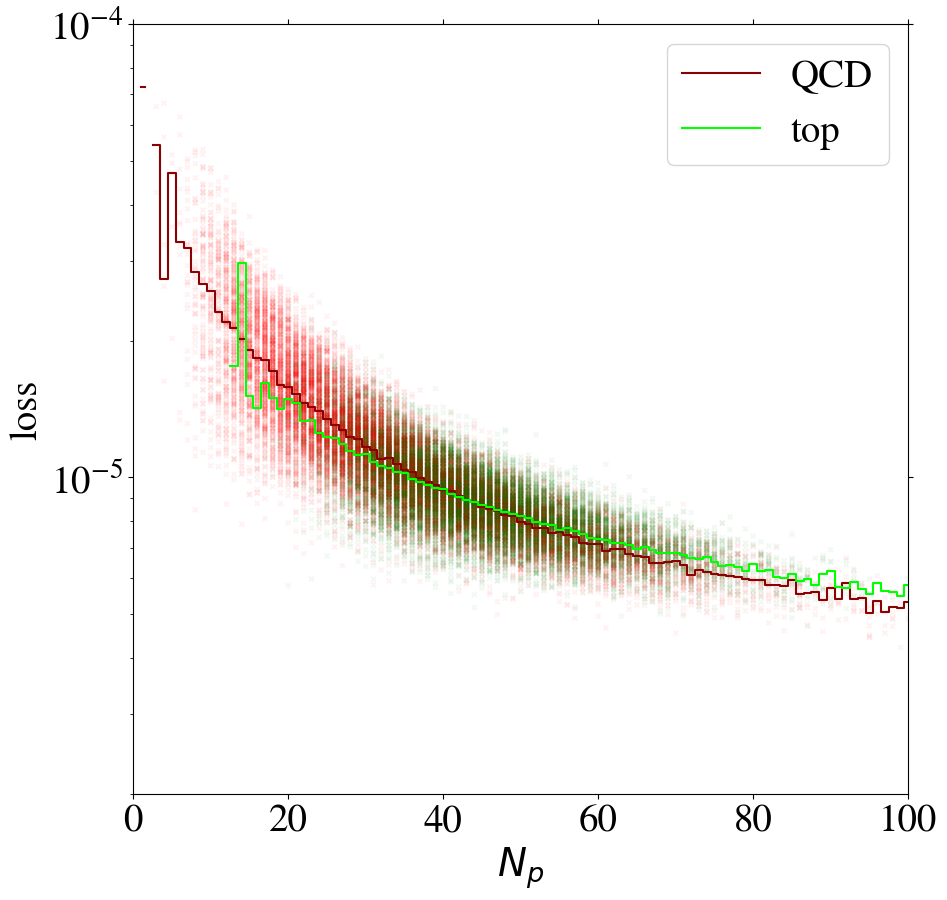}
\caption{Reconstruction loss of individual test jets (only half of the test set is shown as points in the scatter plot) as a function of non-zero pixels for an AE trained on QCD jets (left) and top jets (right) using the $\text{R}_2$ intensity remapping. QCD (top) jets are shown in red (green). The solid lines show the median loss for a given number of non-zero pixels.}
\label{fig:pixnum_MSE_prepr}
\end{figure}

To understand the implications of the intensity remapping with respect to the complexity bias,  
we show the loss of $\text{R}_2$ images as a function of the number of non-zero pixels $N_p$ in Fig.~\ref{fig:pixnum_MSE_prepr} for the AE trained on QCD (left figure) or top images (right figure) using the MSE loss function. Comparing to Fig.~\ref{fig:pixnum_MSE}, the bias for the QCD-trained AE is reduced a lot and even reversed in direction. For the AE trained on top jets, the bias is also reversed, but additionally increased. This can be understood since the inverse AE sees only few of the training images with small $N_p$, the complexity of which has been increased by the intensity remapping.

Training on QCD images, the AE still learns to reconstruct images with low pixel number rather efficiently. Looking at the reconstruction of top images by the direct tagger in the same plot, it still cannot extrapolate well to this unseen data, so that the direct tagger will perform well (see next section). Training on top images with $\text{R}_2$ remapping, the AE has lost its ability to simply interpolate the QCD images with small $N_p$, because their complexity has increased w.r.t.\ the $\text{R}_0$ case. Hence, this AE is a working inverse tagger. However, comparing the average loss for top and QCD jets in the right panel of Fig.~\ref{fig:pixnum_MSE_prepr}, only in a region below 40 active pixels, top images are on average reconstructed better than QCD images. Moreover, the difference is small compared to the width of the distribution and the slope of the bias. Most of the inverse tagging performance is based on the QCD images with small $N_p$, due to the reversed bias and still not due to a well-performing AE.

\begin{figure}[t]
    \begin{minipage}{0.49\textwidth}
        \includegraphics[width=\linewidth]{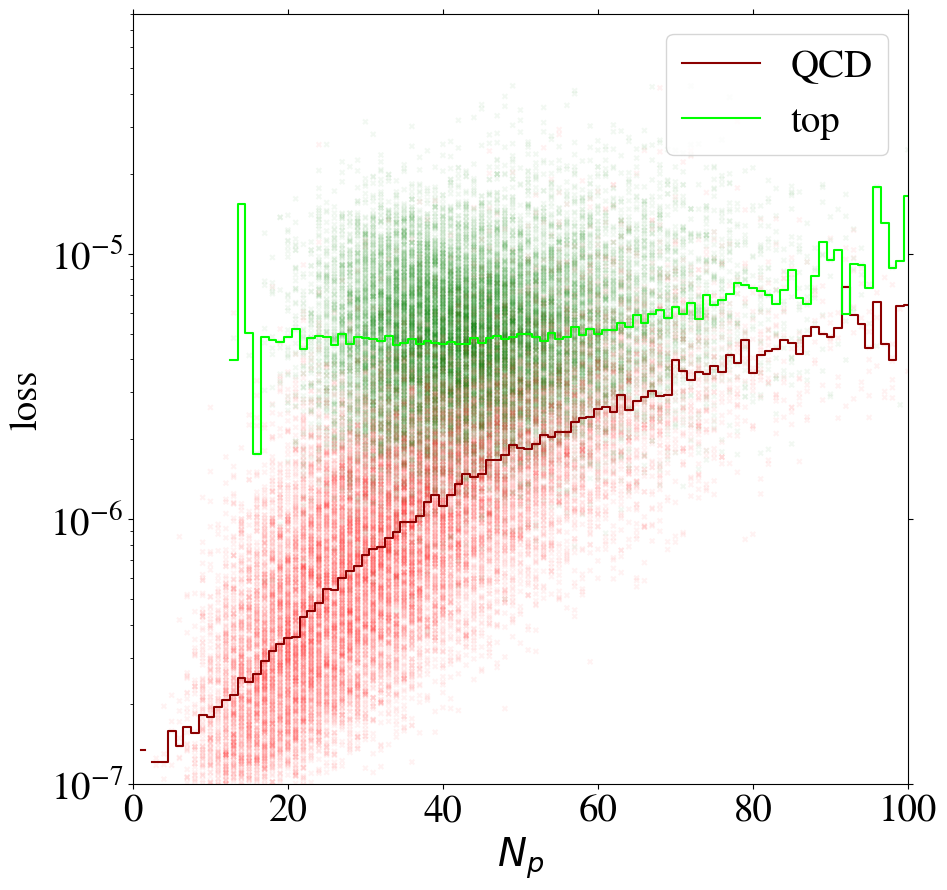}
    \end{minipage}
    \begin{minipage}{0.49\textwidth}
        \includegraphics[width=\linewidth]{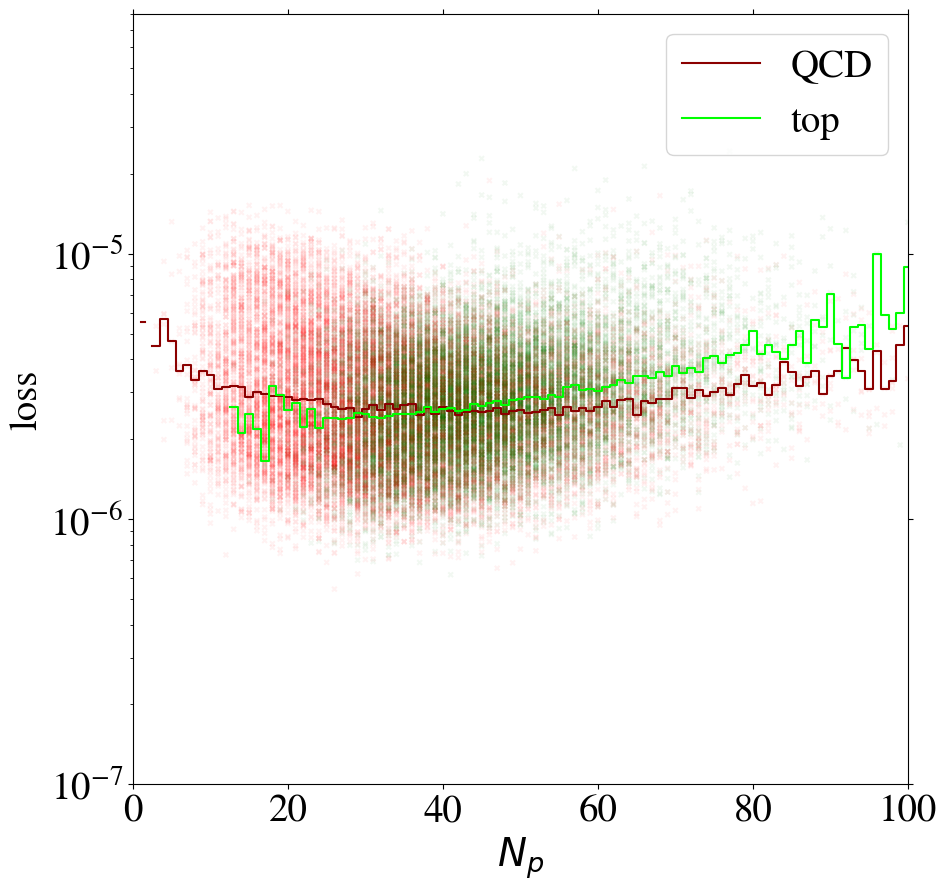}
    \end{minipage}
    \caption{Same as Fig.~\ref{fig:pixnum_MSE} but for KMSE instead of MSE loss (the $\text{R}_0$-remapping is used). }
    \label{fig:pixnum_FIL}
\end{figure}

Using the KMSE loss, Fig.~\ref{fig:pixnum_FIL} shows that the loss as a function of $N_p$ for the direct tagger remains very similar to what we have seen in the MSE case. However, we observe that QCD jets with low $N_p$ are not automatically learned to be reconstructed well when training the inverse tagger on top images. Moreover, for $N_p<35$ top images are again reconstructed on average better than QCD images. Using both a remapping and the KMSE loss function for training, the behavior gets more pronounced. The corresponding distributions are shown in Fig.~\ref{fig:pixnum_all_appendix} in Appendix~\ref{app:more_results} for all combinations of remappings and loss functions.

\subsection{Tagging performance}
\label{sec:Tag_performance}

The performance for the direct and inverse taggers using the different intensity remappings introduced in Section~\ref{subs:preproc} are summarized in terms of their ROC curves in Fig.~\ref{fig:dir_rev_top_MES}. Derived performance measures are shown in Table~\ref{preproc_table} in Appendix~\ref{app:more_results}. For the direct tagger, QCD jets are background ($\epsilon_B=\epsilon_{\rm QCD}$) and top jets are signal ($\epsilon_S=\epsilon_{\rm top}$), so that the area under the curve plotted in Fig.~\ref{fig:dir_rev_top_MES} directly corresponds to the AUC. For the inverse tagger, signal and background interchange their role ($\epsilon_S=\epsilon_{\rm QCD}$ and $\epsilon_B=\epsilon_{\rm top}$) and the area under the plotted curves should be as small as possible, since it corresponds to 1-AUC.

Plotting the ROC curves in this way, the area between the ROC curves of the direct and the inverse tagger, called ABC in the following, can be interpreted as the background specific learning achieved by the AE. Furthermore, the difference between the two AUC values for the direct and the inverse tagger, called $\Delta {\rm AUC}$ in the following, is a measure for how biased the AE is. For an AE with small ABC or large $\Delta {\rm AUC}$ one cannot expect a model-independent tagging capability. On the other hand, a large ABC and a small $\Delta$AUC do not guarantee model-independence. It is only an indication that a step towards model-independence has been made. 

Using the standard preprocessing without intensity remapping ($\text{R}_0$), we find a small ${\rm ABC}=0.28$ and a large $\Delta {\rm AUC}=0.55$. As discussed in Section~\ref{sec:tagging_performance}, the direct tagger only works due to a large complexity bias, and the inverse tagger fails. 

\begin{figure}[t]
    \centering
    \includegraphics[width=0.8\linewidth]{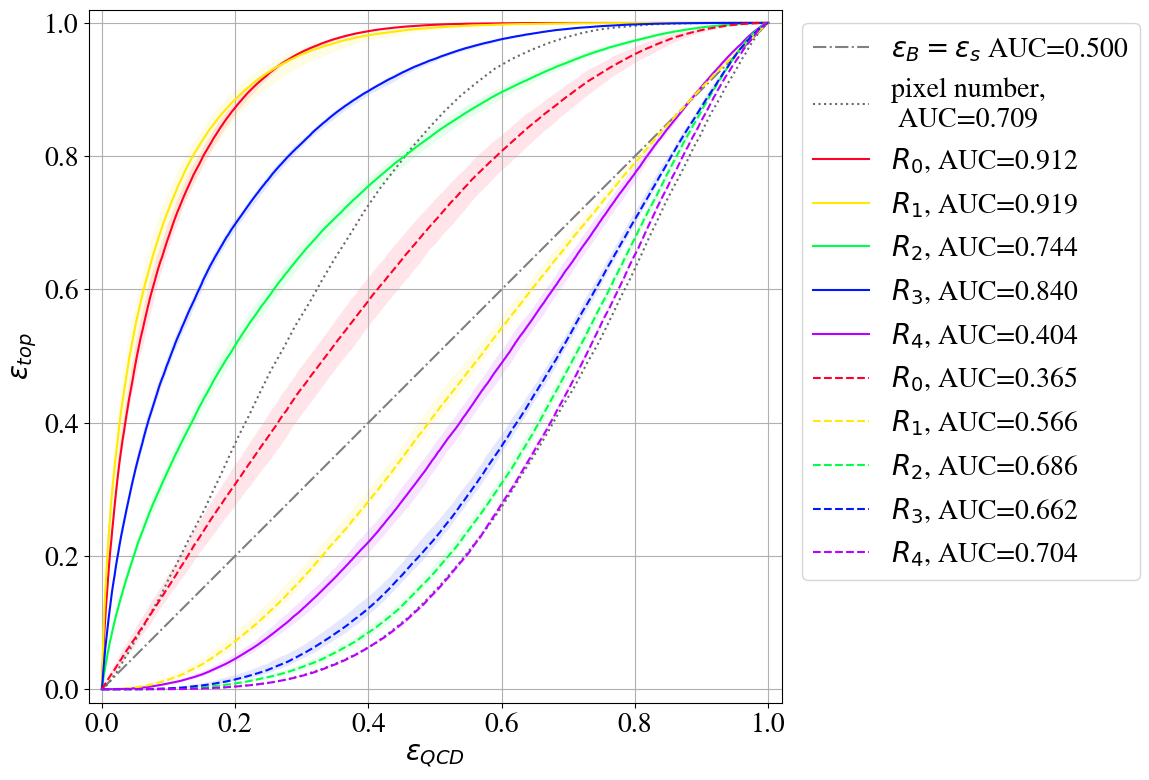}
    \caption{ROC curves of the direct (solid) and inverse (dashed) taggers using the different intensity mappings $\text{R}_0$ to $\text{R}_4$. The dotted lines represent the ROC curves of the two taggers based on the number of the non-zero pixels in the image.}
    \label{fig:dir_rev_top_MES}
\end{figure}

For $\text{R}_1$-remapping we see a minor decrease in the direct tagging performance, but a significant improvement in the inverse tagging performance with ${\rm ABC}=0.48$. This improvement is strong enough to enable inverse tagging although $\Delta {\rm AUC}=0.35$ is still large. In contrast, $\text{R}_2$-remapping shows little bias ($\Delta {\rm AUC}=0.06$) and an ${\rm ABC}=0.43$. Both direct and inverse tagging have a modest AUC close to 0.7. $\text{R}_3$ has a little more bias ($\Delta {\rm AUC}=0.18$) but the best ${\rm ABC=0.50}$. $\text{R}_4$-remapping shows that too much highlighting of dim pixels is also counterproductive. It leads to $\Delta {\rm AUC}=-0.3$ and a very poor ${\rm ABC}=0.11$.

Although these results are a promising first step towards a more model-independent anomaly tagging, we want to recall that the inverse tagging is mainly caused by the correlation of the AE loss with the number of non-zero pixels $N_p$. The AE is still not able to reconstruct a top jet on which it has been trained better than a QCD jet if both have the same $N_p$. Moreover, in particular the tagging performance of the inverse taggers is still very poor. To illustrate this, we also show the two taggers which only use the distribution of $N_p$ for both classes, as shown in Fig.~\ref{fig:pixnum_tagger}, to tag anomalies with either a large or a small $N_p$. None of our inverse taggers can beat the performance of the trivial tagger looking for few non-zero pixels. This shows once more that the inverse tagging for the intensity remappings mostly relies on the correlation of the loss with the number of pixels.

\begin{figure}[t]
    \centering
    \includegraphics[width=0.8\linewidth]{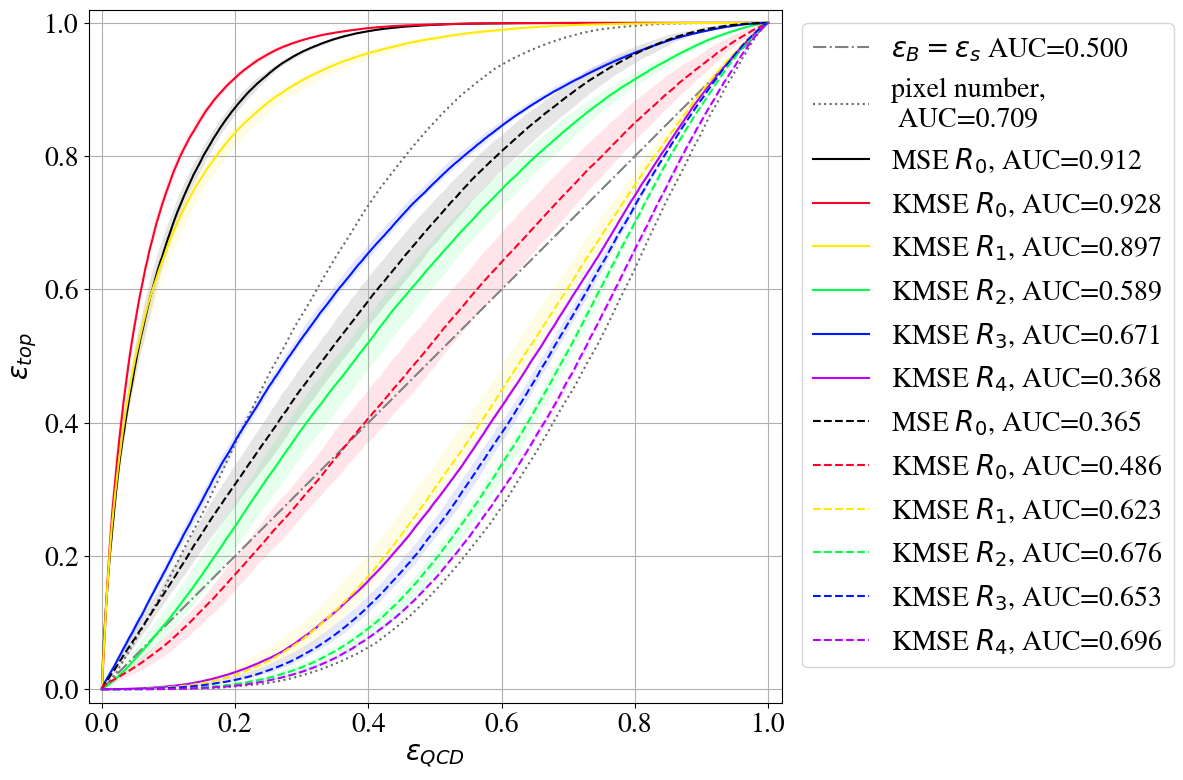}
    \caption{ROC curves of the direct (solid) and inverse (dashed) taggers using the different intensity mappings $\text{R}_0$ to $\text{R}_4$ for the KMSE and $\text{R}_0$ for the MSE loss function. The dotted lines represent the ROC curves of the two taggers based on the number of the non-zero pixels in the image.}
    \label{fig:dir_rev_top_FIL}
\end{figure}

Fig.~\ref{fig:dir_rev_top_FIL} shows the ROC curves for direct and inverse taggers using the KMSE loss and the different intensity remappings (see also Table~\ref{preproc_table_FIL} in Appendix~\ref{app:more_results} for additional performance measures). The KMSE loss is also used as the anomaly score. On the one hand, the direct $\text{R}_0$-tagger using the KMSE loss performs similarly to the MSE one. On the other hand, also the KMSE autoencoder is not able to provide a working inverse tagger, although the performance of the inverse tagger is significantly better than the AE with MSE loss. With ${\rm ABC}=0.41$ and $\Delta {\rm AUC} =0.44$ it is less biased towards reconstructing QCD better.

For the intensity remappings, we find ${\rm ABC}=0.52$ and $\Delta {\rm AUC}=0.27$ for $\text{R}_1$, hence the KMSE loss improves the overall performance. The remapping $\text{R}_2$ results in ${\rm ABC}=0.27$ and $\Delta {\rm AUC} =-0.09$, i.e.\ the bias is still small but the performance is reduced. The remapping $\text{R}_3$ shows almost no bias ($\Delta {\rm AUC}=0.02$) but also reduced ${\rm ABC}=0.32$. The remapping $\text{R}_4$ still fails completely. In contrast to the MSE loss, using KMSE the inverse taggers based on $\text{R}_1$ to $\text{R}_4$ all show a performance very close to the pixel number tagger. While we have investigated the ability of the KMSE-based AE to reconstruct more jet features in Section~\ref{sec:AE_performance}, the tagging performance shows that these features are not necessarily correlated with only one of the two classes. 

\section{Conclusion}
\label{sec:conclusion}

In this work we have investigated a specific autoencoder architecture, based on a convolutional neural network, for tagging top jets in a background of QCD jets - and vice versa - using the reconstruction loss as the anomaly score. We observe a rather limited performance of the autoencoder with respect to its image reconstruction capabilities, see Section~\ref{sec:ignored}. Nevertheless, we confirm findings from the literature~\cite{Heimel_2019,Farina_2020} that such an autoencoder is a powerful top tagger. These apparently contradicting observations are a consequence of a strong complexity bias of the autoencoder-based tagger as discussed in Section~\ref{sec:simplicity_bias}. The sparsity of the jet images in combination with the underlying physics of QCD and top jets allow for a more successful reconstruction of the QCD jet images with respect to the mean squared error of all pixels, no matter which training data (i.e.\ QCD jets or top jets) are used.

Having made this observation in a very specific benchmark scenario, it is nevertheless quite obvious that it might generalize to other autoencoder architectures and most probably also to some applications with completely different data. Similar observations have been recently made for anomaly detection in natural images~\cite{nalisnick2019deep, schirrmeister2020understanding, kirichenko2020normalizing, ren2019likelihood, serra2019input, tong2020fixing}. A poor autoencoder can be a good anomaly tagger if there is a strong bias which favours the reconstruction of the background data for a given anomaly example. Hence, a good tagging performance for a specific example does not imply a functional model-independent unsupervised tagger. On the other hand, a perfect autoencoder is useless as a tagger if it can also interpolate to reconstruct anomalies which have not (in the semi-supervised case) or rarely (in the unsupervised case) been seen during training. 

As the top-tagging example shows, a biased autoencoder is not necessarily a bad thing. If, for example, one is only interested in anomalies with more complex jet structures, the bias helps tagging anomalies, and the setup is certainly also robust in an unsupervised approach working with a training sample including signal contamination~\cite{Heimel_2019,Farina_2020}. However, model-independence will always be lost to a certain extent. For example, dark-matter jets from a strongly interacting sector are impossible to tag as anomalies in this case, as discussed in \cite{thorben,ivan}.

How model-independent an autoencoder based tagger is, is difficult to answer. After all, the autoencoder tagger should exactly be designed to find the sort of anomaly which is not a priori thought of. One way to approach the problem is to test the tagger on as many physics cases as there are available. However, this might be a tedious task, if many physics cases are at hand or if they would have to be designed for this purpose. We propose to at least evaluate the autoencoder by inverting a given test task, i.e.\ interchanging the role of background and signal and investigating the corresponding performance. Possible performance measures are the difference of the corresponding AUCs or the area between the ROC curves in a plot like Fig.~\ref{fig:dir_rev_top_MES} in Section~\ref{sec:Tag_performance}.

Having understood at least some of the limitations of the original setup, we have investigated two modifications to improve the autoencoder. An intensity remapping for the jet images, as introduced in Section~\ref{subs:preproc}, and the kernel MSE loss function, as introduced in Section~\ref{subs:FIL}, are designed to help the autoencoder learn more relevant features within the jet images. Our modifications are shown to be promising first steps, both with respect to the autoencoder performance (Section~\ref{sec:AE_performance}) and with respect to the model-independence of the tagging performance (Section~\ref{sec:Tag_performance}). However, the true progress is hard to quantify and better performance measures are needed in the future. Moreover, the intensity remapping is also affecting the bias of the autoencoder for reconstructing one of the two classes better than the other, irrespective of the training data. In particular, the inverse tagging, i.e.\ finding QCD jets in a top jet background, is to some extent due to a reversed bias, which is helpful for this particular test case but cannot be claimed as a model-independent advantage.

Future directions include the investigation of improved autoencoder architectures which we have not touched upon at all in this work. It would also be interesting to explore how anomaly detection based on the latent space of (variational) autoencoders or even completely different architectures is impacted by the complexity bias or other related biases when trained on the sparse jet images.

To summarize, this work provides valuable insights into the interplay of the autoencoder performance and anomaly tagging as well as first steps for improvements. However, we want to stress that a powerful and truly model-independent autoencoder for unsupervised anomaly tagging on jet pictures still needs to be developed. 

\bigskip
\noindent
\textbf{Note added:} The submission of this paper has been coordinated with \cite{dillon2021, talk1} and with \cite{gershtein2021}, which address the challenges of using autoencoders for anomaly detection in complementary ways.     

\section*{Acknowledgements}
We are grateful to Anja Butter, Tilman Plehn and the participants of the Machine Learning mini workshops organized within the Collaborative Research Center TRR 257 ``Particle Physics Phenomenology after the Higgs Discovery'' for useful discussions. This work has been funded by the Deutsche Forschungsgemeinschaft through the CRC TRR 257 under Grant 396021762 - TRR 257 and the Research Training Group GRK 2497 ``The physics of the heaviest particles at the Large Hadron Collider'' under grant 400140256 - GRK 2497. Simulations were performed with computing resources granted by RWTH Aachen University.

\begin{appendix}
\section{Jet simulation and neural network architecture}
\label{sec:appendix}

The benchmark dataset from Ref.~\cite{Butter_2018} is publicly available \cite{kasieczka_gregor_2019_2603256}. The jets are obtained for a center-of-mass energy of 14 TeV using \textsc{Pythia8}~\cite{Sj_strand_2015} and fast detector simulation with the default ATLAS detector card of \textsc{Delphes}~\cite{de_Favereau_2014}. Multiple interactions and pile-up are ignored. The jets are clustered with \textsc{FastJet} \cite{Cacciari_2012} using the anti-$k_T$ algorithm~\cite{Cacciari:2008gp} with a jet radius of $R=0.8$. Moreover, the jets fulfil $p_T\in [550 $ GeV$, 650 $ GeV$]$ and $|\eta|<2$. For top jets, a parton-level top and its decay partons are required within $\Delta R = 0.8$ of the jet axis. The 200 jet constituents leading in $p_T$ are stored. Jets with less constituents are padded with zeros.

According to the preprocessing in Ref.~\cite{Macaluso:2018tck}, the jets are first centered in the $\eta$-$\phi$-plane. Afterwards, they are rotated such that the principle axis is vertical and finally mirrored along both axes to obtain the hardest component in the first quadrant. After these steps, the jets are converted into two-dimensional jet images with $40 \times 40$ pixels corresponding to a two-dimensional transverse-momentum histogram of the jet constituents in the rotated $\Delta \eta$-$\Delta \phi$-plane. Finally, each jet image is divided by its total transverse-momentum, i.e.\ the pixel intensities sum to one.

Here, we explain in more detail the AE architecture given in Fig.~\ref{fig:AE3_arch}. Each convolutional layer has a stride of 1 in both directions and a $4\times4$ kernel. Padding is performed in all convolutional layers such that the dimension of the output is the same as of the input. First, we use two convolutional layers with 10 and 5 filters for feature extraction. We reduce the dimension using average pooling with stride 2 and a $2\times2$ kernel, and add two more convolutions with 5 filters each. The feature map of the last convolution is flattened into a vector of 2000 nodes that are fully connected to 100 nodes in the next layer, which in turn are fully connected to the bottleneck layer with 32 nodes. These layers together form the encoder part of the AE. For the decoder we first add two fully connected layers with 100 and 400 nodes, respectively. The output of the latter is reshaped into a $20\times20$ feature map. Afterwards, we perform two convolutions (with 5 filters each) followed by a $2 \times 2$ up-sampling layer to restore the dimensions of the image. We complete the decoder by adding three more convolutional layers with 5, 10 and 1 filter, respectively, resulting in a $40\times40$ dimensional output, matching the dimensions of the input. 
We apply the parameterized ReLU activation function in all hidden layers (convolutional and fully connected) with the corresponding $\alpha$ parameter initialized from a uniform distribution in the interval $[-1, 1]$. For the final layer, i.e.\ the last convolutional layer resulting in the AE reconstruction, no activation function is applied.

The training is done on the designated training set consisting of 100k images using the \textsc{Adam} optimizer \cite{kingma2014adam} and a batch size of 500. As we observe a saturation of the reconstruction loss for this amount of training data, we conclude that it is sufficient. After each epoch of training, validation is performed on 40k images. Early stopping is employed if no improvement of the validation loss is achieved in 10 consecutive epochs. Early stopping always terminates the training before the maximum of 1200 epochs is reached, and we use the weights of the epoch with the lowest validation loss for testing. To improve the training with the KMSE loss function, we reduce the learning rate to $10^{-4}$ and the batch size to 64.

To establish an upper limit for the unsupervised learning and to test our framework, we first use a supervised approach, i.e.\ we apply a convolutional neural network to this classification problem. The architecture of the CNN is based on the one used in \cite{Kasieczka_2019}. The corresponding ROC curve in Fig.~\ref{fig:stdAE} is obtained by training for 100 epochs on 100k QCD and 100k top jet images, validating on 40k QCD and 40k top images and using the model with the lowest validation loss. The performance of our CNN is comparable to the results from the literature~\cite{Kasieczka_2019, bernreuther2020casting}.

\section{Intensity distributions after remapping}
\label{app:remapping}

\begin{figure}
    \begin{center}
        \includegraphics[width=0.75\textwidth]{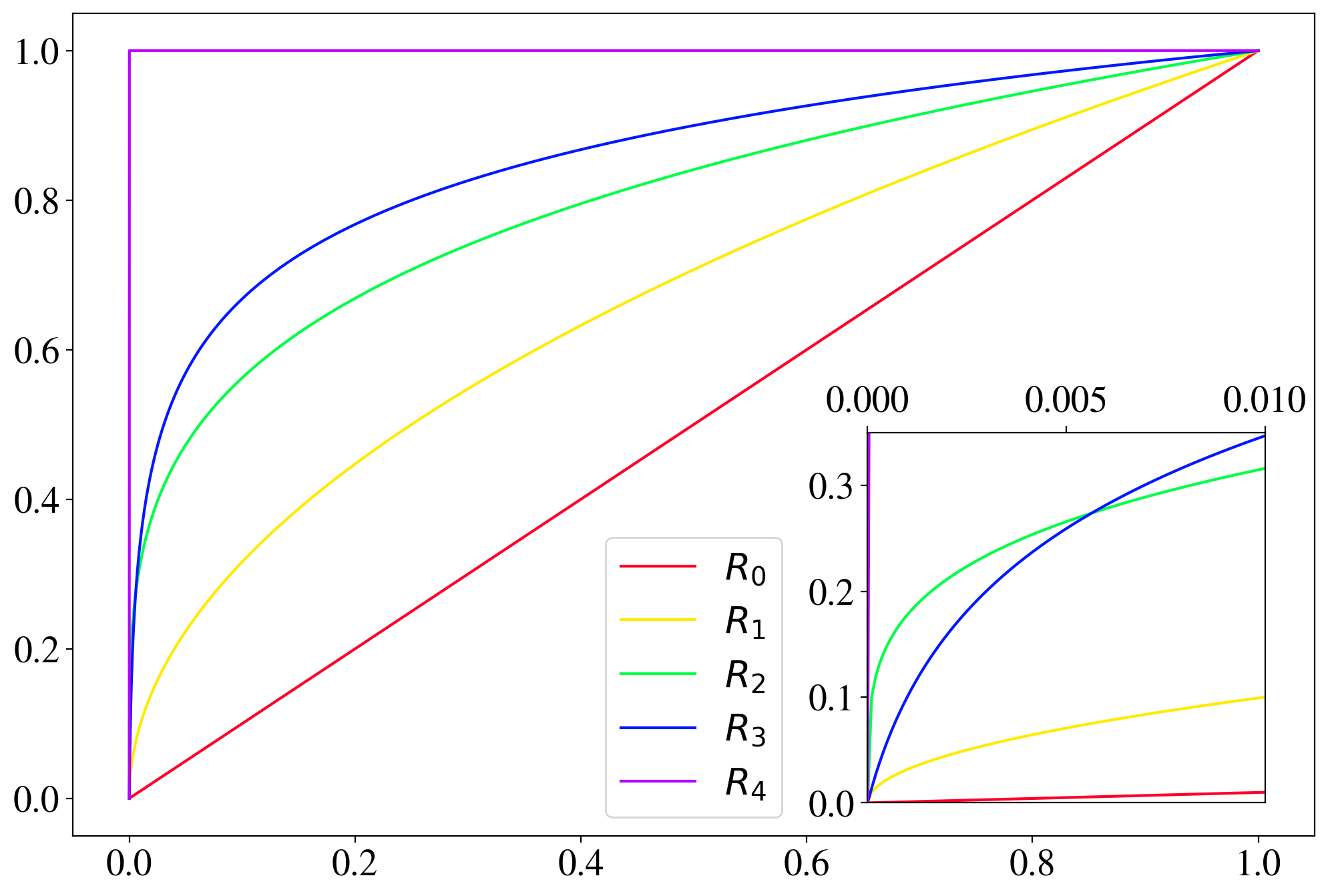}
    \end{center}
    \caption{\label{fig:rep_fct} Remapping functions for highlighting dim pixels, see Eq.~(\ref{eq:preprocessings_app}).}
\end{figure}

In this appendix we exemplify and quantify the effects of the intensity remapping as introduced in Section~\ref{subs:preproc}. The remapping functions, 
\begin{equation}
\text{R}_1(x)=\sqrt{x} \, , \quad \text{R}_2(x)=\sqrt[4]{x}\, , \quad  \text{R}_3(x)=\frac{\log(\alpha x+1)}{\log(\alpha+1)} \, \quad \text{and} \quad \text{R}_4(x)=\Theta(x)\,,
\label{eq:preprocessings_app}
\end{equation}
are displayed in Fig.~\ref{fig:rep_fct}. The identity mapping, corresponding to the original images, is denoted as $\text{R}_0(x)=x$.  The strength of the highlighting of the dimmer pixels is different for the different functions. Moreover, the derivative of the logarithmic remapping, $R_3$, does not diverge at the origin, in contrast to the root functions.

\begin{figure}
    \centering
    \includegraphics[width=0.9\linewidth]{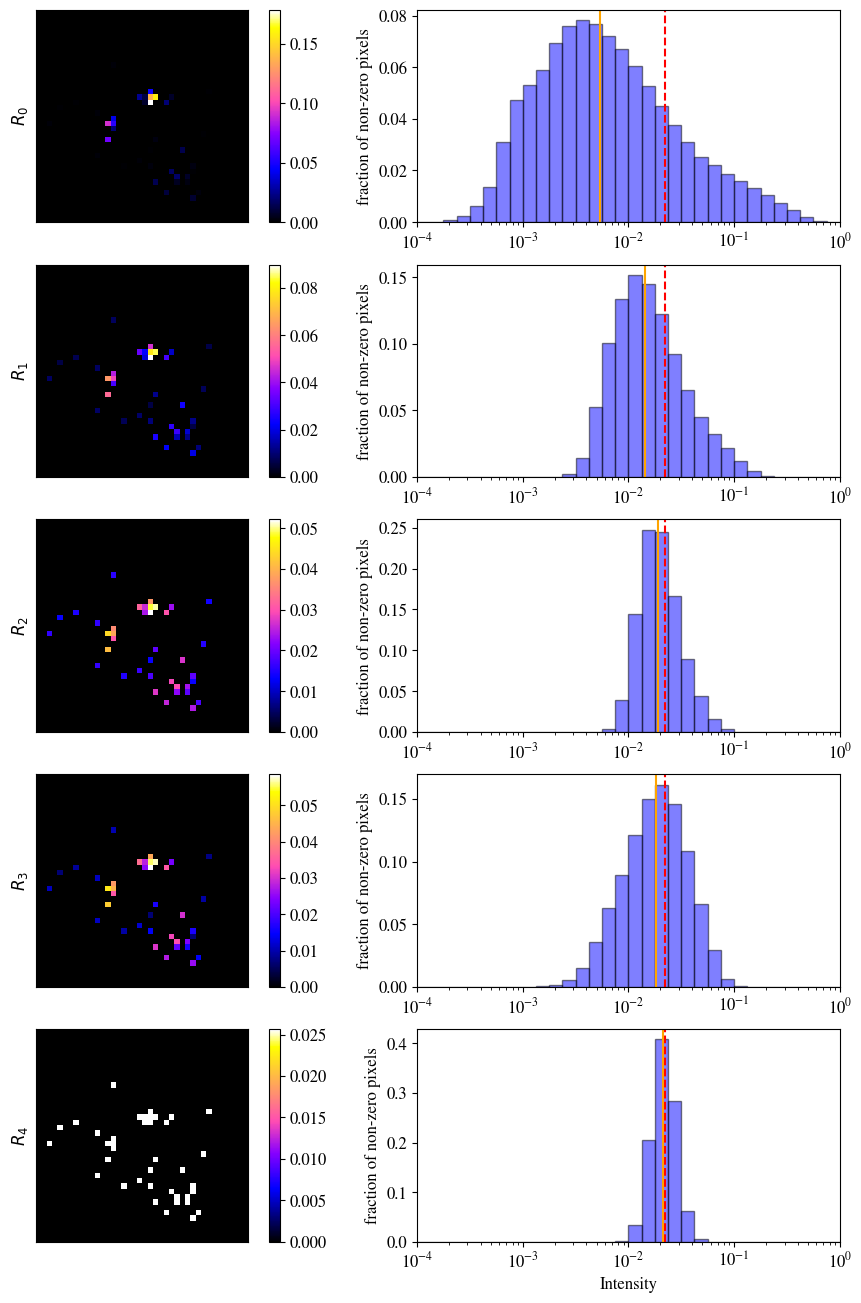}
    \caption{\label{fig:reproc} Example of a top jet image (left), and the distribution of intensities of non-zero pixels for 100k top jet images (right) for different intensity remappings (from top to bottom $\text{R}_0$, $\text{R}_1$, $\text{R}_2$, $\text{R}_3$ and $\text{R}_4$). The orange and red-dashed vertical lines are the median and the average of each distribution, respectively.}
\end{figure}

In the left column of Fig.~\ref{fig:reproc}, we show the effect of the remapping for an exemplary top jet.
The dim region in the bottom right of the $\text{R}_0$ image (upper panel) may be the third sub-jet in the top jet structure. Without highlighting it is barely visible and will most likely be ignored by the AE. By highlighting these dim pixels, the AE can more easily learn the information and features stored in them. Learning more distinctive features of the training set, the AE is also expected to perform better at distinguishing anomalies.

To quantify the effect of the highlighting on the whole set of images, we show the intensity distribution of the non-zero pixels for the 100k top jet images of the training set in the right column of Fig.~\ref{fig:reproc}. In the original images (upper panel) the distribution of the intensities of the non-zero pixels is rather broad, with a median that is more than an order of magnitude below the average value. The intensity remapping compresses the distribution on a logarithmic scale. The distributions after the $\text{R}_1$ and $\text{R}_3$ remapping span across around 1.5 orders of magnitude, and the distribution after the $\text{R}_2$ remapping lies nearly fully within one order of magnitude. Due to the normalization to sum to one, the distribution of pixel intensities for the $\text{R}_4$-images reflects the distribution of the overall number of non-zero pixels. Note that the remappings do not change the number of non-zero pixels. Because of the final normalization, the average intensity of the non-zero pixels is also unchanged.

\section{Further results}
\label{app:more_results}

In this appendix, we complete the display of the results discussed in the main part of this work. 

We first collect the performance measures for the direct and inverse taggers (see Section~\ref{sec:Methods}, Figs.~\ref{fig:dir_rev_top_MES} and \ref{fig:dir_rev_top_FIL}) in Tables~\ref{preproc_table}  and \ref{preproc_table_FIL}. The measures E10 and E100 are defined as $\epsilon_S(1/\epsilon_B = 10)$ and $\epsilon_S (1/\epsilon_B = 100)$, respectively.
\begin{table}[h!]
\centering
\begin{tabular}{ |c|c|c|c|c|c|c| }
    \hline
    background & signal & R & AUC [\%] & E10 [\%] & E100 [\%] & $1/\epsilon_B$ \\
    & & & & & & $(\epsilon_S=0.3)$ \\ 
    \hline
    \hline
    \multirow{5}{*}{QCD} & \multirow{5}{*}{top} &R0 & $91.2^{+0.4 }_{-0.4 }$ & $67.8^{+2.1 }_{-1.9 }$ & $17.5^{+1.9 }_{-1.4 }$ & $46^{+3 }_{-3 }$\\ 
\cline{3-7} & & R1 & $91.9^{+0.6 }_{-1.4 }$ & $72.0^{+2.6 }_{-4.9 }$ & $22.7^{+3.0 }_{-5.2 }$ & $64^{+12 }_{-19 }$\\ 
\cline{3-7} & & R2 & $74.4^{+0.8 }_{-1.0 }$ & $33.0^{+1.4 }_{-1.5 }$ & $6.0^{+0.5 }_{-0.6 }$ & $12^{+1 }_{-1 }$\\ 
\cline{3-7} & & R3 & $84.0^{+0.3 }_{-0.5 }$ & $49.2^{+0.9 }_{-1.0 }$ & $10.4^{+0.9 }_{-0.8 }$ & $23^{+1 }_{-1 }$\\ 
\cline{3-7} & & R4 & $40.4^{+1.1 }_{-1.1 }$ & $0.9^{+0.1 }_{-0.1 }$ & $0.0^{+0.0 }_{-0.0 }$ & $2.2^{+0.1 }_{-0.1 }$\\ 
\hline 
\multirow{5}{*}{top} & \multirow{5}{*}{QCD} &R0 & $36.5^{+2.4 }_{-2.0 }$ & $6.6^{+0.9 }_{-1.0 }$ & $0.6^{+0.2 }_{-0.2 }$ & $2.2^{+0.2 }_{-0.2 }$\\ 
\cline{3-7} & & R1 & $56.6^{+0.7 }_{-1.1 }$ & $23.4^{+0.9 }_{-1.6 }$ & $8.6^{+1.2 }_{-1.6 }$ & $6.1^{+0.3 }_{-0.6 }$\\ 
\cline{3-7} & & R2 & $68.6^{+0.4 }_{-0.6 }$ & $42.1^{+1.0 }_{-1.0 }$ & $20.9^{+1.3 }_{-1.3 }$ & $31^{+5 }_{-4 }$\\ 
\cline{3-7} & & R3 & $66.2^{+0.6 }_{-0.9 }$ & $37.3^{+1.4 }_{-2.2 }$ & $18.0^{+1.2 }_{-2.4 }$ & $19^{+3 }_{-4 }$\\ 
\cline{3-7} & & R4 & $70.4^{+0.2 }_{-0.2 }$ & $45.2^{+0.5 }_{-0.5 }$ & $25.2^{+0.3 }_{-0.4 }$ & $50^{+3 }_{-2 }$\\ 
\hline 

\end{tabular}
\caption{Performance measures for AE-based taggers for different remappings of the jet images trained using MSE. Each value is an average of values for 4 autoencoders trained using the same hyperparameters but different initialization. The mentioned uncertainties denote the envelope of each value.}
\label{preproc_table}
\end{table}

\begin{table}[h!]
\centering
\begin{tabular}{ |c|c|c|c|c|c|c| }
    \hline
    background & signal & R & AUC [\%] & E10 [\%] & E100 [\%] & $1/\epsilon_B$ \\
    & & & & & & $(\epsilon_S=0.3)$ \\ 
    \hline
    \hline
    \multirow{5}{*}{QCD} & \multirow{5}{*}{top} &R0 & $92.8^{+0.1 }_{-0.1 }$ & $75.5^{+1.1 }_{-0.9 }$ & $18.5^{+3.0 }_{-1.4 }$ & $54^{+9 }_{-4 }$\\ 
\cline{3-7} & & R1 & $89.7^{+0.4 }_{-1.0 }$ & $66.8^{+1.5 }_{-3.1 }$ & $18.0^{+1.5 }_{-2.2 }$ & $48^{+6 }_{-7 }$\\ 
\cline{3-7} & & R2 & $58.9^{+2.3 }_{-2.4 }$ & $10.4^{+2.3 }_{-2.1 }$ & $0.7^{+0.3 }_{-0.2 }$ & $4.2^{+0.5 }_{-0.4 }$\\ 
\cline{3-7} & & R3 & $67.1^{+0.7 }_{-0.5 }$ & $18.8^{+1.0 }_{-0.9 }$ & $2.0^{+0.2 }_{-0.2 }$ & $6.3^{+0.2 }_{-0.2 }$\\ 
\cline{3-7} & & R4 & $36.8^{+0.3 }_{-0.2 }$ & $0.4^{+0.1 }_{-0.0 }$ & $0.0^{+0.0 }_{-0.0 }$ & $1.9^{+0.1 }_{-0.1 }$\\ 
\hline 
\multirow{5}{*}{top} & \multirow{5}{*}{QCD} &R0 & $48.6^{+2.5 }_{-3.0 }$ & $13.3^{+2.0 }_{-1.5 }$ & $1.1^{+0.7 }_{-0.4 }$ & $3.5^{+0.3 }_{-0.4 }$\\ 
\cline{3-7} & & R1 & $62.3^{+1.3 }_{-2.1 }$ & $33.2^{+2.4 }_{-3.0 }$ & $16.3^{+3.9 }_{-3.3 }$ & $14^{+4 }_{-4 }$\\ 
\cline{3-7} & & R2 & $67.6^{+1.1 }_{-0.5 }$ & $41.1^{+2.5 }_{-1.1 }$ & $21.8^{+3.8 }_{-1.7 }$ & $32^{+14 }_{-6 }$\\ 
\cline{3-7} & & R3 & $65.3^{+0.6 }_{-0.9 }$ & $37.2^{+1.1 }_{-1.8 }$ & $18.0^{+1.0 }_{-2.6 }$ & $20^{+2 }_{-4 }$\\ 
\cline{3-7} & & R4 & $69.6^{+0.3 }_{-0.3 }$ & $43.1^{+0.4 }_{-0.4 }$ & $23.3^{+0.7 }_{-0.6 }$ & $38^{+3 }_{-3 }$\\ 
\hline 

\end{tabular}
\caption{Performance measures for AE-based taggers for different remappings of the jet images trained using KMSE. Each value is an average of values for 4 autoencoders trained using the same hyperparameters but different initialization. The mentioned uncertainties denote the envelope of each value.}
\label{preproc_table_FIL}
\end{table}

\begin{figure}
    \centering
    \includegraphics[width=0.24\linewidth]{fig/pixnum_loss/pixnum_loss_QCDR0.png}
    \includegraphics[width=0.24\linewidth]{fig/pixnum_loss/pixnum_loss_topR0.png}
    \includegraphics[width=0.24\linewidth]{fig/pixnum_loss/pixnum_loss_QCDR0FIL.png}
    \includegraphics[width=0.24\linewidth]{fig/pixnum_loss/pixnum_loss_topR0FIL.png}
    \includegraphics[width=0.24\linewidth]{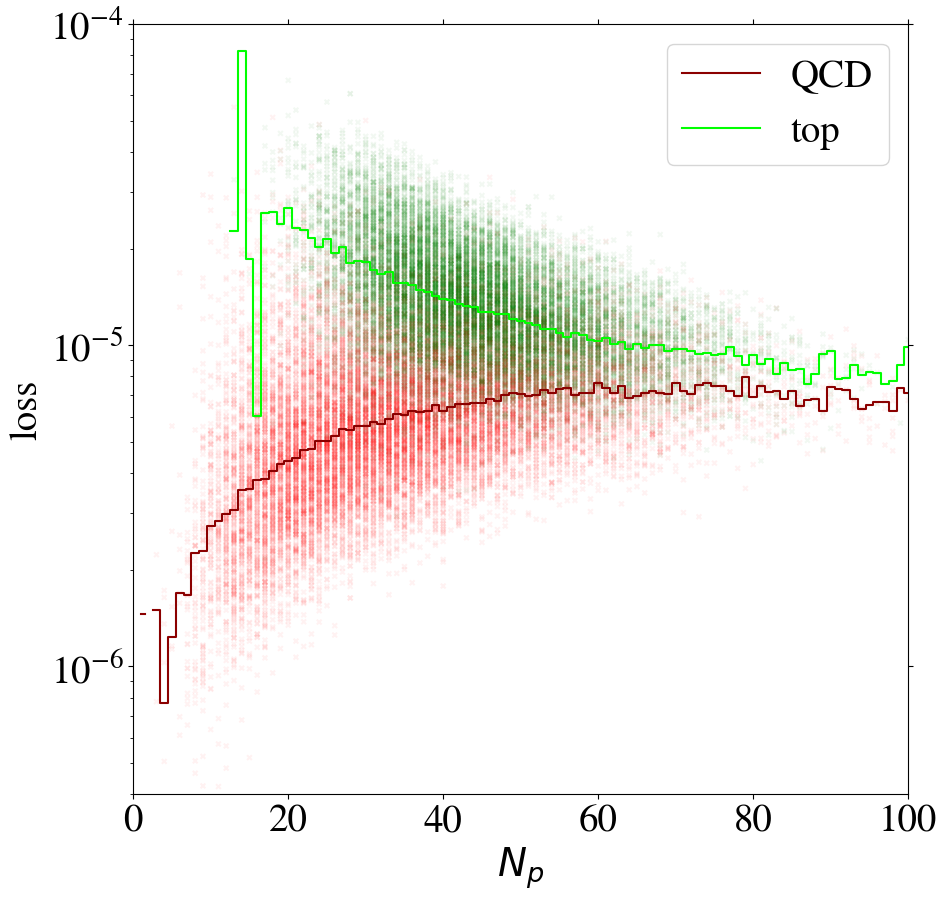}
    \includegraphics[width=0.24\linewidth]{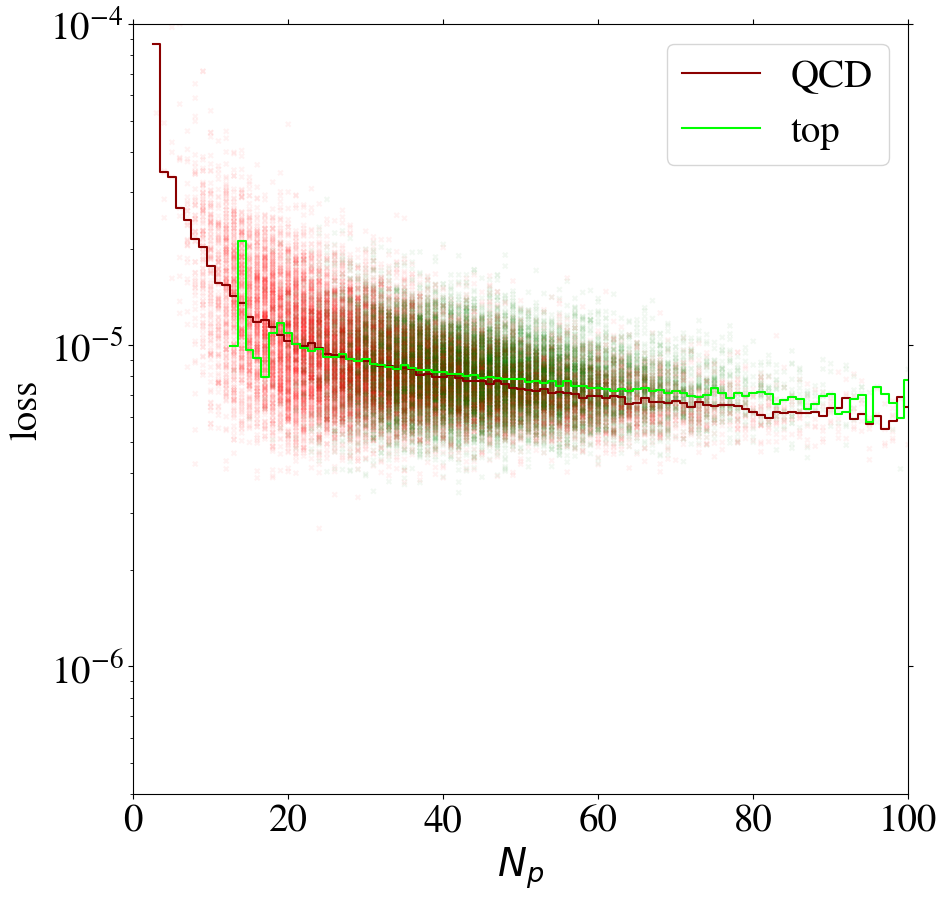}
    \includegraphics[width=0.24\linewidth]{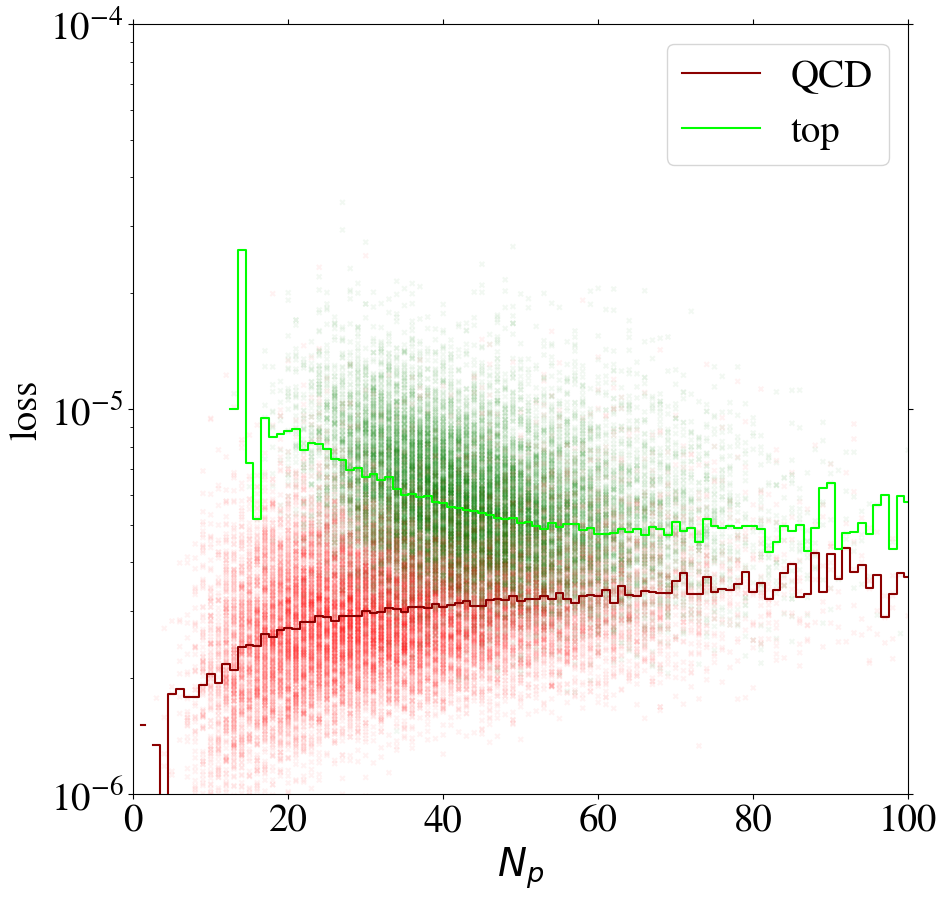}
    \includegraphics[width=0.24\linewidth]{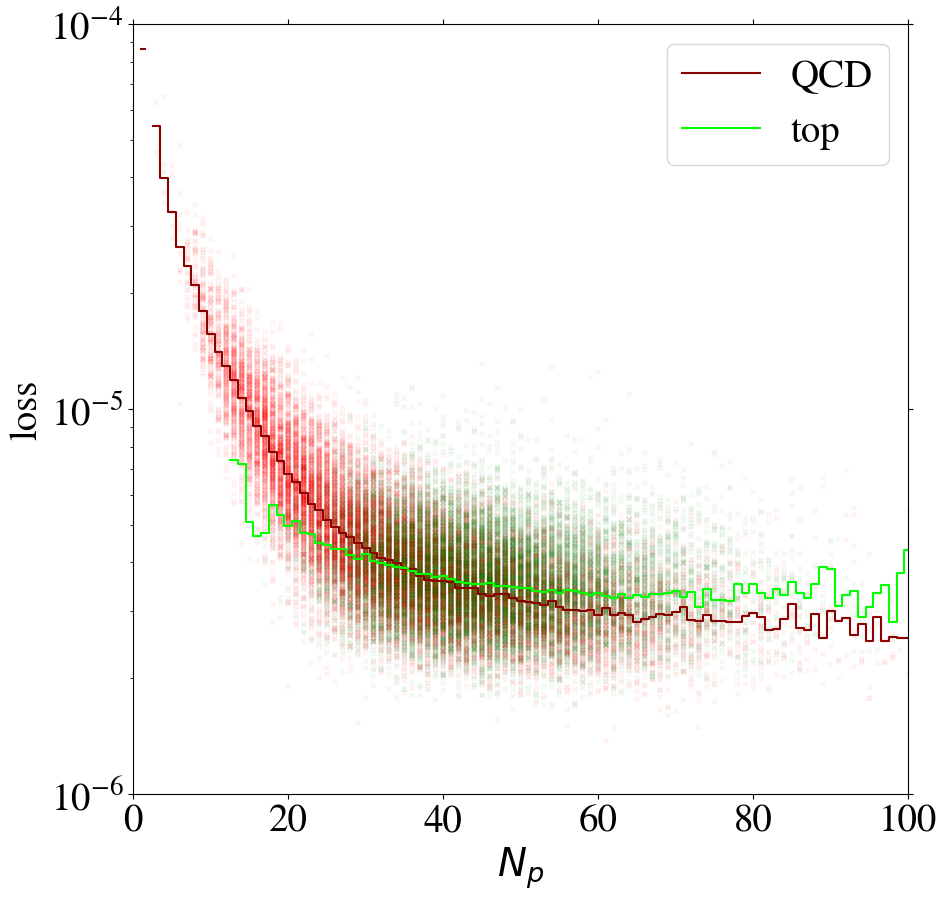}
    \includegraphics[width=0.24\linewidth]{fig/pixnum_loss/pixnum_loss_QCDR2.png}
    \includegraphics[width=0.24\linewidth]{fig/pixnum_loss/pixnum_loss_topR2.png}
    \includegraphics[width=0.24\linewidth]{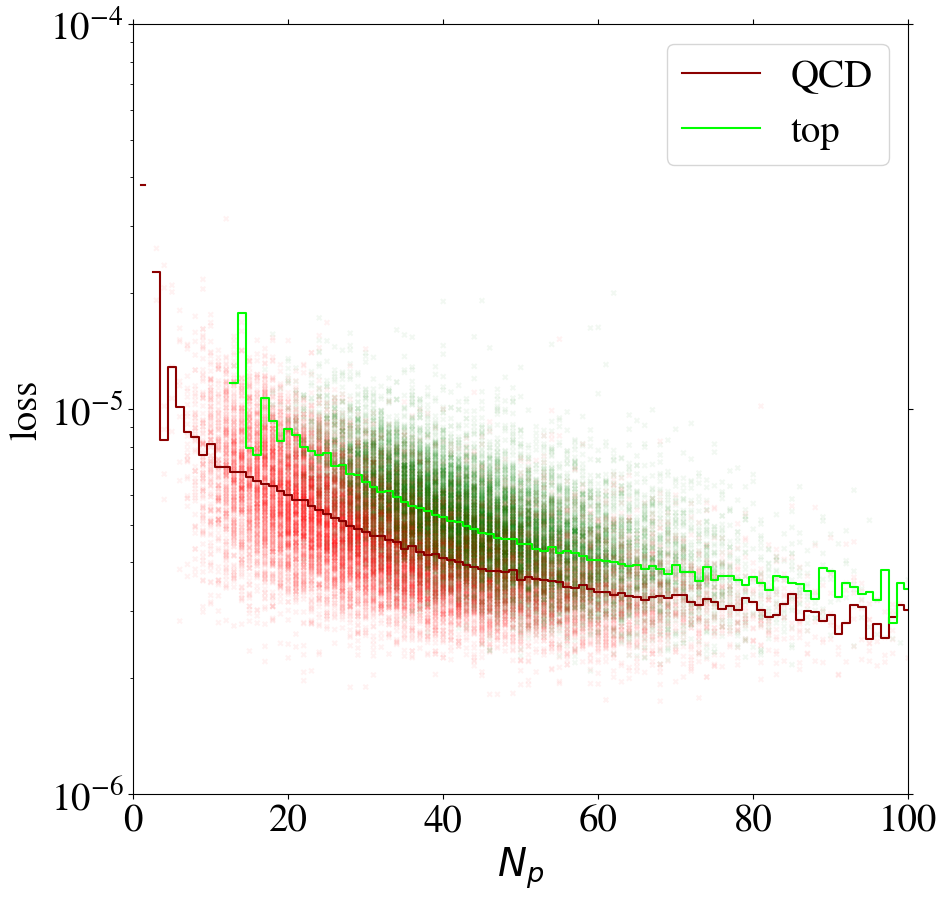}
    \includegraphics[width=0.24\linewidth]{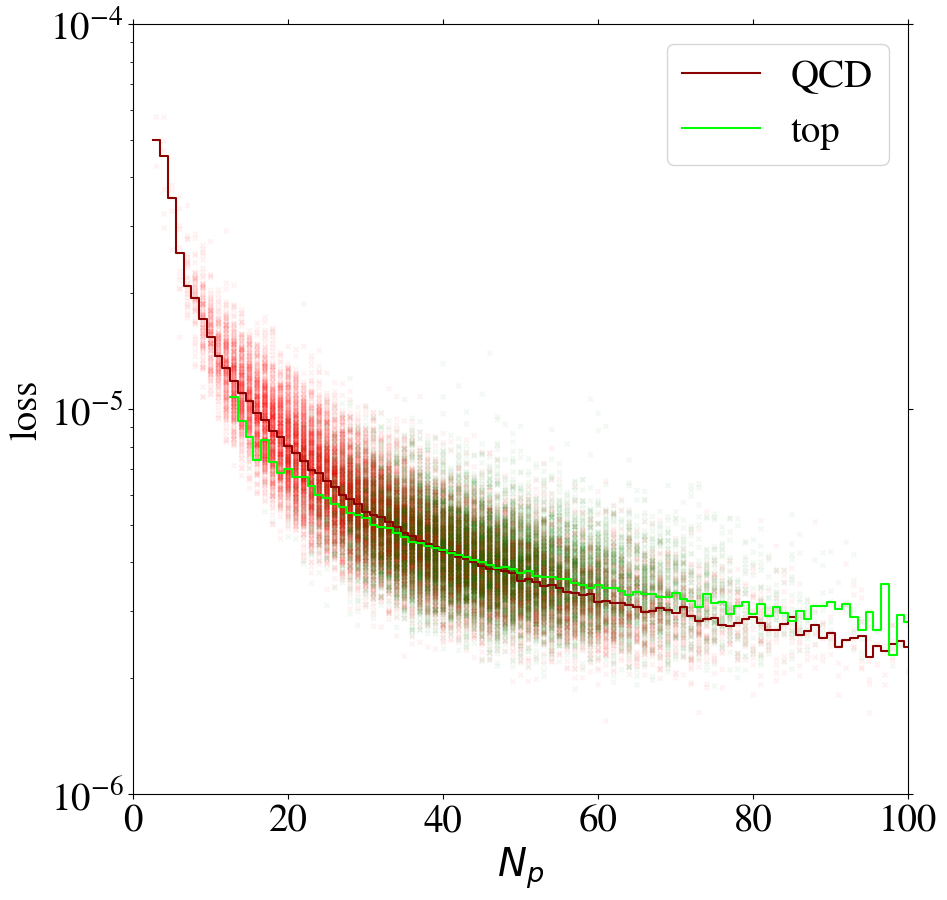}
    \includegraphics[width=0.24\linewidth]{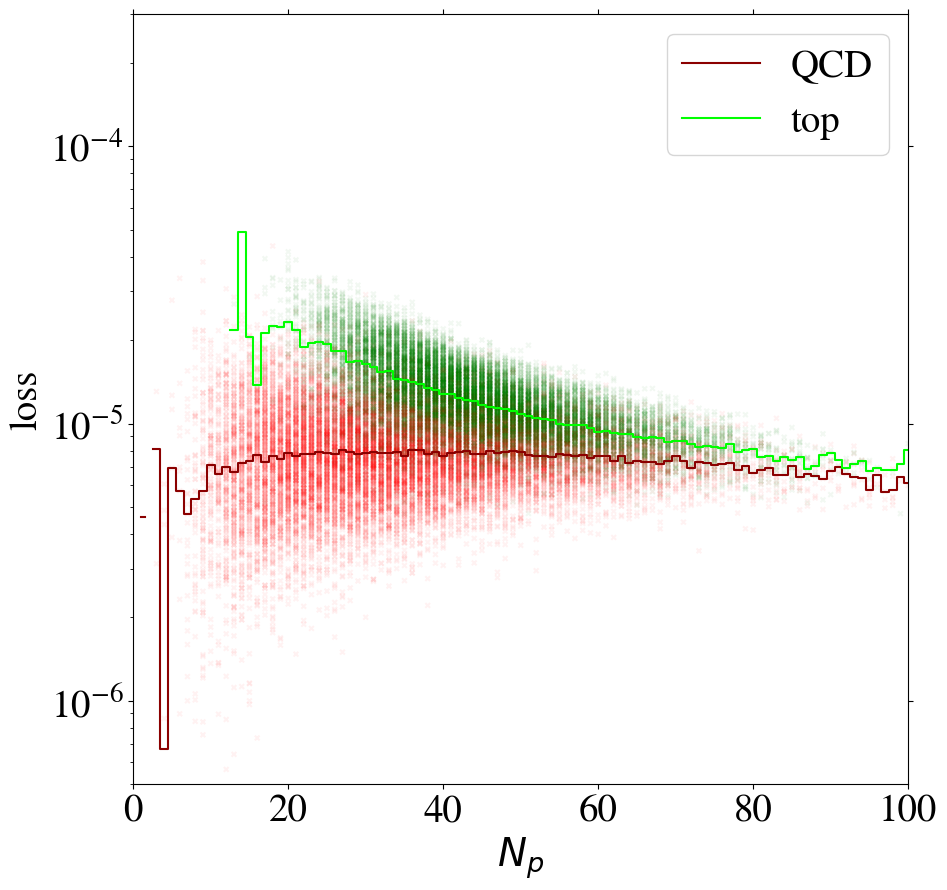}
    \includegraphics[width=0.24\linewidth]{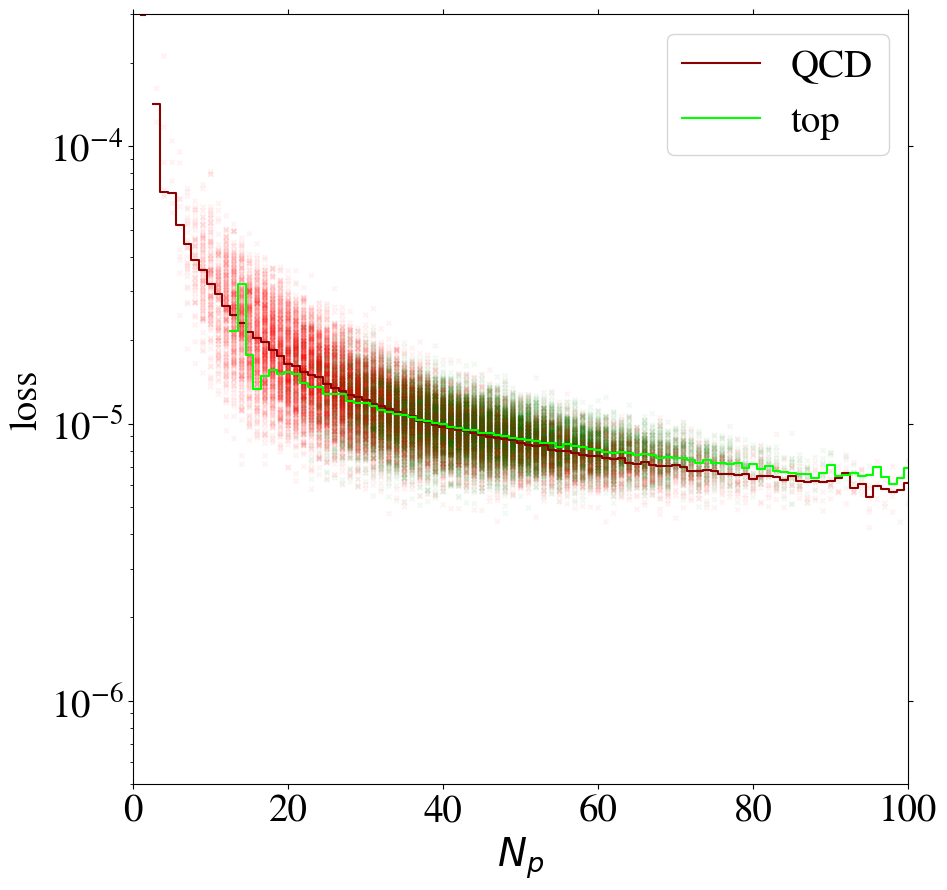}
    \includegraphics[width=0.24\linewidth]{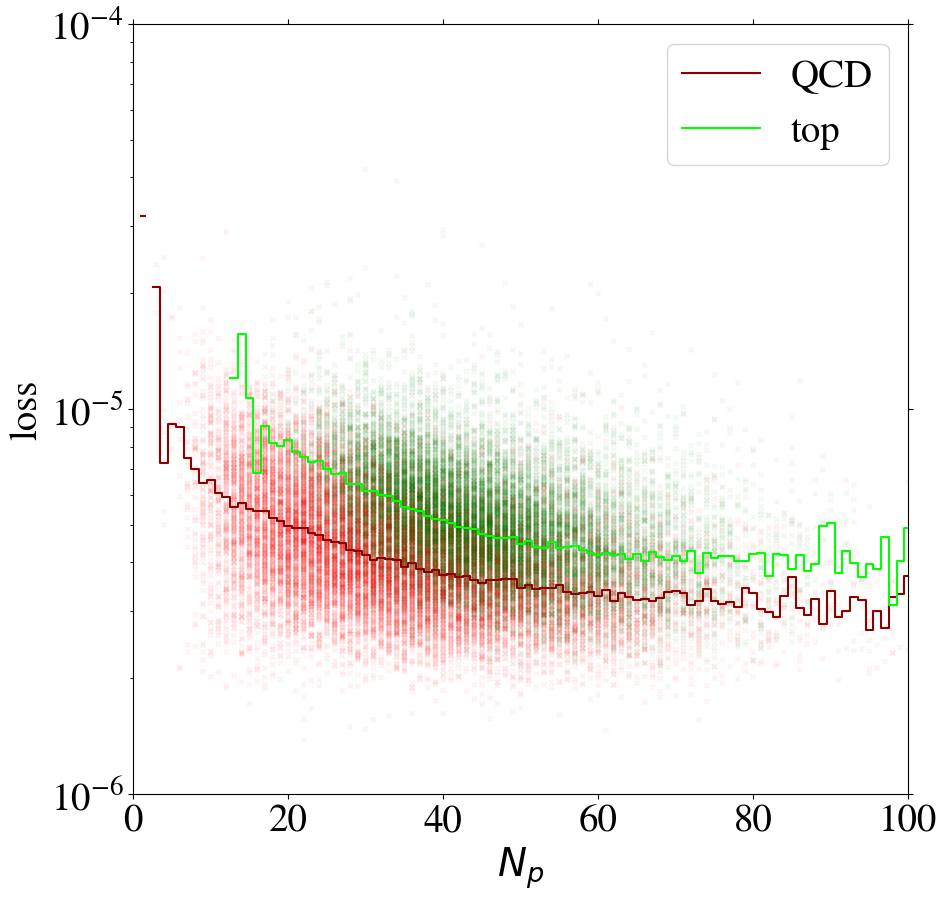}
    \includegraphics[width=0.24\linewidth]{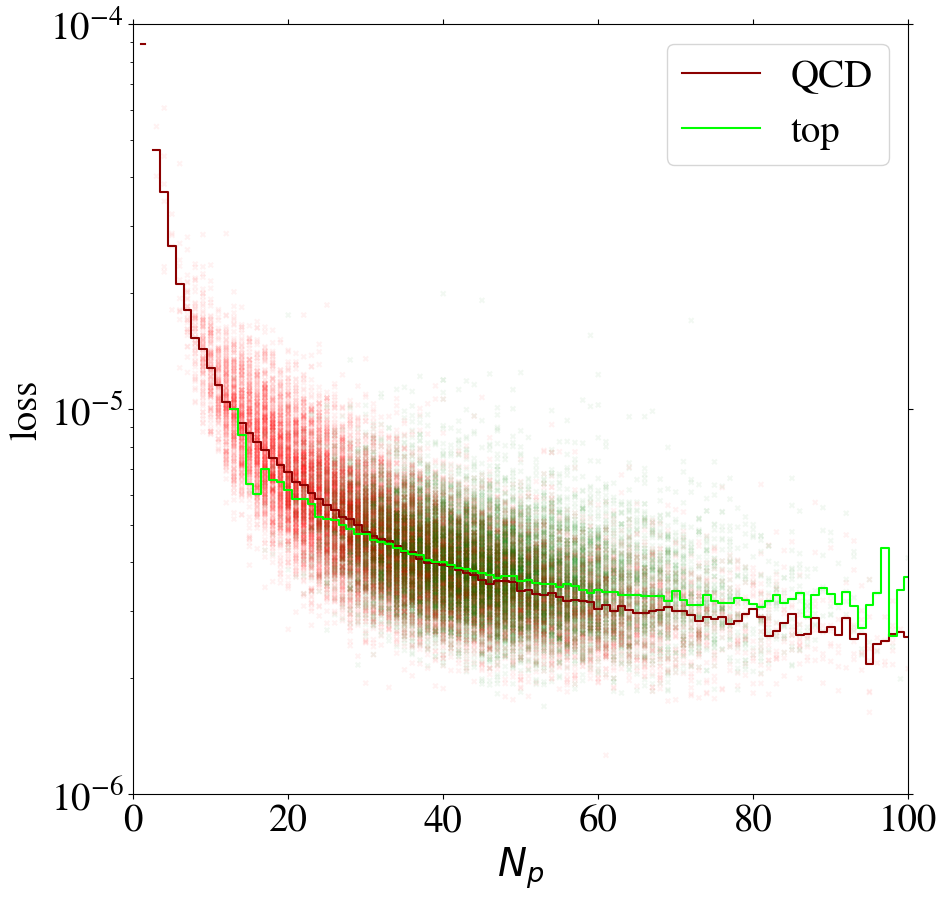}
    \includegraphics[width=0.24\linewidth]{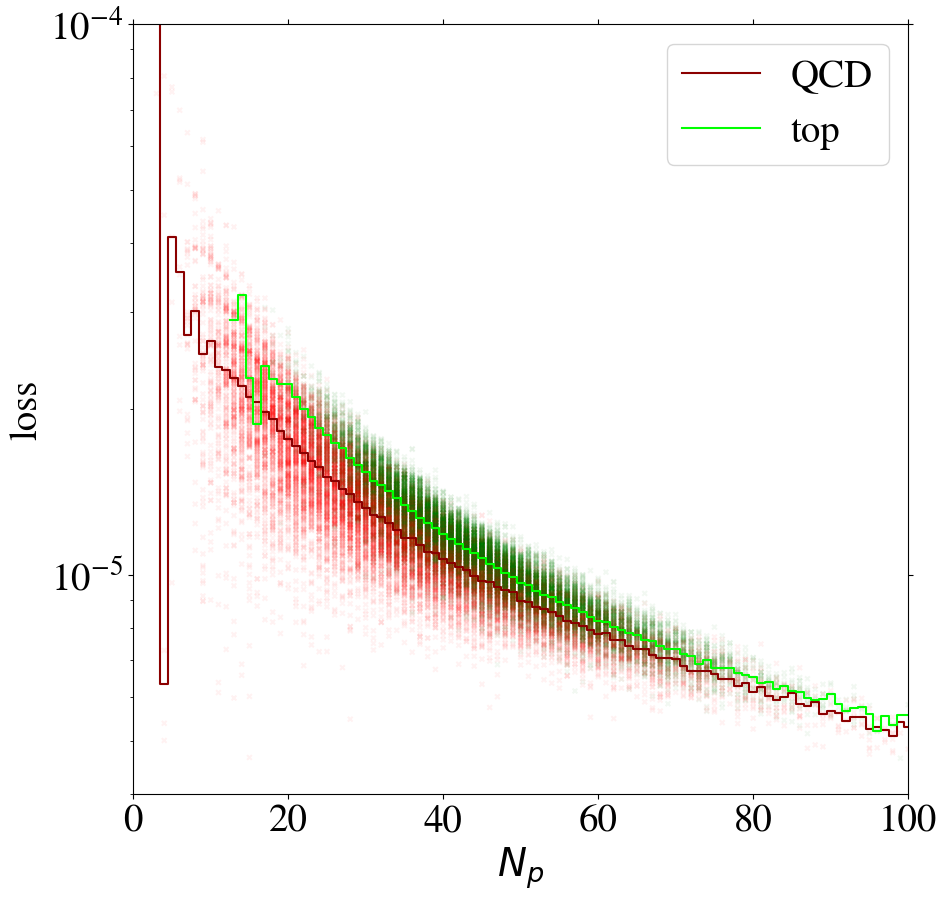}
    \includegraphics[width=0.24\linewidth]{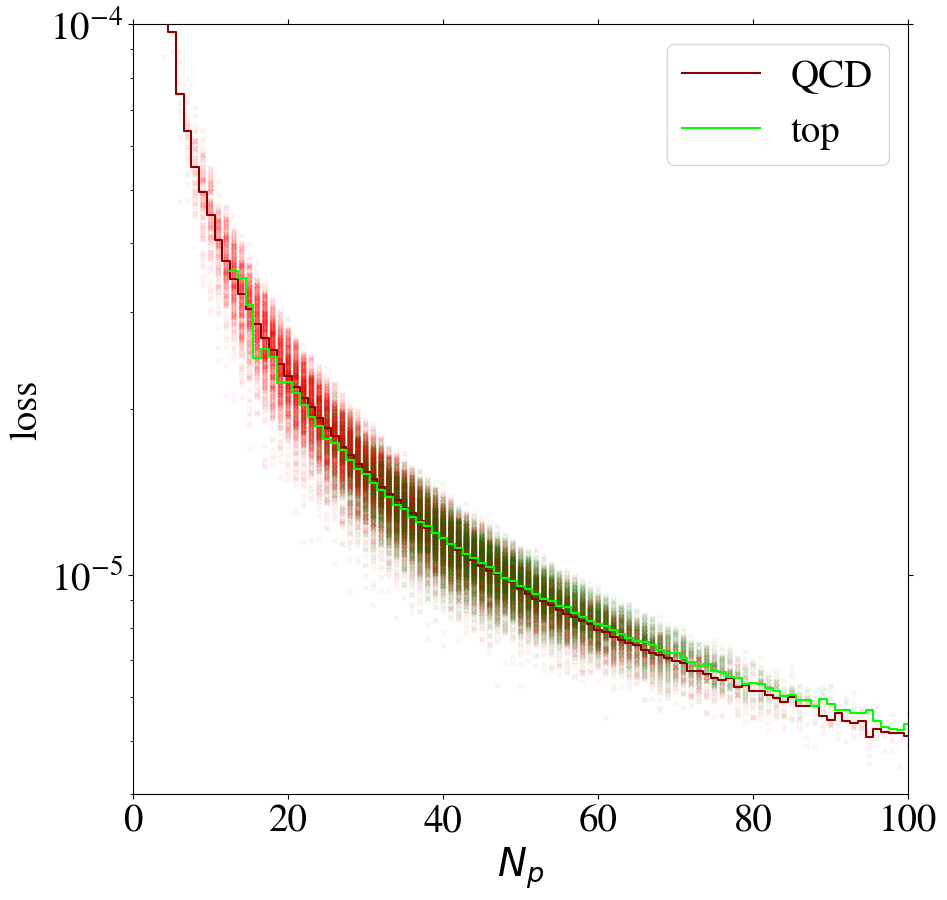}
    \includegraphics[width=0.24\linewidth]{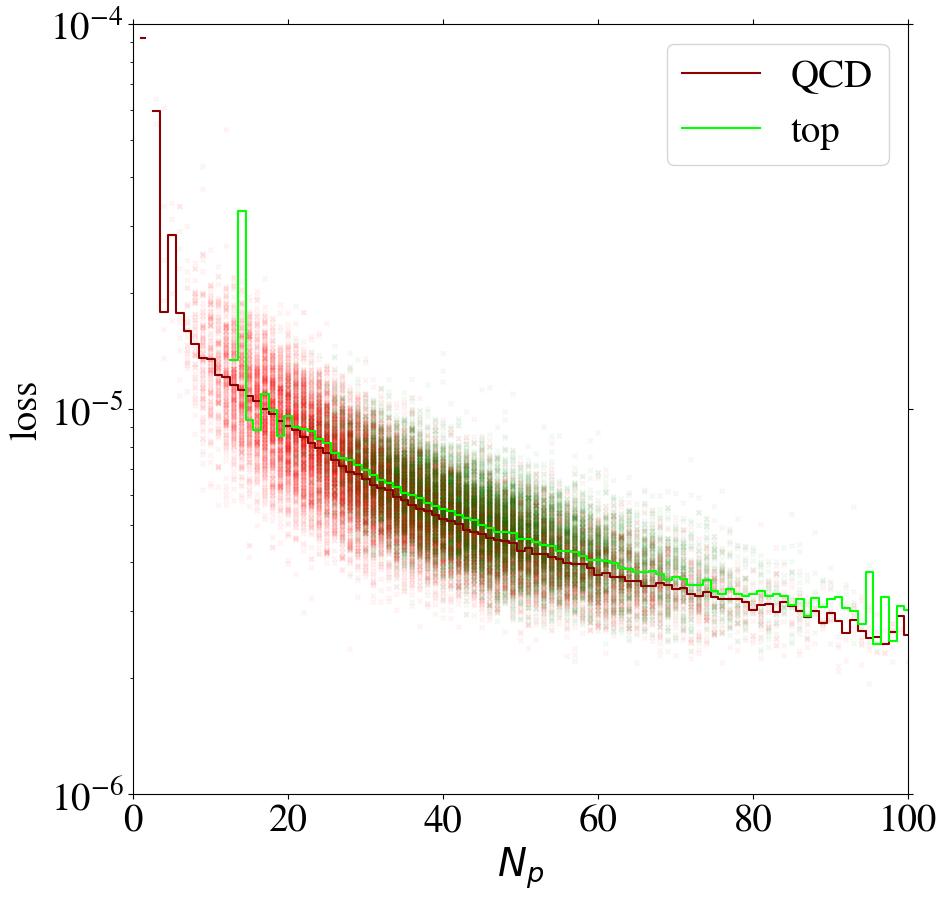}
    \includegraphics[width=0.24\linewidth]{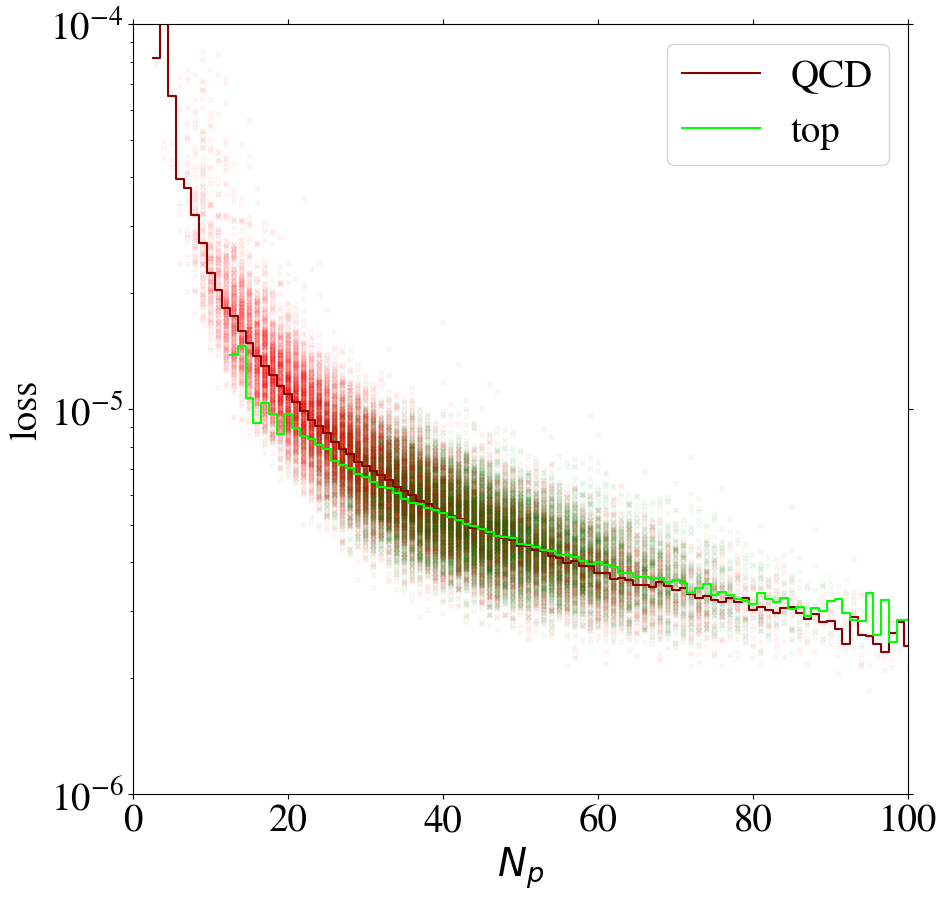}
    \caption{\label{fig:pixnum_all_appendix} Reconstruction loss of individual test jets as a function of the number of non-zero pixels $N_p$, cf.\ Fig.~\ref{fig:pixnum_MSE}. The first and second columns represent the results for an AE trained on QCD and top jet images, respectively, using MSE loss, while the third and fourth columns show the corresponding results using the KMSE loss function. The rows correspond to the different image remappings $\text{R}_0$, $\text{R}_1$, $\text{R}_2$, $\text{R}_3$ and $\text{R}_4$ from top to bottom.}
\end{figure}

\begin{figure}
	\centering
	\includegraphics[width=0.24\linewidth]{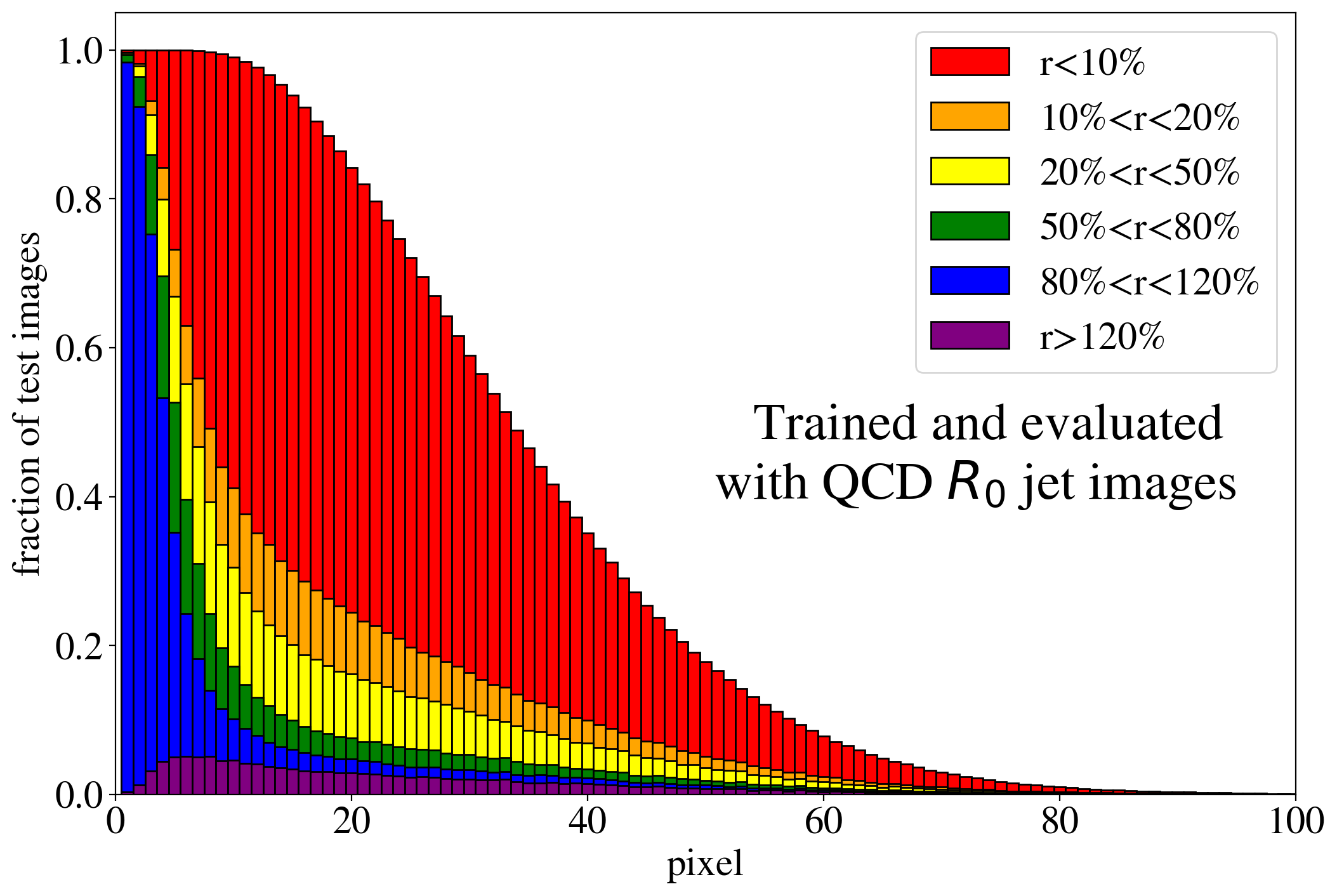}
	\includegraphics[width=0.24\linewidth]{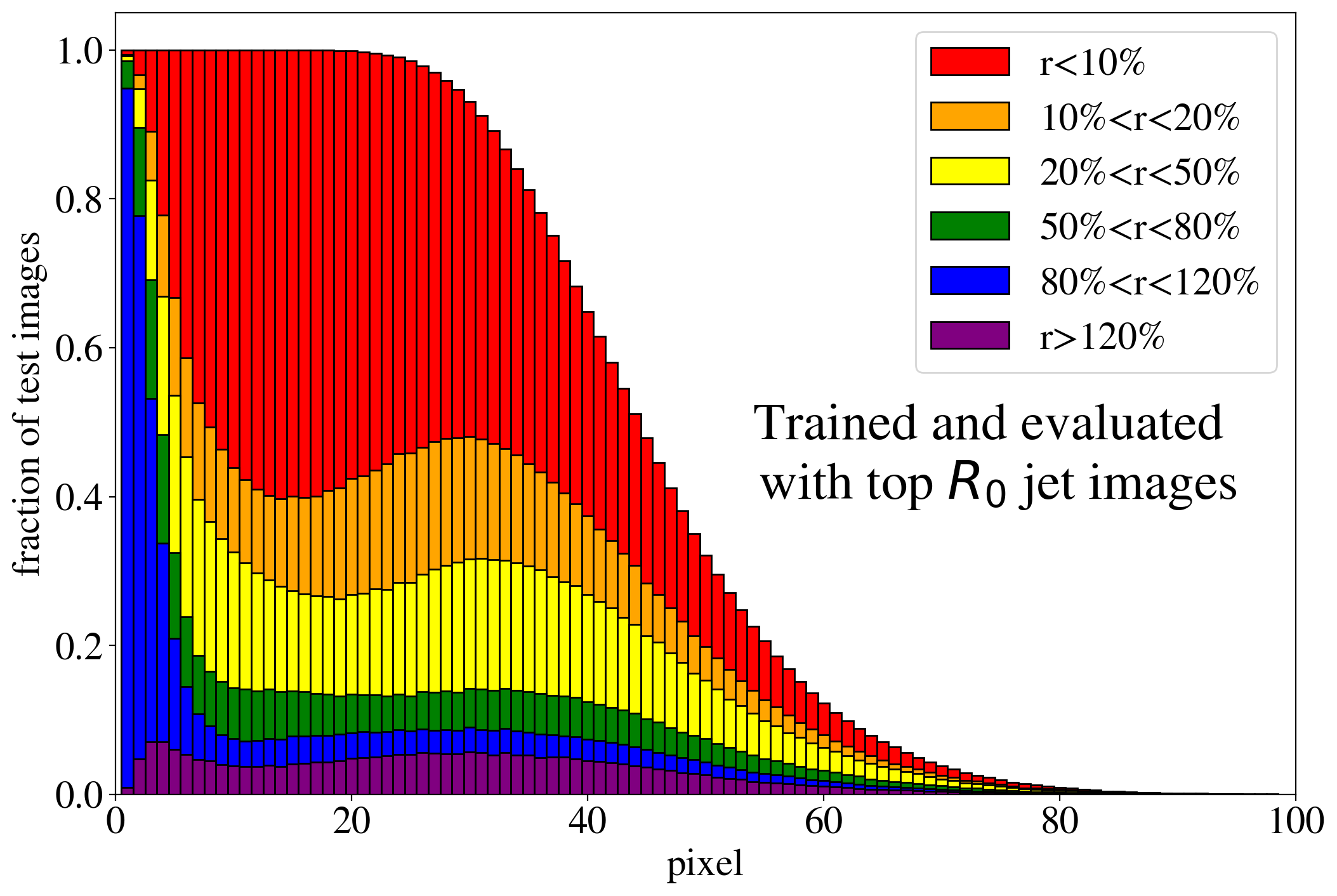}
	\includegraphics[width=0.24\linewidth]{fig/spectra/spectra_order_QCDR0FIL.png}
	\includegraphics[width=0.24\linewidth]{fig/spectra/spectra_order_topR0FIL.png}
	\includegraphics[width=0.24\linewidth]{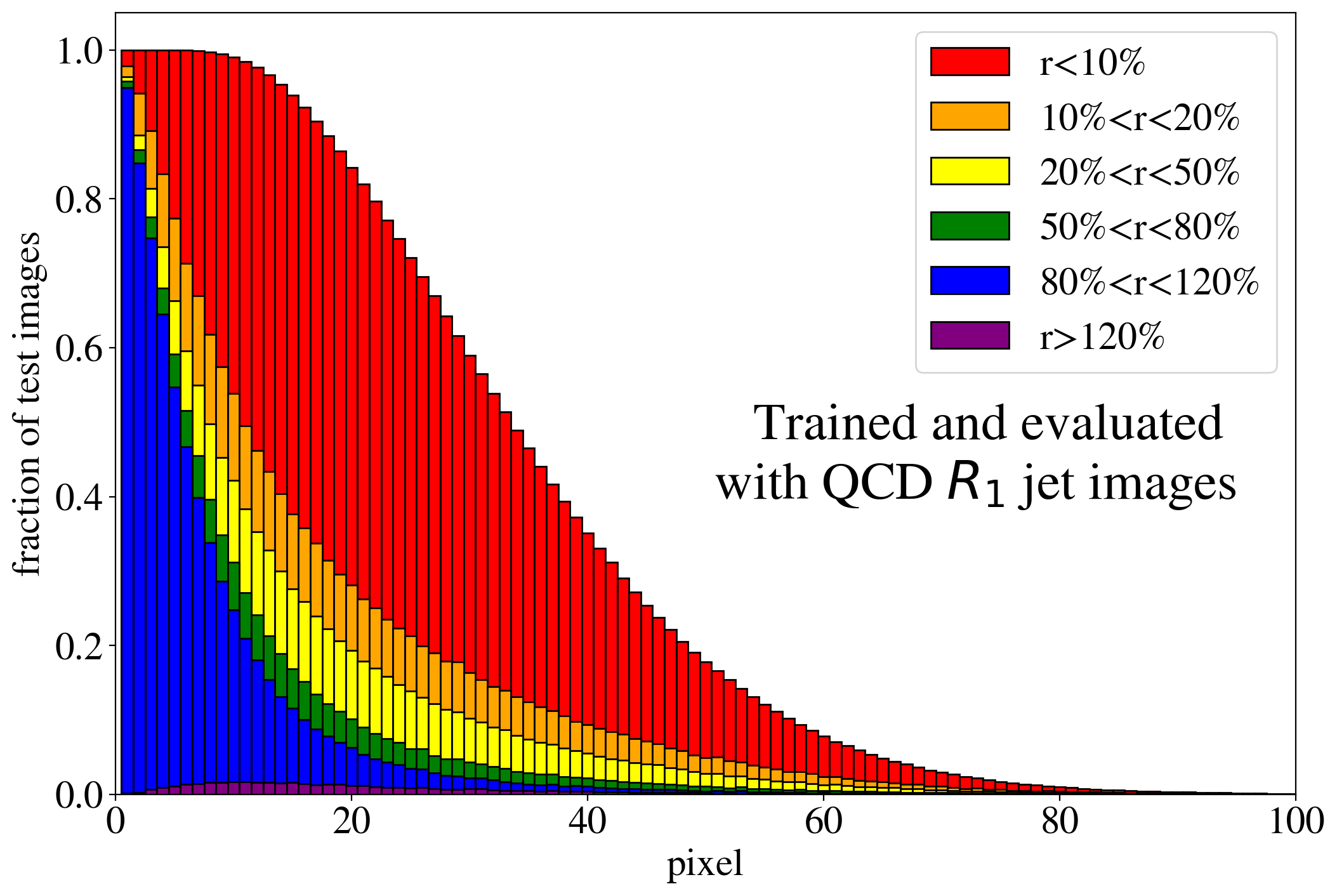}
	\includegraphics[width=0.24\linewidth]{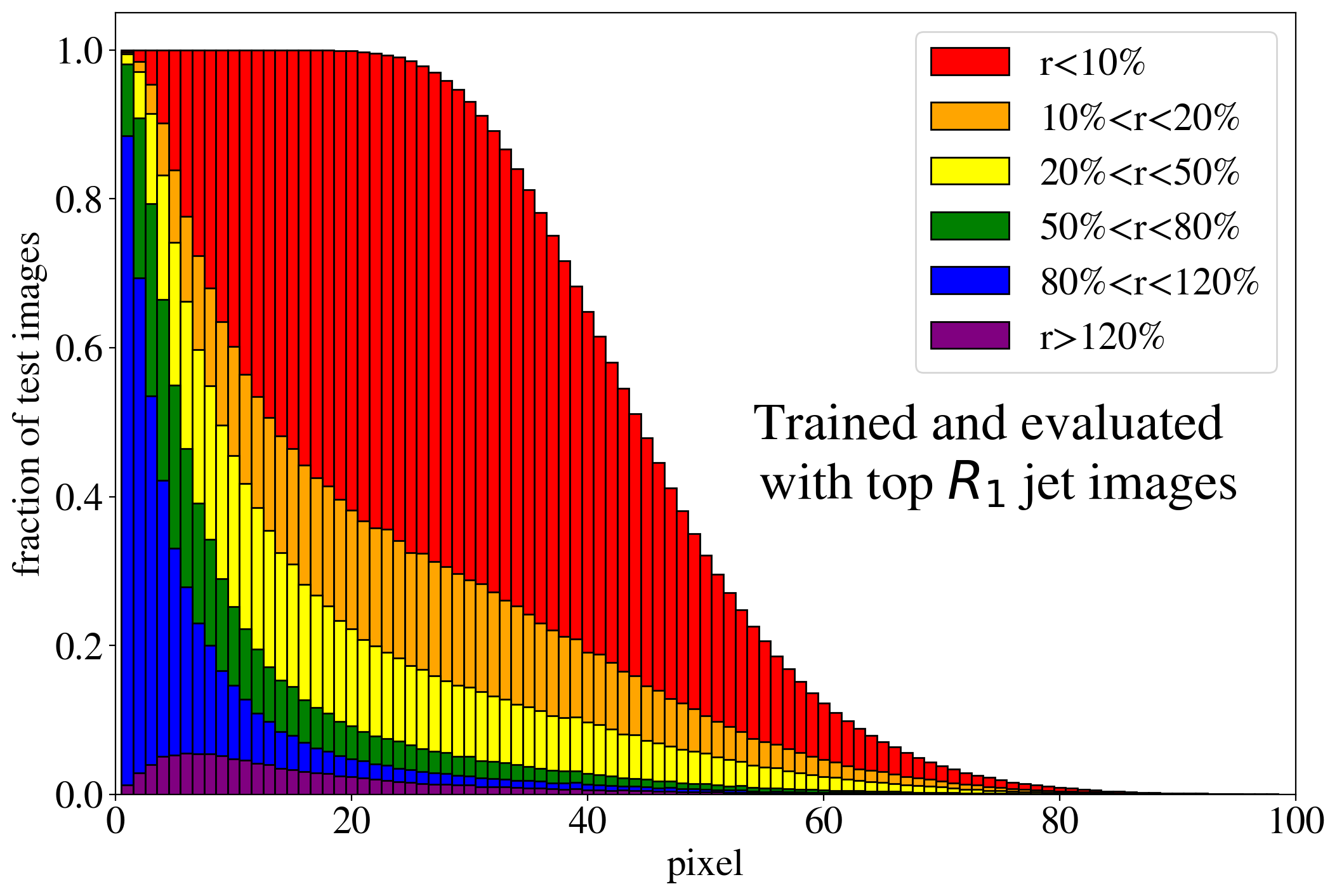}
	\includegraphics[width=0.24\linewidth]{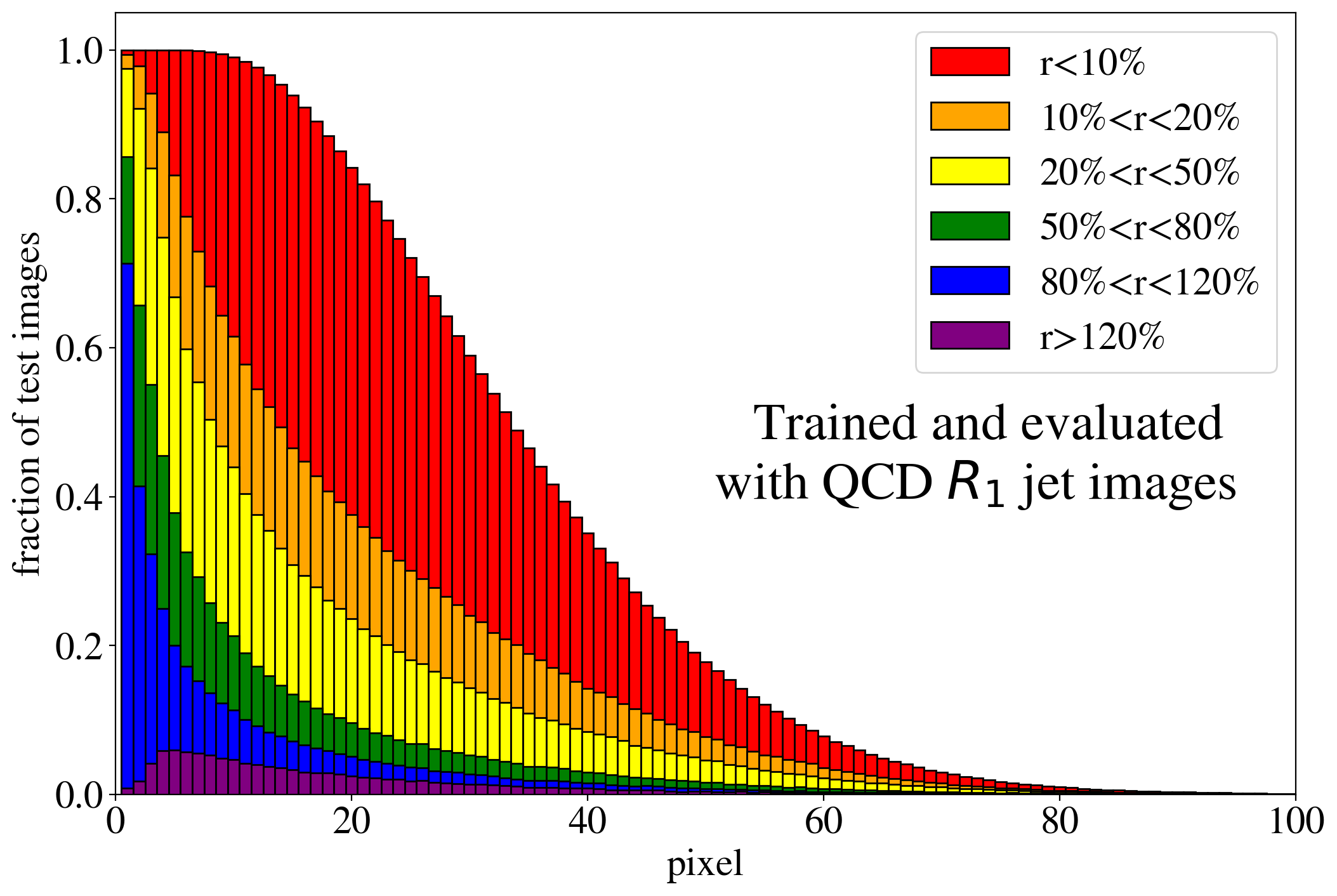}
	\includegraphics[width=0.24\linewidth]{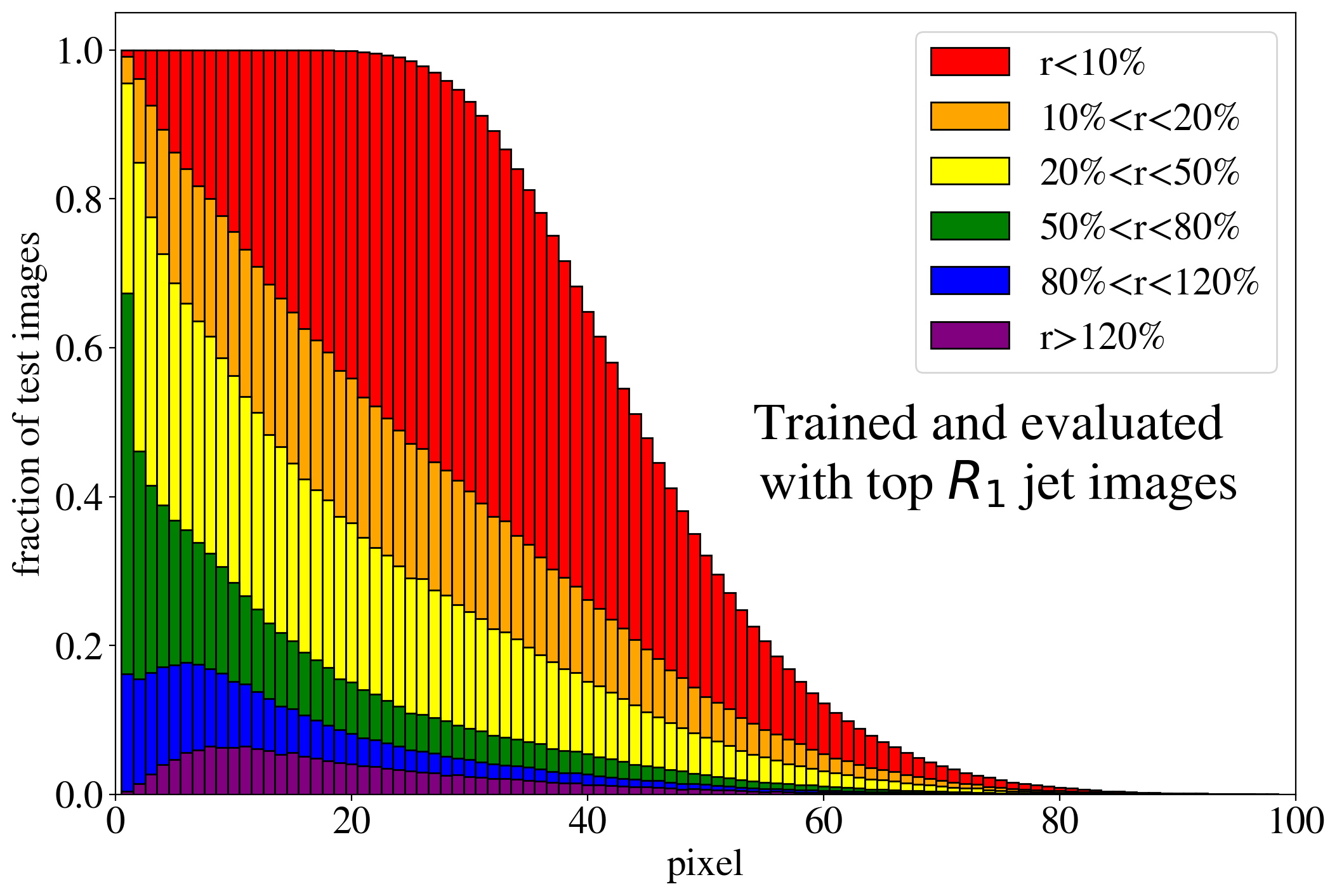}
	\includegraphics[width=0.24\linewidth]{fig/spectra/spectra_order_QCDR2.png}
	\includegraphics[width=0.24\linewidth]{fig/spectra/spectra_order_topR2.png}
	\includegraphics[width=0.24\linewidth]{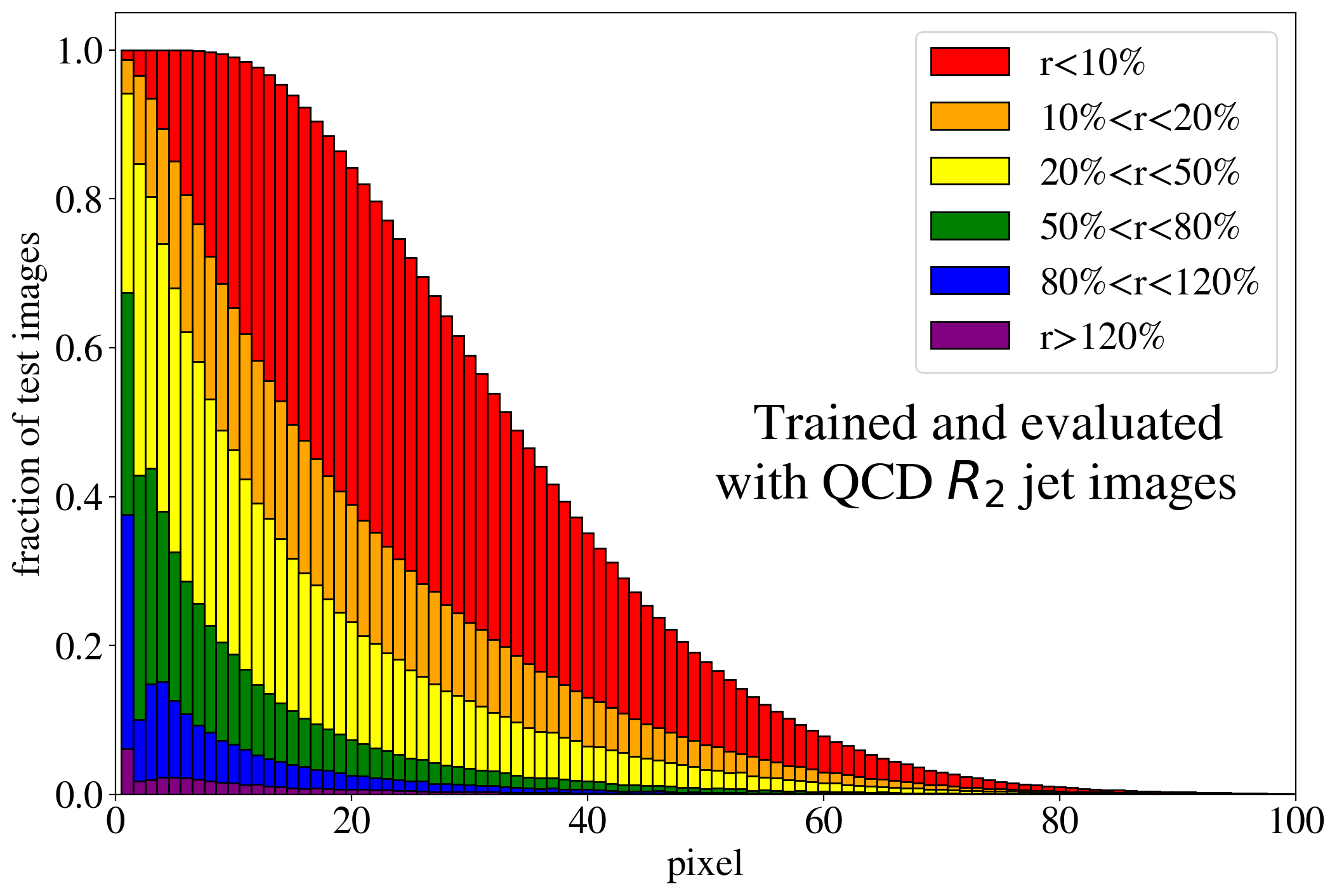}
	\includegraphics[width=0.24\linewidth]{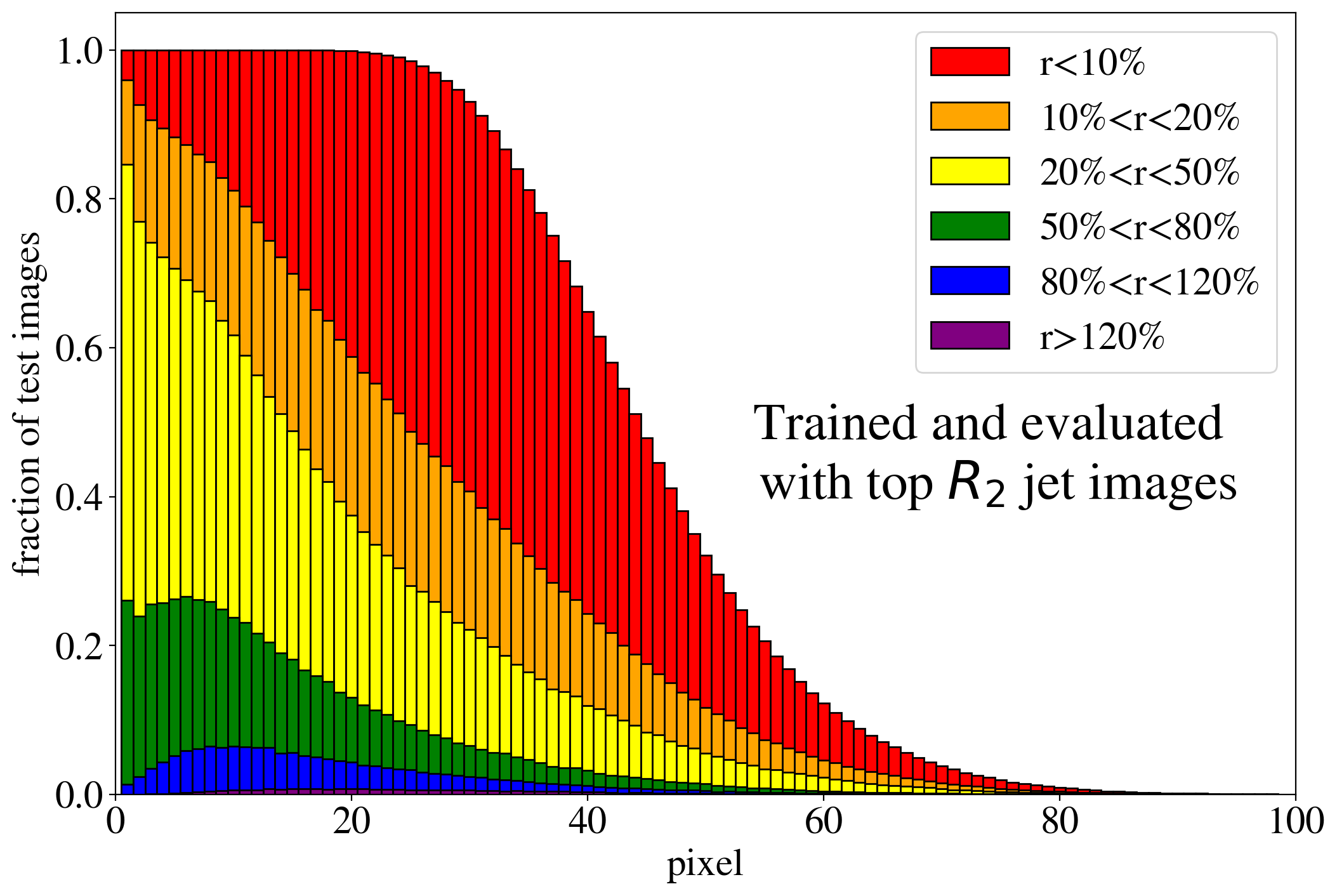}
	\includegraphics[width=0.24\linewidth]{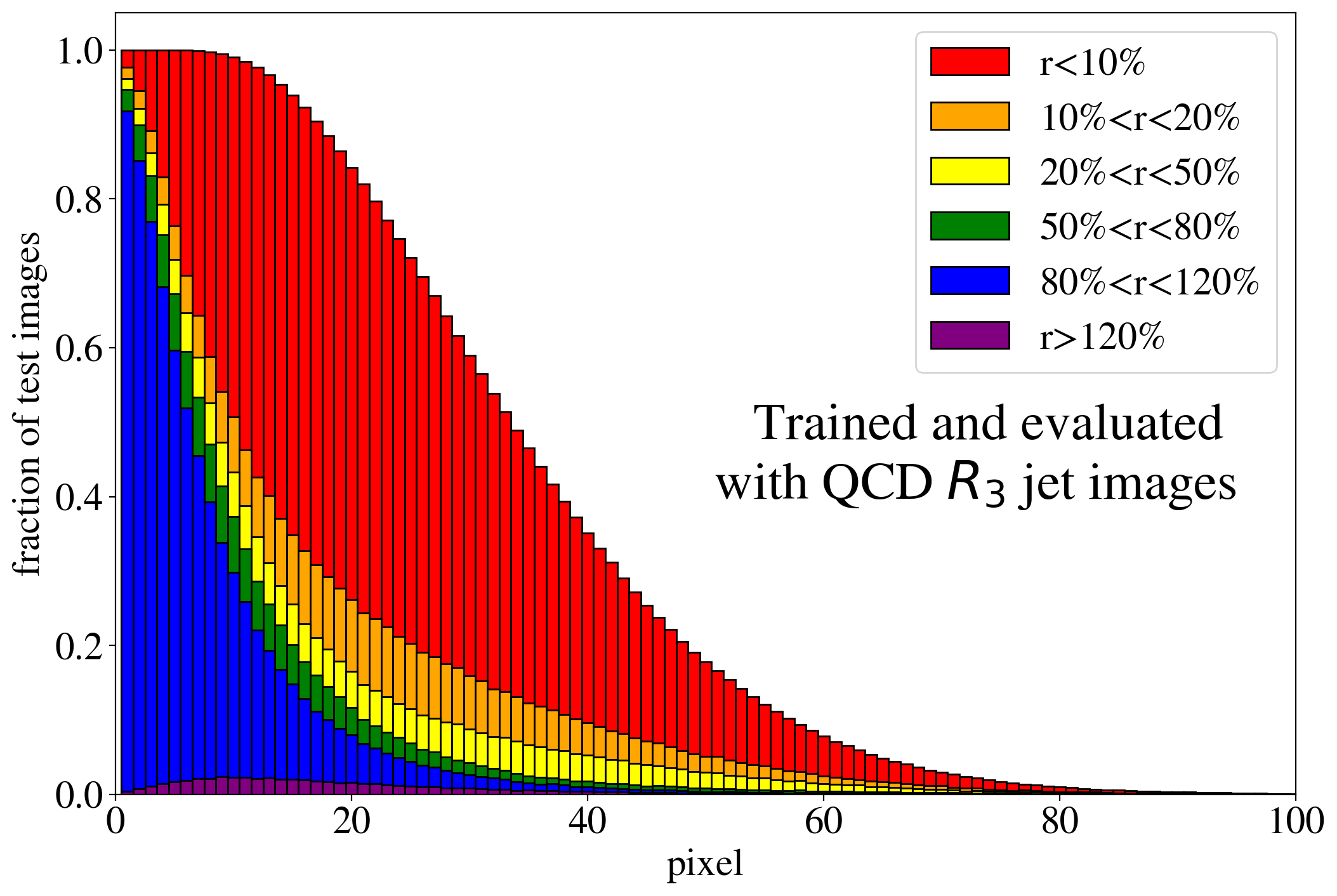}
	\includegraphics[width=0.24\linewidth]{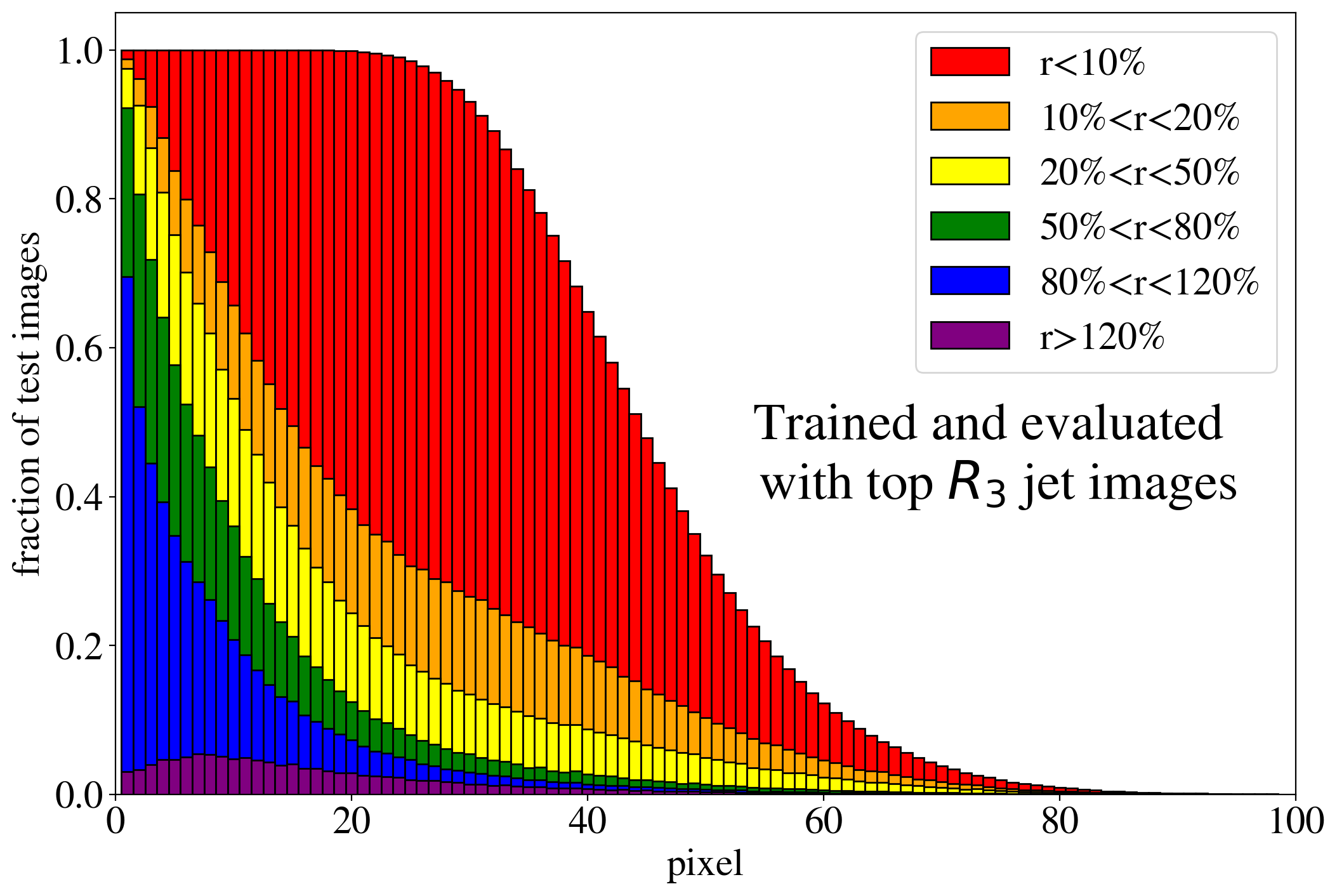}
	\includegraphics[width=0.24\linewidth]{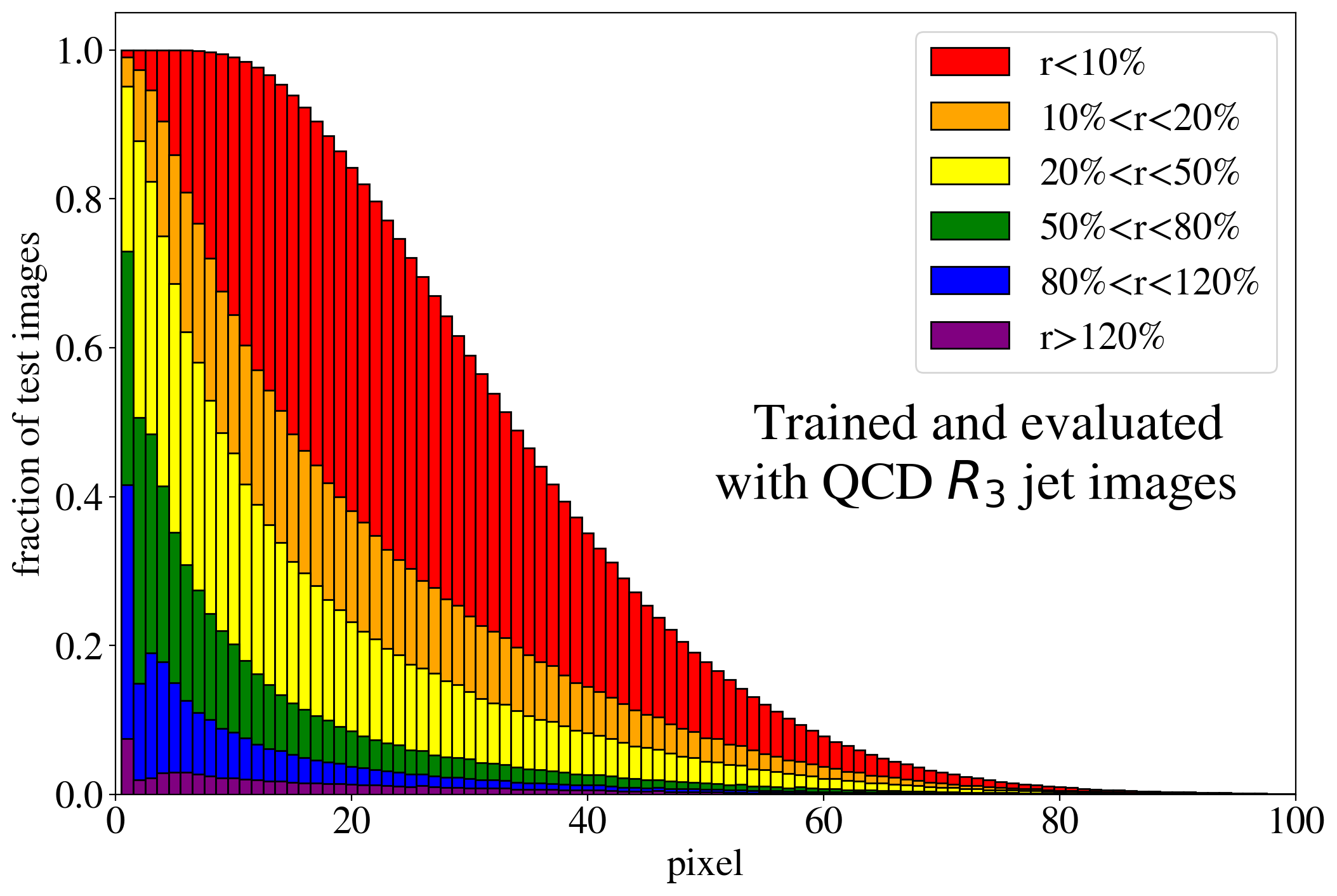}
	\includegraphics[width=0.24\linewidth]{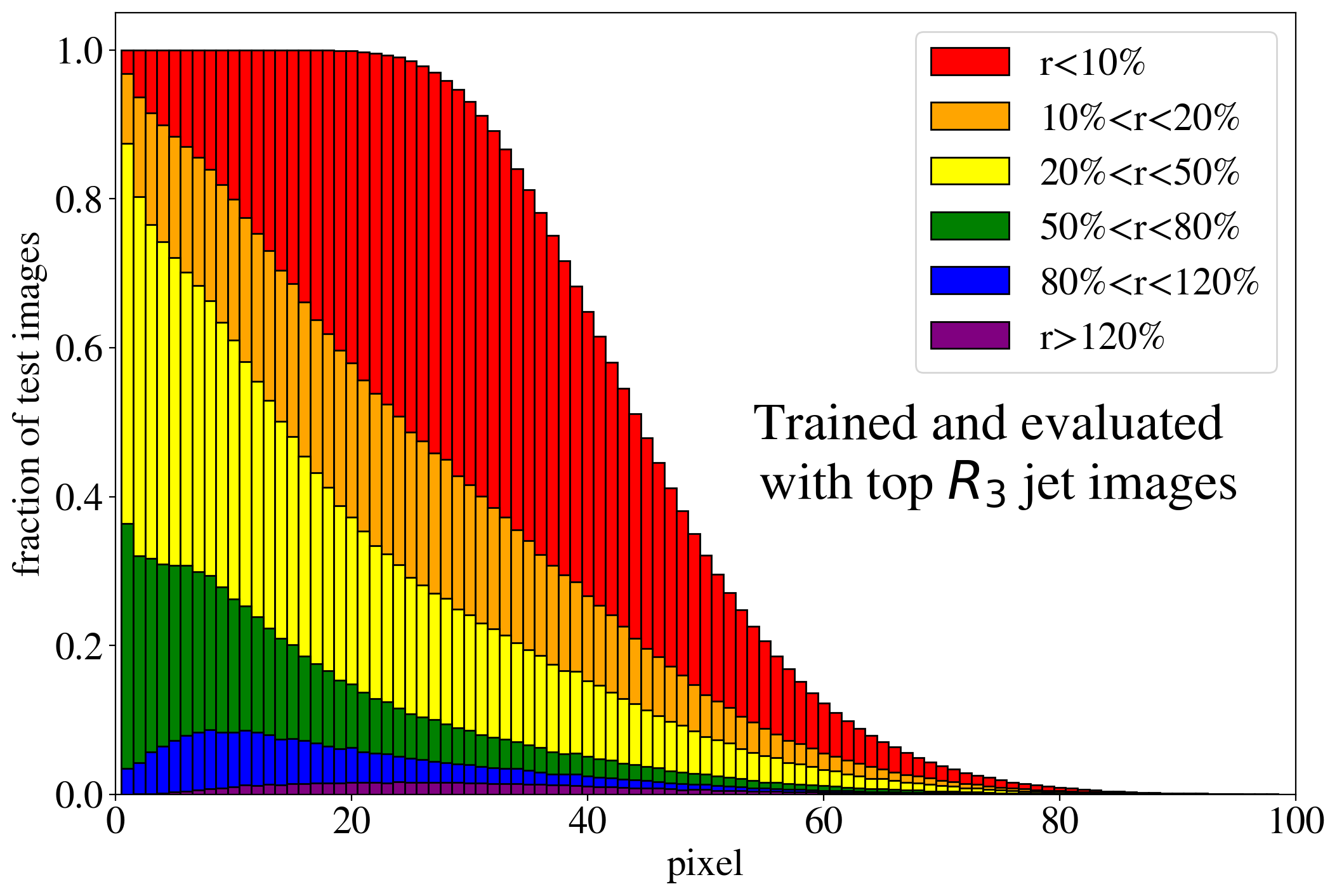}
	\caption{\label{fig:order_all_appendix} Stacked histogram for different categories of the ratio $r$ of the reconstructed and the input intensity of the non-zero pixels. Pixels are ordered by intensity from left to right for each of the 40k test jets, cf.\ Fig.~\ref{fig:igno_pix}. The first and second columns represent the results for an AE trained on QCD and top jet images, respectively, using MSE loss, while the third and fourth columns show the corresponding results using the KMSE loss function. The rows correspond to the different image remappings $\text{R}_0$, $\text{R}_1$, $\text{R}_2$, and $\text{R}_3$ from top to bottom.}
\end{figure}

In Figs.~\ref{fig:pixnum_all_appendix} and \ref{fig:order_all_appendix} we finally show the reconstruction loss for individual jet images as a function of the number of non-zero pixels
and the distribution of the ratio of reconstructed and input intensity for the leading pixels, respectively, for the different remappings and for the MSE and KMSE loss functions. 

\end{appendix}

\clearpage
\bibliographystyle{JHEP_improved}
\bibliography{bibliography.bib}

\end{document}